\RequirePackage{lineno}
\documentclass[11pt,tightenlines,floatfix,prd,epsfig,nofootinbib,
superscriptaddress,
bibtex,nolongbibliography]{revtex4}

\usepackage{enumitem}
\usepackage[normalem]{ulem}
\usepackage{array}
\newcolumntype{P}[1]{>{\raggedright\arraybackslash}p{#1}}
\usepackage{graphicx}
\usepackage{wrapfig}
\usepackage{amsmath}
\usepackage{amssymb}
\usepackage{mathptmx} 
\usepackage{hyperref}
\usepackage{pdfpages}
\usepackage{bm}
\usepackage{color}
\usepackage{xcolor}
\usepackage[title]{appendix}
\usepackage[utf8]{inputenc}
\usepackage{longtable}
\hypersetup{
    colorlinks,
    linkcolor={red!50!black},
    citecolor={blue!50!black},
    urlcolor={blue!80!black}
}
\usepackage{xspace}

\newcommand\redout{\bgroup\markoverwith
{\textcolor{red}{\rule[.4ex]{2pt}{1.2pt}}}\ULon}

\newcommand{\mee}       {$m_{\beta\beta}$}

\newcommand{\BBz}       {$0\nu\beta\beta$}

\newcommand{\BB}        {$\beta\beta$}

\newcommand{\qval}      {$Q_{\beta\beta}$}                
 
\newcommand{\thalf}     {\ensuremath{T_{1/2}}}

\newcommand{\nuc}[2]    {$^{#1}$\textrm{#2}} 

\newcommand{\LEG}       {LEGEND}
\newcommand{\Ltwo}      {{\LEG-200}}
\newcommand{\Lthou}     {{\LEG-1000}}

\newcommand{\MJ}        {\textsc{Majorana}}
\newcommand{\DEM}       {\textsc{Demonstrator}}

\newcommand{\Gerda}     {\textsc{Gerda}}

\newcommand{\GF}        {\textsc{Geant4}}

\newcommand{\be}        {\begin{equation}}
\newcommand{\ee}        {\end{equation}}

\begin{document}

\title{Fundamental Symmetries, Neutrons, and Neutrinos (FSNN):  \\[.3cm] 
Whitepaper for the 2023 NSAC Long Range Plan
\\[1.cm]}

\author{B.~Acharya}
\affiliation{Oak Ridge National Laboratory, Oak Ridge, TN, USA}

\author{C.~Adams}
\affiliation{Physics Division, Argonne National Laboratory, Lemont, IL, USA}

\author{A.~A.~Aleksandrova}
\affiliation{Kellogg Radiation Laboratory, California Institute of Technology, Pasadena, CA, USA}

\author{K.~Alfonso}
\affiliation{Department of Physics, Virginia Tech, Blacksburg, VA, USA}

\author{P.~An}
\affiliation{Department of Physics, Duke University, Durham, NC, USA}

\author{S.~Bae{\ss}ler}
\affiliation{University of Virginia, Charlottesville, Virginia, USA}
\affiliation{Oak Ridge National Laboratory, Oak Ridge, TN, USA}

\author{A.~B.~Balantekin}
\affiliation{University of Wisconsin, Madison, Madison, WI USA}

\author{P.~S.~Barbeau}
\affiliation{Department of Physics, Duke University, Durham, NC, USA}
\affiliation{Triangle Universities Nuclear Laboratory, Durham, NC, USA}

\author{F.~Bellini}
\altaffiliation{INFN Sezione di Roma, Sapienza University of Rome}

\author{V.~Bellini}
\affiliation{National Institute of Nuclear Physics - Sezione di Catania, Catania, Italy}

% ...


\maketitle
\newpage

\tableofcontents

\newpage

\section{Executive Summary}
\label{sect:executivesummary}
Through the exploration of fundamental symmetries, and by using nuclei, neutrons, and neutrinos, nuclear physics addresses some of the most profound questions in science.  
Why does the universe contain so much more matter than antimatter?  
Are neutrinos their own antiparticles
and where do their masses come from?
What objects make up the dark matter that is responsible for most of the universe's mass?  
Does nature contain more forces than the four we know about? 
Our Standard Model of nature's particles and forces is incomplete because it does not answer these questions; new physics, from beyond the Standard Model (BSM) is needed. 
With that physics not appearing at the high energy frontier, 
it has become imperative to realize the potential of the burgeoning program of 
precision nuclear-physics measurements.

Since the last long-range plan, research in fundamental symmetries, neutrons, and neutrinos has put us on a path toward answering some of these supremely important questions.  
To ensure US leadership in the enterprise and to increase the diversity of its workforce the community has come to a consensus on its priorities and made the following recommendations: \\

\noindent
{\bf    1) We recommend the timely construction of ton-scale neutrinoless double beta decay experiments, each using a different isotope, and continued support of the broader research program. }

\begin{itemize}
\item The US-led ton-scale program of discovery science could elucidate the nature of neutrinos, the origin of neutrino mass, 
and the mechanism for  generating   matter in the Early Universe. The 2021 portfolio review affirmed the readiness of CUPID, LEGEND, and nEXO experiments to proceed to construction.  Full realization of this program will require international partnerships.

\item A robust research program that includes ongoing efforts in theory and experiment as well as a diverse R\&D program into multiple promising isotopes and technologies is required both to support the ton-scale program and to prepare for experiments with sensitivity beyond the inverted mass ordering.
\end{itemize}

\noindent
{\bf 2)    We recommend a suite of targeted experiments aimed at challenging the Standard Model and uncovering new phenomena.}

Nuclear physics provides unique opportunities to probe the fundamental structure of the electroweak interaction, search for electric dipole moments with unprecedented sensitivity, and investigate neutrino masses and interactions. Realizing these compelling scientific opportunities requires that we

\begin{itemize}
\item  Expeditiously complete high-impact, larger-scale experimental campaigns: nEDM@SNS, the world’s most ambitious search for the neutron electric dipole moment (EDM), and MOLLER@JLab, soon to provide the most precise low energy measurement of a purely leptonic weak neutral current interaction.

\item Strengthen small and mid-scale university and laboratory programs, so that we fully profit from the BSM discovery potential of the precision frontier.   Investigations of non-unitarity in the quark-mixing matrix through neutron and nuclear beta decay and of new sources of time reversal violation through EDMs are high priorities.

\item Pursue emerging ideas, technologies and programs.  Innovative but still not fully realized projects include next-generation measurements of the absolute neutrino mass (Project 8), lepton flavor universality tests in the weak interactions (PIONEER), a search for new neutral current interactions (SoLID), and BSM searches enabled by FRIB and quantum sensing.
\end{itemize}

The success of this program demands robust research support commensurate with the tremendous discovery potential. 
FSNN science is broad and diverse, with unique needs. The community strongly
leverages facilities managed by other programs, offering an exceptional return on
investment. Without a central managing facility, however,  
deliberate efforts and measures  must be taken to protect the level of research support and other resources, such as
beamtime, devoted to mid- and small-scale FSNN projects. Erosion of research support delays science, which threatens US
leadership, and reduces the ability of the FSNN community both to respond to new ideas and to develop
the capabilities (such as brighter UCN sources and isotope harvesting at FRIB)
needed to push the precision frontier beyond this Long Range Plan.\\

\noindent
{\bf 3)  We recommend new investments aimed at enlarging and supporting the nuclear theory efforts in FSNN.}

Theory plays a central role in assessing the discovery potential of FSNN experiments, extracting their implications for fundamental physics, and developing new experimental directions. FSNN experiments require multi-scale theoretical analyses at energies that range from very small nuclear level splittings to the electroweak scale and beyond.  Such analyses require theoretical expertise in phenomenology, effective field theory, lattice QCD, and nuclear many-body physics. An enhanced theoretical research program is thus essential for taking full advantage of the exciting physics opportunities discussed in Recommendations 1) and 2).  

The following investments will have the highest impact on the health of the entire FSNN community: 

\begin{itemize}
\item  Increased support of collaborative efforts such as Theory Hubs, Topical Collaborations, and Physics Frontier Centers, to tackle the multi-scale  problems that pervade FSNN science.  

\item  The creation of a faculty bridge program administered by a national consortium to develop a diverse community of FSNN theorists, with procedures that create and sustain an equitable, welcoming, and inclusive culture.
\end{itemize}

\noindent
{\bf 4)  We recommend enhanced investment in the growth and development of a diverse workforce to maximize our opportunities for scientific discovery and increase its impact in society.}	

Recruiting and maintaining a diverse workforce requires treating all community members with respect and dignity.  Diversity leads to stronger teams.  The nuclear-physics research program plays an important role in developing a diverse STEM workforce for the critical needs of the nation.  FSNN science is a particularly strong training ground that allows students to take part in many kinds of experimental and theoretical work. A more diverse base can help us communicate the importance and excitement of nuclear research to the broader public. Creating and maintaining an inclusive, equitable, productive working environment for all members of the community is a necessary part of this development. 	 

\begin{itemize} 			
\item We recommend more resources and training programs to help the community recognize and reduce bias and establish enforceable conduct standards, supporting the recent initiatives by the APS and DNP.  The enforcement of such standards is the combined responsibility of all universities and laboratories, theoretical and experimental collaborations, conference organizers, and individual investigators supported by the nuclear physics research program. 

\item We recommend the development and expansion of programs that enable participation in research by students from under-represented communities and cultures, and by faculty from minority-serving institutions at national labs and/or research universities.
			 								
\item  We recommend the development and expansion of programs to recruit and retain diverse junior faculty and staff at universities and national laboratories through bridge positions, fellowships, traineeships, and other incentives. 

\item  We recommend that federal grants include resources to support living wages for graduate research assistants and postdocs.

\item We recommend increased resources for collecting data,  which will include the membership in our community and the career trajectories of our students and postdocs.  With these data, we can monitor progress and learn to attract a more diverse group of young researchers.
\end{itemize}

In addition to making the recommendations above, 
the FSNN community  has endorsed cross-cutting initiatives related to 
computational nuclear physics (see Section~\ref{sec:computing}),  
nuclear data (see Section~\ref{sect:ND}),  and  
quantum computing/quantum sensing (see Section~\ref{sect:QIS}).

The rest of this white paper provides context for our recommendations.

\section{Introduction \& scientific questions driving the field}
\label{sect:introduction}
The  Standard Model (SM)  of strong and electroweak interactions, while extremely successful,  
leaves a number of questions about the observed universe without 
answers. 
These include:  
Why are there more baryons than anti-baryons?
What is the origin and nature of neutrino masses? 
What is dark matter (and dark energy)? 
The SM also does not address theory-driven questions 
related to the large gap between the weak and the Planck scales, the absence of Charge-Parity (CP) violation in the strong interactions,  the origin of quark 
and lepton generations, and the possible unification of forces.  
The solutions to both observational and theoretical puzzles almost certainly require new particles and 
undiscovered interactions, a fact that  motivates searches for new fundamental physics across many energy scales. 
As the  2015 NSAC Long Range Plan~\cite{LRP2015} recognizes,
nuclear physics plays a prominent role in  this enterprise
through a  ``targeted program of fundamental symmetries and neutrino research that opens new doors to physics beyond the Standard Model (BSM)''.

BSM physics has escaped detection so far 
because the new particles are  either very heavy, or light and weakly coupled. 
There are two traditional routes in the search for new physics in laboratory experiments. 
One is to increase the  energy of particle accelerators  to  directly produce new particles 
-- working at what is known as the ``energy frontier". 
The other route is to  perform very precise and sensitive measurements 
in low-energy systems, working at the ``precision/intensity frontier'' to indirectly access the virtual exchange of heavy particles 
or directly excite light and weakly coupled particles. 
Searches at both frontiers are  needed  to discover and unravel 
the underlying new dynamics.  An example:
current high-energy colliders are the most powerful direct  probe of 
new particles with masses near the TeV scale, possibly associated with the electroweak symmetry breaking mechanism,  but 
precision frontier experiments can indirectly access even higher mass scales and  provide the strongest probes of 
lepton (L) and baryon (B) number violation, 
CP violation,   flavor violation in the quark and lepton sectors, neutrino properties,  and dark sectors.

Nuclear-science research in fundamental symmetries, neutrons, and neutrinos (FSNN) 
plays a prominent role at the precision/intensity frontier. 
The research involves many probes that 
naturally fall into three classes, with each pushing the boundary of BSM sensitivity in a qualitatively different way and at a different mass scale:

\begin{itemize}

 \item {\it Searches for rare or SM-forbidden processes} that break approximate or exact symmetries of the SM. These include  
neutrinoless double beta  ($0\nu \beta \beta$) decay; 
permanent electric dipole moments (EDMs) of the neutron, atoms, and molecules;  
$\mu \to e$ conversion in nuclei;
neutron-antineutron oscillations; 
and searches of time-reversal (T) violation in neutron processes. 
Typically, these experiments probe very high mass scales (for example EDMs probe physics up to $10^3$ TeV and $0\nu \beta \beta$ decay reaches 
$10^{12}$ TeV)  and  also have  sensitivity to dark sectors. 
A discovery in any of these searches would be paradigm-shifting.

\item  {\it High precision measurements of SM-allowed processes}, for example 
$\beta$-decay (of mesons, the neutron, and nuclei), parity-violating electron scattering, and the muon lifetime and anomalous magnetic moment.  
Measurements of these processes probe both new physics at 10--100 TeV and light masses, 
and they become powerful discovery tools when sufficiently precise predictions of the SM are available.

\item  {\it Experiments that explore properties of known and hypothetical 
light weakly-coupled particles} such as active neutrinos (through absolute mass measurements and neutrino scattering), sterile neutrinos,  axions, dark photons,  etc. 
These experiments provide powerful ways to explore BSM dark sectors. 

\end{itemize}

\begin{figure}[t]
\centering
\includegraphics[width=.9\textwidth]{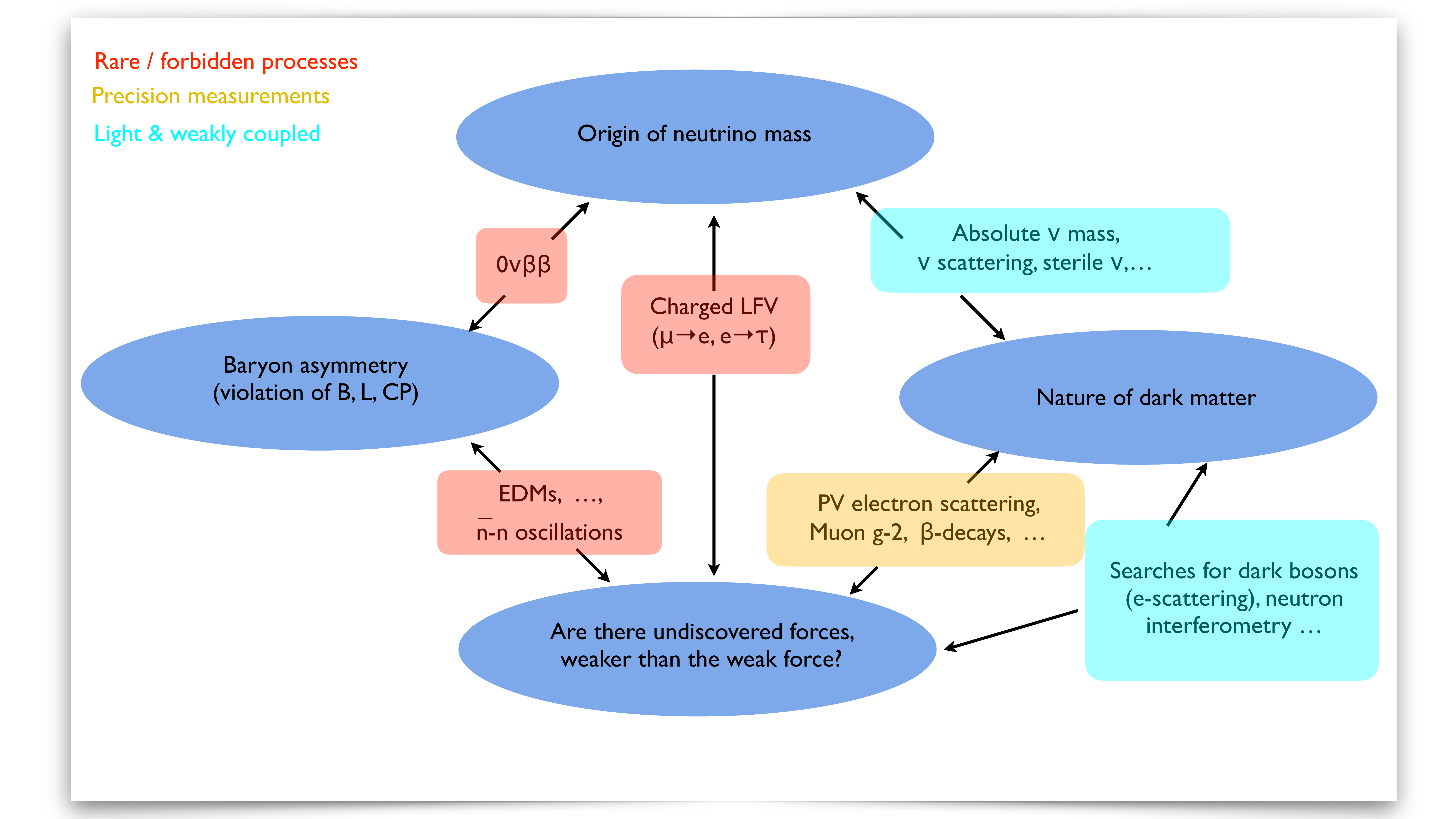}
\caption{The Nuclear Science ``targeted program" of research in Fundamental Symmetries, Neutrons, and Neutrinos 
addresses four interconnected questions about fundamental interactions and the observed universe.}
\label{Fig:figQ}
\end{figure}

In summary, then, FSNN experiments and the theory required to interpret them are a crucial part of the drive to find new physics and explain the universe we inhabit.  Our field has great discovery potential; in some cases, e.g.\ in the effort to understand the symmetry breaking needed for baryogenesis and in the search for ultralight and dark particles, FSSN experiments allow us to explore physics that is beyond the reach of colliders.  And by combining experiments we can increase their power.  Together, for example, experiments on the EDMs of the neutron and a variety of atoms and molecules tell us much more about CP violation than any of the experiments would alone.  Figure \ref{Fig:figQ} illustrates the enormous scientific questions that our field seeks to address and the experimental programs that connect them.  The importance of the questions and the proven ability of our community to creatively address them make it vital that this enterprise be supported.

One of the key strengths of the FSNN program is its interdisciplinary nature and strong connections to other fields in science, in particular in high energy physics and atomic and molecular physics in both theory and experiment. The FSNN field will continue to rely on and contribute to advances in exascale computing as well as Artificial Intelligence and Machine Learning techniques, cutting edge quantum information science and sensing technologies, the production of isotopes at nuclear accelerators, and support for the availability of nuclear data. This field broadly makes excellent use of facilities built for other purposes, including underground facilities, accelerators for other fields in nuclear physics, and neutron sources. While this approach is very economical, support is needed to ensure FSSN research is a priority and can be successfully executed at these facilities. Finally, the field of FSNN is an exceptional training ground for the next generation of scientists due to the broad skillsets needed for exploratory research, and there are real opportunities to expand and diversify the pool and ensure equity in opportunities for every member of the community. 

Next we summarize the achievements of the FSNN community in the previous LRP period, and overview the research opportunities that have been identified for the coming decade. 
Subsequent sections of this white paper will describe the various components of the FSNN research portfolio in more detail, highlighting the discovery potential of the experimental probes, their power in combination, and the theoretical input needed to maximize their impact. Discussion of cross-cutting initiatives and needs to support this research and its community follow.

\section{Progress since the last Long Range Plan}
\label{sect:progress}
Since the last Long Range Plan, the Fundamental Symmetries community has improved our understanding of the Standard Model and laid the groundwork for discovery in the next decade. These accomplishments, which span the full extent of our diverse field, are summarized below.

\subsection{Searches for neutrinoless double beta decay}

The discovery of neutrinoless double beta decay would result in a fundamental shift in our understanding of neutrinos and of the creation of matter in the 
Universe.   Experimental groups in many countries are looking for the decay.
Following the release of the 2015 LRP, a 
subcommittee report to NSAC~\cite{NSAC-BB-Report-2015} listed 
recommendations related to R\&D 
for some of the key US experimental programs and indicated goals they should accomplish. 
These goals and others have now been achieved. 

\begin{itemize}

\item
Half life limits now exceed $10^{26}$~yr, ten times longer than those existing in 2015. The constraints on \mee\ now reach near the top of the inverted-ordering mass region and, for some isotopes and nuclear matrix element calculations, even extend a bit into that region. 

\item The CUORE~\cite{ADAMS2022CUORE}, EXO-200~\cite{Anton_2019EXO200}, \Gerda~\cite{Agostini_2020GERDA}, KamLAND-Zen~\cite{KamLAND-Zen2022}, \MJ\ \DEM~\cite{Arnquist2022MJDzero}, and  NEXT~\cite{NEXT2nu2022} projects have established experimental programs demonstrating that experiments at the ton scale are feasible. 

\item   \Ltwo~\cite{LEGEND-pCDR} is taking data at LNGS. 

\item CUPID-Mo~\cite{CUPID-MOzero, CUPID-MOzero2022} and CUPID-0~\cite{CUPID-0Se} demonstrated energy resolution, radio-purity, and alpha rejection of scintillating bolometers.

\item
SNO+ has measured all of its detector-related 
backgrounds~\cite{SNO:2022trz,Inacio:2022vjt} and shown in bench top studies that it can load up to 3\% Te by mass in its scintillator with an acceptable light yield\cite{Auty:2022lgh}. 

\item SuperNEMO~\cite{Arnold_2010} has operated its demonstrator. 

\item Both the nEXO~\cite{nEXOBaTag2019} and NEXT~\cite{NEXTBaTag2019,NEXTBaTag2021} collaborations 
made substantial progress in isolating and detecting a lone Ba ion within a dense Xe environment.

\end{itemize}

The  DBD Topical Theory Collaboration~\cite{DBD} 
led to concerted theoretical effort in $0\nu \beta \beta$ decay, 
involving theorists with expertise in phenomenology, effective field theory (EFT), lattice QCD, and nuclear structure. 
Much of the US-led progress in {\it ab-initio} 
matrix elements is linked to this Topical Collaboration: 

\begin{itemize} 
 \item 
 EFT methods for lepton number violation (LNV) beyond the Standard Model~\cite{Cirigliano:2018yza} 
 and nuclear operators \cite{Cirigliano:2018hja,Cirigliano:2020dmx}, 
lattice QCD computations of pion-level matrix elements from TeV-scale LNV~\cite{Nicholson:2018mwc,Feng:2018pdq,Tuo:2019bue,Detmold:2022jwu}, and
 {\it ab initio} $0\nu\beta\beta$ nuclear-matrix-element calculations~\cite{Yao:2019rck,Belley:2020ejd,Novario:2020dmr,Yao:2020olm,Wirth:2021pij} all progressed tremendously.

\item There was great progress in the theory and phenomenology of leptogenesis mechanisms 
and in the simultaneous analysis of cosmological data, collider data, and $0\nu \beta \beta$ 
decay~\cite{Deppisch:2017ecm,Li:2020flq,Li:2021fvw,Harz:2021psp,Graesser:2022nkv}.
  
\end{itemize}

\subsection{Searches for electric dipole moments}
\label{sec:prog:edm}

A permanent electric dipole moment of a particle or system would imply the presence of a new source of CP violation, which could explain the matter-antimatter asymmetry in the Universe.

\begin{itemize}
 \item The nEDM@SNS experiment, which will use unique cryogenic techniques to make the most precise search for the neutron's EDM, moved from R\&D to construction of the apparatus, starting with the cryostats and the magnetic field system. Assembly and testing has now begun at ORNL’s SNS~\cite{Ahmed2019}.

\item The LANL nEDM experiment achieved the polarized UCN density required for goal sensitivity~\cite{Ito:2017ywc,Wong:2022ixk}.  A magnetically shielded room  was installed and the magnetic fields characterized. Precession chambers, electrodes, and UCN valves are ready and magnetometers are under development.

 \item Numerous atomic EDM experiments, using methods ranging from vapor cells to optical lattices, improved sensitivity to hadronic CP-violation via nuclear Schiff moments in atoms such as $^{199}$Hg~\cite{Graner:2016ses}, $^{225}$Ra~\cite{Parker2015RaEDM,Bishof2016RaEDM}, and $^{129}$Xe~\cite{Allmendinger2019XeEDM,Sachdeva2019XeEDM}, and a new experiment reported a limit on the $^{171}$Yb EDM~\cite{Zheng2022YbEDM}.  

 \item Work with radioactive pear-shaped nuclei, which are extremely sensitive to hadronic CP violation, has made major progress~\cite{RadMolWhitepaper2023}: the Ra EDM work mentioned above, the first 
 spectroscopy on a radioactive molecule, RaF~\cite{GarciaRuiz2020Nature}, and the first  
 control of radium-containing molecular ions~\cite{Fan2021RaPoly,Yu2021RaOCH3}.
 
 \item Limits on the electron EDM were improved by an order of magnitude by the ACME~\cite{ACME2018} and JILA~\cite{Cairncross2017PRL,Roussy2022HfFEDM} experiments, which leverage internal molecular electric fields.
 The YbF~\cite{Hudson2011YbFEDM} and NL-EDM eEDM~\cite{Aggarwal2018BaF} experiments 
 made major 
 improvements in laser-cooling~\cite{Alauze2021YbFUpgrade} and trapping~\cite{Aggarwal2021SrFTrap}.  

 \item
 Atomic electron-EDM experiments with Cs~\cite{Tang2018EDMSublevels} and Fr~\cite{Hayamizu2022FrEDM} continued their  
 push to leverage 
 quantum science methods.  
 Several new molecular approaches are 
 under development, including laser-cooled polyatomics~\cite{Kozyryev2017PolyEDM,Anderegg2023CaOHSP} and matrix-isolated diatomics~\cite{Li2022BaFEDM3}.
 
 \item Molecular eEDM methods are being expanded to search for 
 hadronic CP violation, both through 
 nuclear Schiff moments and magnetic quadrupole moments, in several active experiments, including CeNTREX~\cite{Grasdijk2021TlF}, YbOH~\cite{Kozyryev2017PolyEDM,Pilgram2021YbOHOdd}, and YbF~\cite{Ho2023YbFMQM} and several others in initial stages of development.

\item 
The phenomenology of EDMs was connected with 
physics at the energy frontier \cite{Brod:2013cka,Cirigliano:2016nyn,Cirigliano:2019vfc,Brod:2022bww}. The ways in which the EDM program and LHC complement each other in exploring the origin of CP violation
and the Universe's matter-antimatter asymmetry are now much better understood~\cite{Gritsan:2022php,Ramsey-Musolf:2019lsf,Wang:2022dkz}.

\item 
Lattice QCD calculations of the nucleon EDM 
have appeared \cite{Bhattacharya:2015esa,Gupta:2018lvp,Dragos:2019oxn,Alexandrou:2020mds,Bhattacharya:2021lol,Liang:2023jfj,Abramczyk:2017oxr,Bhattacharya:2022whc,FlavourLatticeAveragingGroupFLAG:2021npn}, paving the way for results with quantified uncertainties. 
At the nuclear level, we have new Schiff moment 
computations \cite{Engel:2013lsa,Dobaczewski:2018nim,Yanase:2020agg}.  Progress in \textit{ab initio} techniques promises 
\textit{ab initio} calculations of Schiff moments soon 
\cite{Alarcon:2022ero}.

\end{itemize}

\subsection{Parity-violating electron scattering}
\label{sect:pvesh}

Parity violation in electron-nucleon scattering is a powerful tool for both for uncovering BSM physics and for examining nuclei. 

\begin{itemize}
  \item Qweak at Jefferson Lab (JLab) carried out a high-precision elastic electron-proton scattering parity-violating (PV) asymmetry measurement~\cite{Qweak:2018tjf}, providing the most precise low-energy determination of the weak mixing angle and setting constraints on new semi-leptonic multi-TeV scale PV physics.
  \item Also at JLab, elastic electron-nucleus PV asymmetry measurements (PREX (Pb-208)~\cite{PREX:2021umo} and CREX (Ca-48)~\cite{CREX:2022kgg}) provided the most accurate constraints on 
  neutron skins, challenging models of neutron-rich matter and facilitating the next-generation experiments MOLLER and SoLID.
\end{itemize}

\subsection{Precision beta decay with nuclei}

Nuclear beta decay provides precision tests of the Standard Model and probes BSM physics.

\begin{itemize}
    
  \item The $0^+ \rightarrow 0^+$ superallowed-beta-decay data set was refined.  Updated theoretical corrections revealed some tension in CKM unitarity and tightened constraints on exotic scalar 
  currents~\cite{Hardy:2020qwl}.
  
  \item There was progress in tests of CKM unitarity in mirror nuclei, in half-lives ($^{37}$K~\cite{Shidling2014} and $^{21}$Na~\cite{Shidling2018} at TAMUTRAP, $^{25}$Al~\cite{Long2017}, $^{11}$C~\cite{Valverde2018}, $^{13}$N~\cite{Long2022}, $^{15}$O~\cite{Burdette2020},  and $^{29}$P~\cite{Long2020} at Notre Dame);  $Q_{EC}$-values ($^{11}$C~\cite{Gulyuz2016}, $^{21}$Na, and $^{29}$P~\cite{Eibach2015} at NSCL); and $\beta$-asymmetries ($^{37}$K ~\cite{Fenker2018} with TRINAT).

  \item Cyclotron Radiation Emission Spectroscopy (CRES), which promises dramatic improvements in sensitivity to exotic couplings through precision spectroscopy, was demonstrated in $^{6}$He and $^{19}$Ne~\cite{Byron:2022wtr}. 
  
  \item High-precision angular-correlation measurements to improve limits on exotic scalar and tensor currents were performed in $^{8}$Li~\cite{sternberg2015,burkey2022} and $^{8}$B at the BPT at Argonne's ATLAS facility~\cite{gallant2022}, in $^{6}$He at Washington~\cite{mueller2022}, and in $^{37}$K at TRINAT~\cite{Fenker2018}.
  
 \item   A new dispersion-theoretical calculation of the  inner radiative corrections in neutron and nuclear beta decay 
 reduced the associated theoretical uncertainty~\cite{Seng:2018yzq}. 
New  nuclear-structure-dependent effects were discovered~\cite{Gorchtein:2018fxl}.

\end{itemize}
  
\subsection{Precision beta decay with neutrons}

The neutron is the simplest nucleus that undergoes beta decay and provides a particularly clean laboratory for BSM searches.  

\begin{itemize}
  \item The UCN$\tau$ collaboration performed the most precise measurement of the free neutron lifetime at LANL~\cite{UCNt:2021pcg} and set new limits on neutron dark decay~\cite{Tang:2018eln}, leveraging improvements at what is now one of the world's brightest UCN sources~\cite{Ito:2017ywc}.

  \item The UCNA collaboration published its final results for the neutron $\beta$-asymmetry, measured with UCN, thereby resolving tension among previous measurements~\cite{UCNA:2017obv}. It also extracted first and improved limits on the BSM Fierz term ~\cite{Hickerson:2017fzz, UCNA:2019dlk} and set new limits on dark neutron decay~\cite{UCNA:2018hup}.

\item The aCORN experiment at NIST completed two runs and published a new measurement of the electron-antineutrino  angular correlation ($a$  coefficient) in free neutron  $\beta$-decay with final uncertainty 1.7\%~\cite{Darius:2017arh, Hassan:2020hrj}.

  \item The RDK II experiment completed a second run, following the first observation of radiative neutron decay with improved measurements of the branching ratio and a first precision measurement of the photon energy spectrum~\cite{RDKII:2016lpd}.
  
  \item A new beam-based measurement of the neutron lifetime, the beam lifetime (BL2) experiment at NIST, has begun data-taking, with a sensitivity sufficient to provide critical input with respect to the current discrepancy between beam and bottle measurements.  

  \item The Nab experiment has begun commissioning at the Spallation Neutron Source~\cite{Fry:2018kvq}.

  \item Lattice QCD calculations of the axial-to-vector coupling ratio reached  
  percent-level precision ~\cite{Chang2018,Walker-Loud2020} and new radiative corrections to this ratio were identified 
  \cite{Cirigliano:2022hob}.    At about the same time, the origin of the so-called quenching of the axial coupling in nuclei was found~\cite{Pastore:2017uwc,Gysbers:2019uyb}.
  
\end{itemize}

\subsection{Precision Measurements with muons and mesons}

Precision muon and pion experiments are used to search for charged lepton flavor violation, test lepton flavor universality, and search for BSM corrections to muon properties.

\begin{itemize}
    \item The Muon g-2 Experiment is completing its 6th year of data taking. First results~\cite{Muong-2:2021ojo,Muong-2:2021ovs,Muong-2:2021vma,Muong-2:2021xzz}
    have confirmed the previous measurement from BNL~\cite{Muong-2:2006rrc}; when the two are combined they are in tension with the Standard Model at $4.2\,\sigma$.  The experiment will meet its final precision goal of 140 ppb. 
\end{itemize}

\subsection{Hadronic Parity and Time-Reversal Violation}

Hadronic parity violation provides a unique probe of the weak interaction. 
Searches for time-reversal violation complement EDM experiments on CP violation. 

\begin{itemize}
  \item NPDGamma has reported the first observation of the parity violating correlation between the neutron spin and the $\gamma$ ray emitted from proton capture~\cite{NPDGamma:2018vhh}. The n-$^3$He collaboration has performed precision measurements of the parity-odd asymmetry in neutron capture on $^3$He~\cite{n3He:2020zwd}.
  \item The Neutron Spin Rotation experiment completed the most sensitive search for parity violating neutron spin rotation in $^{4}$He~\cite{Swanson:2019cld}.  At its current level of sensitivity the null result is consistent with theory, but paves the way for a new measurement that will yield the first non-zero observation.
  \item The ZOMBIES experiment has performed a proof-of-principle demonstration of a measurement of anapole moments in Z$\sim$40 nuclei~\cite{Altuntas:2018ots}.
   \item  A new theoretical paradigm, based on the $1/N_C$ expansion,   has emerged for hadronic parity nonconservation and the interpretation of experiments in the field~\cite{Phillips2015,Schindler2016,Gardner:2017xyl}.
   \end{itemize}

\subsection{Searches for neutron oscillations}

Neutron oscillations violate baryon number, and their discovery would provide a mechanism for matter, but not antimatter, to evolve in the early universe.

\begin{itemize}
 \item A search for baryon number violation (BNV) via neutron oscillations into mirror neutrons has ruled out an exotic explanation for the neutron lifetime discrepancy~\cite{Broussard:2021eyr}. This is the first step in a staged program toward a high sensitivity search for  neutron-antineutron oscillations in NNBAR~\cite{Addazi:2020nlz}.
\end{itemize}

\subsection{Neutrino mass and sterile neutrinos}

We still don't know the masses of the three flavors of neutrinos.  Sterile neutrinos could be the source of anomalies in neutrino-oscillation experiments and provide a component of the universe's dark matter. 

\begin{itemize}
  \item The KATRIN experiment has produced the world-leading limit on the neutrino mass scale~\cite{KATRIN:2021uub} and is continuing data-taking toward a mass sensitivity goal of $\sim$0.2~eV, and has set competitive limits on eV-scale sterile neutrinos~\cite{KATRIN:2020dpx, KATRIN:2022ith}.
  \item The Project 8 collaboration extracted a first limit on the neutrino-mass scale by using the promising CRES technique~\cite{Project8:2022hun}. 
  \item The BeEST experiment currently sets the most stringent laboratory limits on sub-MeV sterile neutrinos leveraging superconducting tunnel junction sensors ~\cite{Friedrich:2020nze}.
  \item A program of Penning-trap measurements has either ruled out or tentatively confirmed more than a dozen candidates for ultra-low-Q-value nuclear decays of possible interest for neutrino-mass measurements~\cite{Keblbeck:2022twm}.
  \item HUNTER has been fully simulated and major components are being fabricated and assembled including vessel, extreme ultrahigh vacuum pump system, loading magneto-optical trap (MOT), orthotropic oven, x-ray detector, spectrometers, and MOT and electron spectrometer coils~\cite{WOS:000623186300001}. 
\end{itemize}

\subsection{Neutrino interactions}
Studies of neutrino scattering from nuclei provides important input relevant to studies of neutrino oscillations, astrophysical neutrinos, and BSM searches. 
\begin{itemize}
  \item The COHERENT experiment has measured coherent elastic neutrino-nucleus scattering (CEvNS) in CsI~\cite{COHERENT:2017ipa} and Ar~\cite{ COHERENT:2020iec}, and established a program of ongoing CEvNS and inelastic measurements on multiple targets.
  \item There was significant improvement in the {\it ab-initio} calculation of neutrino-nucleus scattering 
on a variety of light nuclei~\cite{Lovato:2020kba}. 

\end{itemize}

\subsection{Neutrinos in astrophysics and cosmology } 
While not strictly nuclear physics, studies of neutrinos produced in the sun or from the cosmos provide relevant input. 

\begin{itemize}
  \item Cosmological probes have constrained the sum of the neutrino mass with an uncertainty approaching the 0.1 eV scale (see, e.g., \cite{Abazajian:2022ofy} for a review), and the number of effective neutrino species with an uncertainty of 10\% or less \cite{Planck:2018vyg,ACT:2020gnv,SPT-3G:2021wgf}.
  \item High-energy extragalactic neutrinos observed at IceCube have provided many new constraints on neutrino properties, most notably through a measurement of the neutrino-nucleon cross section at $\sim$0.1–1\,TeV of center-of-mass-frame energy \cite{IceCube:2017roe,IceCube:2020rnc}. 
\end{itemize}

\subsection{Other precision measurements}
Beyond these targeted searches, the field of FSNN includes a broad program to look for BSM physics using low energy precision techniques. 
\begin{itemize}
 \item Neutron Pendell\"osung interferometry was used for the first time to make a new measurement of the neutron charge radius, measure lattice dynamics, and place new limits on fifth forces~\cite{doi:10.1126/science.abc2794}.
\end{itemize}

\section{Synopsis of current and future projects}
\label{sect:synopsis}

The next LRP period promises exciting opportunities to address some of the most compelling questions in physics, and additional support is needed to maximize the impact of the US FSNN program. 
In Table~\ref{tab:synoppsis} we summarize the projects and initiatives advanced for consideration at the FSNN Town Meeting.  More detail on the motivation and opportunities that form the basis of our recommendations follow in subsequent sections. 

Each project corresponds to one of various stages as of January 1, 2023:  
\begin{itemize}
\item Concept (a general idea exists and is being investigated with theory and simulations), 
\item R\&D (specific aspects are being explored in hardware and/or design is underway), 
\item Planning (a design has been settled on and prepared for a proposal), 
\item Preliminary Design (project officially started, in design phase before construction), 
\item Construction (funding and building), and 
\item Data Taking (commissioning included). 
\end{itemize}
For projects in the DOE Critical Decision (CD) process, the current phase and phase being prepared for are indicated.  The Comment column can include sources of funding already received or where funding is being requested. The Request column indicates the funding category: 
\begin{itemize}
\item Project funding for projects in a DOE CD phase, 
\item Operations for projects undergoing/completed construction and needing an operations budget (including new or anticipated normal operating budgets), 
\item Research funding for base DOE/NSF funds, or 
\item R\&D to capture requests for bigger equipment/demonstrators/large scale R\&D.
\end{itemize}
The location indicates where the experiment will ultimately be staged or multiple sites if not yet determined. 

\footnotetext[1]{Multiple possible locations including LNGS, SURF, CJPL, and SNOLAB}

\setlength{\tabcolsep}{9pt}
    \begin{longtable}{ P{0.16\textwidth} P{0.12\textwidth} P{0.11\textwidth} P{0.135\textwidth} P{0.15\textwidth} P{0.08\textwidth} } 
    \hline\hline
   Project/Initiative & Status (Jan\,1,\,2023) & CD Phase current; prep & Comment & Request & Location \\
    \hline
    \multicolumn{6}{l}{TON SCALE NEUTRINOLESS DOUBLE BETA DECAY} \\
    CUPID & Prelim design & CD-0; CD-1 & DOE & Project funding & LNGS \\ 
    LEGEND-1000 & Prelim design & CD-0; CD-1/3A & DOE + NSF & Project funding & SNOLAB or LNGS \\
    nEXO & Prelim design & CD-0; CD-1 & DOE & Project funding & SNOLAB \\
    \multicolumn{6}{l}{BEYOND TON SCALE NEUTRINOLESS DOUBLE BETA DECAY DURING THIS LRP PERIOD} \\
    CUPID-1T & R\&D & & DOE+NSF & R\&D & Multiple\footnotemark[1] \\ 
    \textsc{Eos}@SNS & Planning & & DOE-NNSA-DNN/DOE-SC & R\&D & SNS \\
    NEXT & R\&D, Construction & - ; CD-0 & DOE (demonstrator) & R\&D, Project funding & SURF/LSC/ SNOLAB \\ 
    Selena & R\&D & & DOE + NSF & R\&D & LSC \\ 
    SNO+ 3\% & Planning & & DOE/NP & Research funding & SNOLAB \\ 
    Theia & R\&D & & NSF + DOE & R\&D & SURF \\
    \hline
    \multicolumn{6}{l}{ELECTRIC DIPOLE MOMENTS} \\
    nEDM at LANL & Construction & & LDRD + NSF + DOE & Research funding  & LANSCE \\
    nEDM@SNS & Construction & & DOE + NSF & Research funding & SNS \\ 
    Ra EDM & R\&D & & DOE NP & Operations & ANL \\ 
    SLAM/Pear Factory & R\&D, Planning & & U.S-led & Research funding & FRIB \\
    \hline
    \multicolumn{6}{l}{PARITY VIOLATING ELECTRON SCATTERING} \\
    MOLLER & Construction & CD-1; CD-2/CD-3 & DOE MIE + NSF MidScale + CFI & Operations & JLab \\
    SoLID & Planning & - ; CD-0 & & Project funding & JLab \\
    \hline
    \multicolumn{6}{l}{PRECISION BETA DECAY WITH NUCLEI} \\
    BPT & Data Taking & & DOE & Operations & ANL \\
    He6-CRES & R\&D & & DOE & Operations & UW \\
    SALER  & Construction &  & DOE & Operations & FRIB \\
    St.\ Benedict & Construction & & NSF & Operations & UND \\
    TAMUTRAP/ LSTAR & Data Taking; Construction & & DOE & Operations & TAMU \\
    \hline
    \multicolumn{6}{l}{PRECISION BETA DECAY WITH NEUTRONS} \\
    BL2 & Data Taking & & DOE + DOC + NSF & Operations & NIST \\ 
    BL3 & Construction &  & NSF & Operations & NIST \\ 
    Nab / pNab & Data~Taking &  & DOE + NSF & Operations & SNS \\
    Space-based lifetime & R\&D &  & NASA + DOE & R\&D & APL \\ 
    UCNA+ & R\&D & & LDRD & Research funding & LANSCE  \\  
    UCNProBe & R\&D & & LDRD+DOE & Research funding & LANSCE \\
    UCN$\tau$+ & R\&D & & LDRD & Research funding & LANSCE  \\ 
    \hline
    \multicolumn{6}{l}{PRECISION MEASUREMENTS WITH MUONS AND MESONS} \\
    Muon g-2 & Data~Analysis &  & DOE-HEP NP/NSF & Research funding & Fermilab  \\ 
    PIONEER & R\&D &  & DOE-HEP NP/NSF & R\&D & PSI \\
    \hline
    \multicolumn{6}{l}{HADRONIC PARITY AND TIME-REVERSAL VIOLATION} \\
    NOPTREX & Construction &  & NSF + DOE + Japan + China & R\&D & JPARC \\ 
    NSR & Construction & & NSF + DOC & Operations & NIST \\
    TREK & R\&D &  & Japan + DOE + NSF & Research funding & JPARC \\
    \hline
    \multicolumn{6}{l}{NEUTRON OSCILLATIONS} \\
    HIBEAM & R\&D & & DOE + EU + Sweden & R\&D & ESS \\
    NNBAR & Concept & & DOE + EU + Sweden & R\&D & ESS \\
    ORNL $nn'$ & R\&D &  & DOE + Sweden & Research funding & HFIR \\
    \hline
    \multicolumn{6}{l}{NEUTRINO MASSES AND STERILE NEUTRINOS} \\
    KATRIN & Data Taking & & DOE NP + Germany & Operations & KIT \\ 
    Project 8 & R\&D & & DOE NP & R\&D & UW \\ 
    BeEST & Data Taking &  & DOE NP & Research funding & LLNL \\
    SuperBeEST & R\&D &  & DOE NP & R\&D & TBD \\
    HUNTER & Construction &  &  NSF + BSF + ISF + private & Operations & UCLA \\
    \hline
    \multicolumn{6}{l}{NEUTRINO INTERACTIONS} \\
    COHERENT & Data~Taking, Construction & & HEP + NP + NSF + Korea & Research funding & SNS \\
    COHERENT@STS & Planning & & & Research funding & SNS \\
    \hline
    \multicolumn{6}{l}{DARK SECTOR SEARCHES} \\
    DarkLight & Construction &  & DOE + NSF + Canada & Research funding & TRIUMF \\ %
    \hline
    \multicolumn{6}{l}{THEORY} \\
    Bridge programs &  Planning &   & DOE & & Universities \& Labs \\ 
    \hline
    \multicolumn{6}{l}{FACILITIES AND CAPABILITIES} \\
    New UCN Source & Concept & & & R\&D & TBD \\
    CN interferometry & Data taking & & DOE + DOC & Operations & NIST \\ 
    UCN interferometry & R\&D & & DOE & Research funding & LANSCE \\ 
    \hline \hline
    \caption{Current and planned projects and initiatives in the Fundamental Symmetries, Neutrons, and Neutrinos Community}
    \label{tab:synoppsis}
    \end{longtable}

\section{Recommendation I:  Lepton Number Violation and Neutrinoless Double Beta Decay}
\label{sect:LNV}
The discovery of neutrinoless double beta (\BBz) decay would reshape our fundamental understanding of neutrinos and of matter in the Universe. 
The search for \BBz\ decay tests whether there is a fundamental symmetry of Nature associated with lepton number, probes the quantum nature of neutrinos, and allows the measurement of their effective mass. It is the only practical way to demonstrate if neutrinos are their own antiparticles, that is, if neutrinos have a Majorana mass. 
The discovery of Majorana neutrinos would open the door to new physics beyond the discovery of neutrino oscillation, and would signify a paradigm shift in our understanding of the origins of mass and matter. The neutrino's non-zero mass impacts the evolution of the Universe from the beginning of time to the formation of large-scale structures in the present epoch, and Majorana neutrinos 
play a key role in scenarios that explain the matter-antimatter asymmetry in the Universe. 

In the following sections we articulate the science case for a multi-experiment ton-scale experimental campaign, 
R\&D for beyond ton-scale experiments, and associated theoretical research.
We draw heavily from the community \BBz\  whitepaper \cite{Adams:2022jwx}, where details and references can be found.

\subsection{Significance of Research}
In \BBz\ decay two neutrons convert into two protons while emitting two electrons and no neutrinos, thus changing the number of leptons by two units. Since lepton number $L$ (or more precisely, the difference between $L$ and baryon number $B$) is conserved in the Standard Model, observation of \BBz\ decay  would be direct evidence of new physics and  would demonstrate that the neutrino mass has a Majorana component~\cite{Schechter:1981bd}, implying in turn that neutrinos are self-conjugate, i.e.\ their own antiparticles. Observation of \BBz\ decay would also point to new mechanisms for generating neutrino masses, quite distinct from the one giving mass to other particles, and possibly originating at very high energy scales. 
Finally,  the observation of a ``matter-creating'' process such as \BBz\ decay would 
 help us understand ``leptogenesis,'' in which the 
 universe's
matter-antimatter asymmetry stems from the decay of heavy Majorana neutrinos~\cite{Davidson:2008bu}. 

\BBz-decay searches will test a wide variety of mechanisms for lepton number violation (LNV) with unprecedented precision.  Such mechanisms range from the high-scale seesaw, in which neutrinos with masses of perhaps $10^{16}$ GeV that occur in Grand Unified Theories are important, all the way down to models in which eV-scale right-handed neutrinos play a role.  They also include processes that affect electroweak physics, the scale of which is about 1 TeV. This range of physics makes the discovery potential of \BBz\ decay nearly unique. At the same time, however, it makes it requires the use of more than one metric to quantify the discovery potential of individual experiments. 

Here we use several quantites to  characterize the reach of experiments. First, following standard practice, we focus on the class of models 
in which \BBz\ decay is mediated by the exchange of the three known light neutrinos, which must be Majorana particles.  The decay rate in such models is proportional to  
$|M_{0\nu}|^2 \, |$\mee $|^2$,  where 
$M_{0\nu}$ is a nuclear matrix element and \mee $ \equiv \sum_{i=1}^{3}  U_{ei}^2 m_i$ is the lepton-number violating parameter, expressed in  terms of neutrino masses $m_i$ and elements $U_{ei}$ of the leptonic mixing matrix.  We know a lot about neutrino masses and mixing parameters from neutrino oscillation data~\cite{ParticleDataGroup:2022pth}.  The combination \mee\ depends on just a few things that we don't know: CP-violating phases in the mixing matrix, the overall neutrino-mass scale, and whether the ordering of masses is ``normal'' (two very light neutrinos and a significantly heavier one)  or ``inverted'' (one light neutrino and two heavier ones).  Because we already have a lot of information, we can set concrete discovery targets.  Provided the nuclear matrix element is not too different from 
current 
theoretical predictions (we address the issue below), ton-scale experiments should observe the decay if the ordering is inverted, which requires that \mee $> 18.4 \pm 1.3$~meV,  and a discovery will be possible if $m_{\rm lightest} > 50$~meV, irrespective of the ordering. In addition, results of these experiments
will have consequences for other measurements of neutrino mass, in the laboratory and in cosmology, and vice versa.

\begin{figure*}[ht]
 \centering
 \includegraphics[width=16cm]{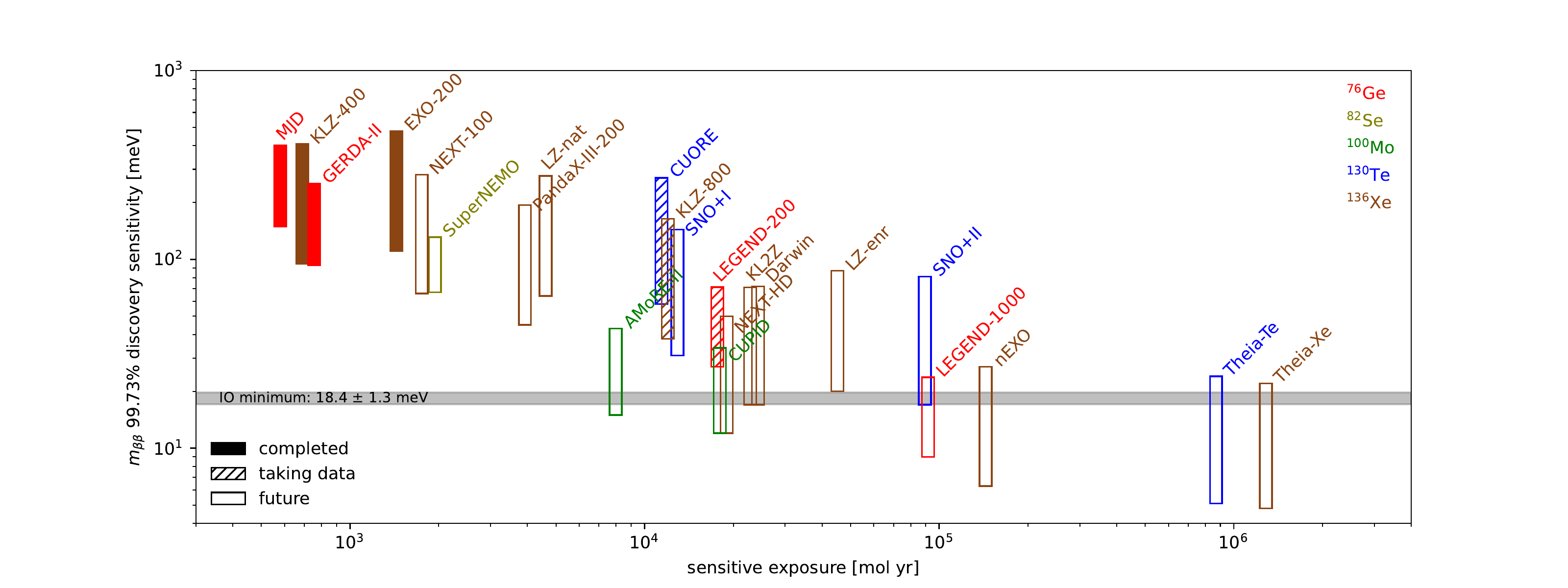} \par
 \includegraphics[width=8cm]{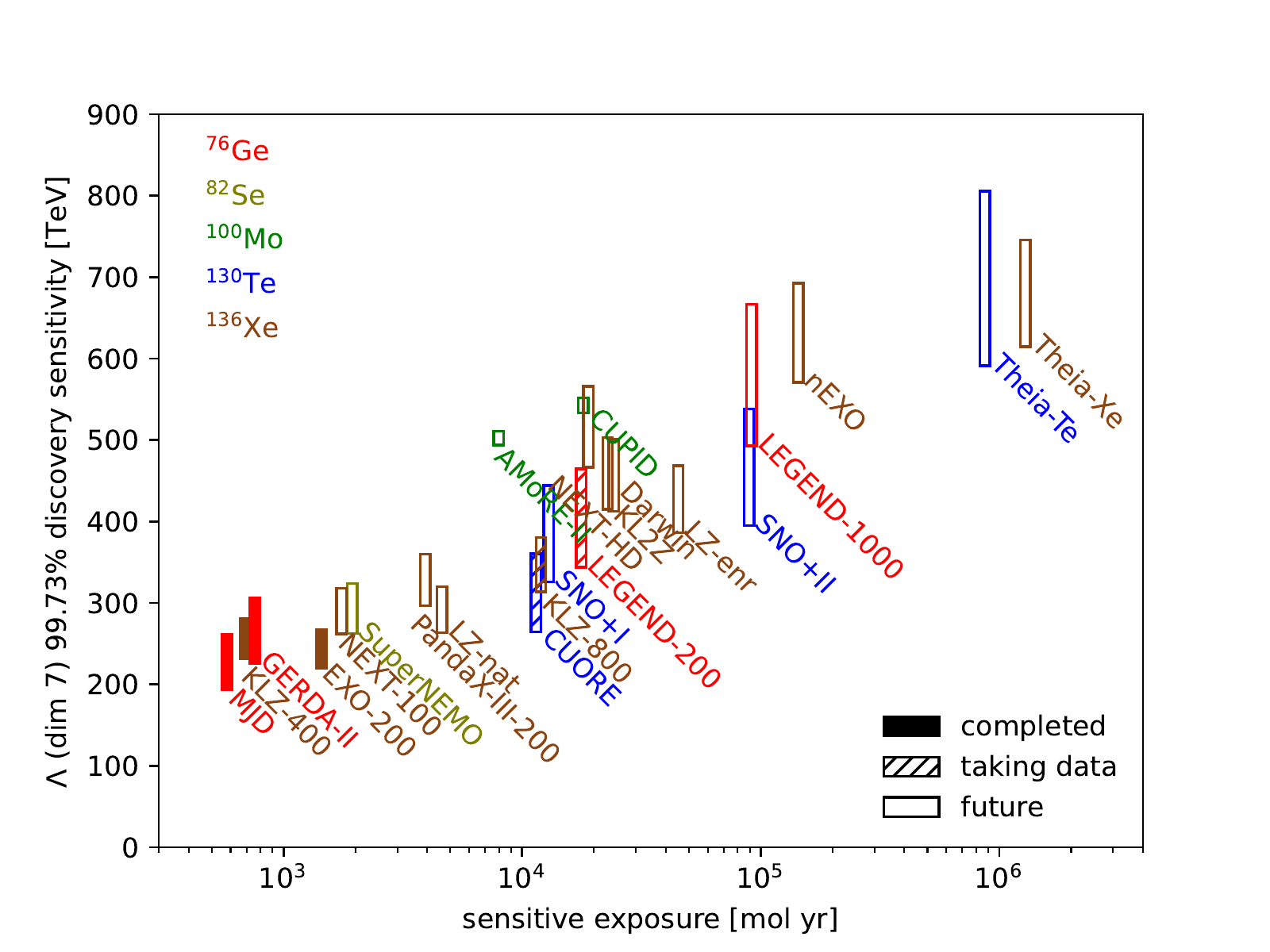}
 \includegraphics[width=8cm]{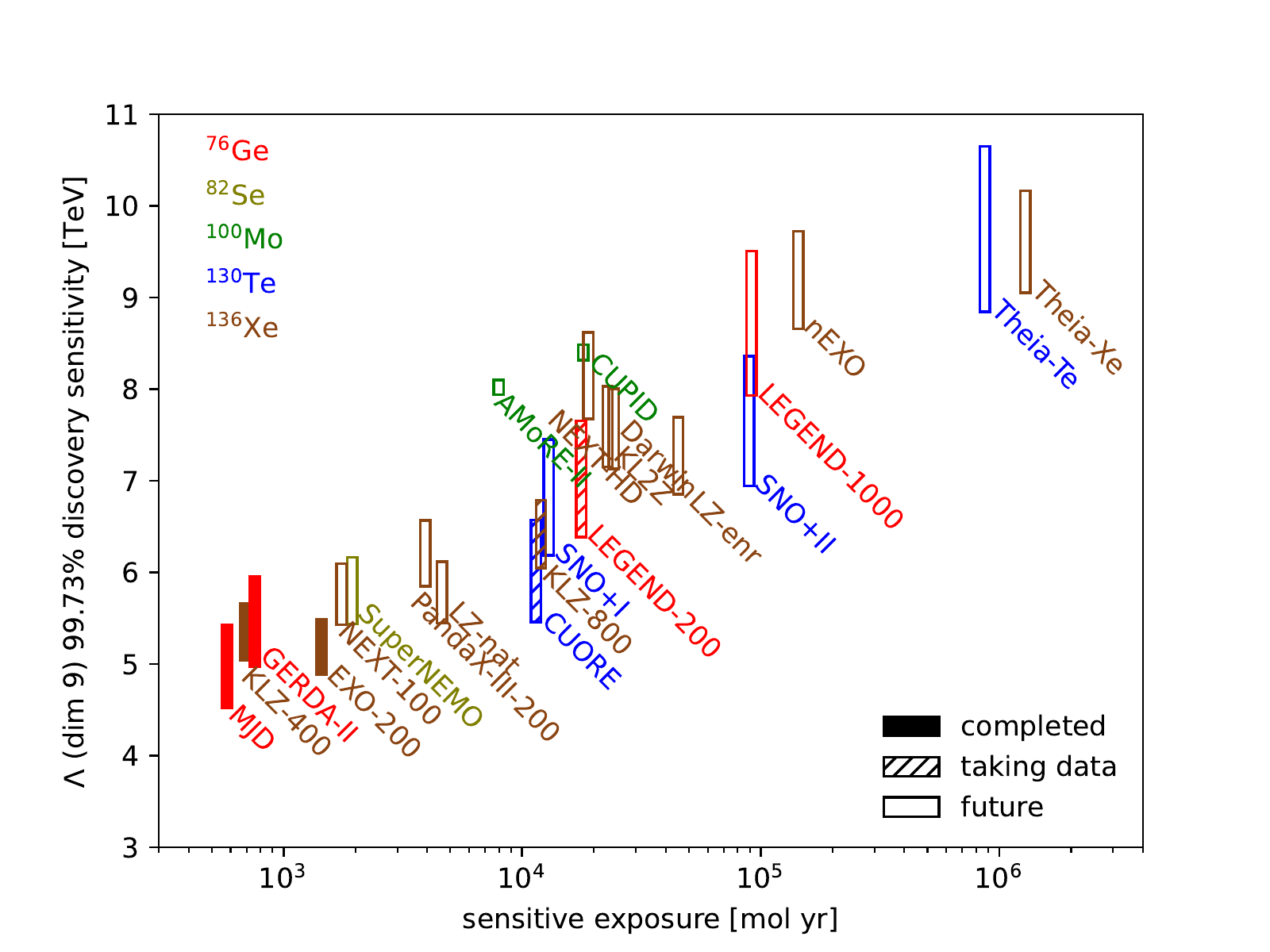}
 \caption{
Discovery sensitivities of current- and next-generation \BBz-decay experiments for 
various mechanisms of LNV, 
dominated by effective operators of dimension 5 corresponding to  light-neutrino exchange (top panel), of
dimension 7 (lower left panel) and of dimension 9 (lower right panel). 
Values of \mee\ larger than those in the bars in the top panel are tested at higher confidence level. 
Values of $\Lambda$ smaller than those in the the bars in the lower panels are tested at higher confidence level. 
At dimension 7 and 9, 
we  show  the reach for a single operator, the one that 
is least suppressed by chiral and electroweak scales~\cite{Cirigliano:2018yza}.
The sizes of the bars reflect the spread of the corresponding nuclear matrix elements (NMEs) and should be understood as a conservative range, not a standard deviation.
The nuclear matrix elements are taken from~\cite{Menendez:2017fdf,Horoi:2015tkc,Coraggio:2020hwx,Coraggio:2022vgy,Mustonen:2013zu,Hyvarinen:2015bda,Simkovic:2018hiq,Fang:2018tui,Terasaki:2020ndc,Rodriguez:2010mn,LopezVaquero:2013yji,Song:2017ktj,Barea:2015kwa,Deppisch:2020ztt,Jiao:2017opc}. Some matrix elements~\cite{Coraggio:2020hwx,Coraggio:2022vgy} include an initial estimate of quenching mechanisms that require further work.
The IO minimum is taken from~\cite{Agostini:2021kba}. Figure adapted from Ref.~\cite{Agostini2022MatterDiscover}. 
}
\label{fig:ExperimentComparision}
\end{figure*}

Although it is common to present the physics reach of \BBz-decay searches in terms of \mee\,  it is important to realize that this parameter is relevant in only one class 
of models for Majorana neutrino masses, those in which LNV originates at a very high mass scale $\Lambda$ and leaves behind $m_{\beta \beta} \sim v_{\rm ew}^2/\Lambda$  (where $v_{\rm ew} \sim 200$~GeV is the Higgs expectation value) as  its only  low-energy footprint.  In many models with Majorana neutrinos, however,  other sources of LNV can cause \BBz\ decay in a way that is not directly related to the exchange of light neutrinos.  In left-right symmetric models, for example, the exchange of heavy neutrinos, heavy $W$ bosons, and charged scalars, all with masses in the TeV range, contribute to the decay alongside the exchange of light Majorana neutrinos.  No matter what the model, the low-energy effects of these heavy particles are captured by a set of  $\Delta L=2$  local operators of odd dimension (seven, nine, ...), which are suppressed by odd powers of the heavy mass scale  $\Lambda$ associated with LNV ($1/\Lambda^3$,  $1/\Lambda^5$, ...). This suppression is analogous to what happens in the familiar Fermi theory of weak interactions, in which the effect of $W$ exchange is captured at low energy  by the  usual $V-A$ current-current  (dimension six) interaction, suppressed by $1/\Lambda_{\rm ew}^2$ with $\Lambda_{\rm ew} = 1/\sqrt{G_F}$. A systematic development of this effective-field-theory approach to LNV at low energies can be found in Ref.~\cite{Cirigliano:2018yza}. In Fig.~\ref{fig:ExperimentComparision} we present the physics reach of current  and future \BBz\ searches in terms of  both \mee\ and the scale $\Lambda$ associated with representative dimension-seven and dimension-nine operators~\cite{Agostini2022MatterDiscover}.  The sensitivity extends to hundreds of TeV, energies that are inaccessible to any other probe.

The discussion above should make it clear that if \BBz\ decay is observed, we will not know the underlying source of LNV  right away.
To determine it, we can study the isotope dependence of the decay rates, and phase space variables 
such as the single electron spectra and the relative angle of the two emitted electrons (see \cite{Graf:2022lhj} and references therein). 
Unraveling the underlying mechanism will probably also require complementary probes of neutrino mass and LNV, as we discuss below. 
Finally, it is worth noting that ongoing and future experiments will produced high-statistics data sets 
 of two-neutrino double beta decay events. Although these events are studied mostly as a background for neutrinoless searches, they can also be used to probe physics beyond the Standard Model, such as Majoron models, right handed currents, and sterile neutrinos~\cite{Blum:2018ljv,Deppisch:2020mxv}.

All the statements above depend to a degree on the values of nuclear matrix elements, which affect not only the rate of light-neutrino exchange but also that of other \BBz\ processes.  At present these matrix elements have an uncertainty that is impossible to estimate precisely but is probably a factor of two or three, enough to affect the planning and interpretation of experiments.  Computing the matrix elements accurately involves the combination of 
new-physics models, QCD, EFT, and nuclear-structure theory.   The last five years have seen tremendous progress, in part because of a recently concluded DOE topical theory collaboration (\url{https://a51.lbl.gov/~0nubb/webhome/}), but more is needed, see Section~\ref{sect:theory}.  With proper organization and support,  
better accuracy and controlled theoretical uncertainty are achievable.
Ref.\ \cite{Cirigliano_2022}, written in response to a request by the NSF, presents the current situation in theory and the road ahead.  We discuss some of the important issues in Section \ref{sect:theory}. 

Even with this remaining uncertainty,  however,
ton-scale \BBz\ experiments will have discovery potential that goes beyond the inverted mass ordering region in \mee\ to encompass a plethora of models across the landscape of particle physics.  As a result, although other kinds of experiments can complement results from \BBz\ decay searches, they will not replace them.   Large scale \BBz\ decay experiments and accompanying theoretical work will be crucial to progress in fundamental physics. We discuss the interplay with other kinds of experiments next.

\subsection{Relationship with other probes of neutrino mass and lepton number violation}

Other experimental efforts can complement \BBz\ decay searches, making them even more important than they would otherwise be.
The most relevant complementary probes are: 

{\it Experiments that Determine the Neutrino Mass Ordering  ---} If the neutrino mass ordering is determined, the inverted mass ordering could either be singled out or become irrelevant. Even in the latter case, the normal ordering still allows high \mee\ values without violating constraints from oscillation experiments. Furthermore, lepton-number-violating processes other than light neutrino exchange are not constrained by oscillations at all. At present, even if the normal mass ordering is what nature has chosen, the probability of a discovery of \BBz\ decay is significant~\cite{Agostini2017}.

{\it Cosmological Probes of the Sum of Neutrino Masses ---} Future work in observational cosmology has the aim of performing a first measurement of $\Sigma \equiv \sum_{i=1}^{3} m_i$. An observation of $\Sigma<100$~meV would effectively rule out the inverted mass ordering. Cosmology, however, does not have the ability to distinguish between Majorana and Dirac neutrinos.  And a three-neutrino normal-ordering scenario with $\Sigma$ near its minimum would not in any way constrain other lepton-number-violating processes that might contribute to
\BBz\ decay.  Moreover, the standard cosmological model contains many parameters that must be deduced, and it must be tested in all ways possible. There are few complementary laboratory experiments that directly test results from cosmology. Laboratory measurements of neutrino properties can provide such tests.

{\it Neutrino-Mass Measurements  ---}{ If a neutrino mass is measured in $\beta$ decay, the observation/non-observation of \BBz\ decay will become even more exciting. A null \BBz\ decay result might indicate Dirac neutrinos or, if LNV is observed in collider experiments, interference among mechanisms or flavor symmetries that make $m_{\beta\beta}$ small.}

{\it Collider Experiments that Observe LNV ---} The LHC or other collider experiments might observe 
processes that change lepton number by two units in the next decade, at a level consistent with a heavy neutrino, 
left-right symmetry, or other BSM physics near the TeV scale.
Such an observation would make \BBz\ searches still more important, because 
we would have learned that lepton number is not conserved and neutrinos are Majorana particles, 
so the combination of light neutrino masses and the exchange of TeV-scale particles would definitely induce \BBz\ decay.
The combination of collider and \BBz\ results would be essential for extracting the underlying LNV physics.

{\it Sterile-Neutrino Experiments ---} If a sterile neutrino is found, it will fit well into the Majorana neutrino paradigm, making \BBz\ decay even more interesting.  The new neutrino
might directly contribute to \BBz\ decay and significantly alter the dependence of the rate on \mee. (See for example Ref.~\cite{Barea2015}.) 
The  sensitivity, reach, and importance of of the \BBz\ 
experiments will remain, however.

\subsection{Introduction to $0 \nu \beta \beta$ experiments}

Several experimental approaches are now available to search with high sensitivity and low backgrounds for \BBz\ decay in a variety of isotopes, covering the entire region of the inverted mass ordering and beyond. Ton-scale experiments using large bolometer arrays (CUPID), high-resolution Ge detectors (LEGEND), and a large-volume liquid-Xe TPC (nEXO), have been identified as the leading next-generation experiments with US leadership. All three experiments are based on international collaborations to leverage the strengths of international partnerships and the world's premier underground-laboratory facilities.  Those experiments are expected to extend the sensitivity to \BBz\ half lives by as much as two orders of magnitude.  During 2021, CUPID, LEGEND, and nEXO were examined in a  portfolio review organized by the DOE Office of Science, Nuclear Physics,
to address the opportunity for discovery of \BBz\ decay by
covering the parameter space associated with the inverted-ordering mass scale. All three experiments were highly rated and judged to be worth pursuing. R\&D challenges facing these three programs that were identified by a 2015 NSAC sub-committee have been resolved and CUPID, LEGEND, and nEXO are now preparing to proceed through the DOE Critical Decision process.  We discuss these experiments in section \ref{subsec:tonscale} below.

If \BBz\ decay is discovered at the ton scale, advanced techniques will be required to probe the decay mechanism via topological information and event identification. If \BBz\ decay is not discovered, detectors that can reach greater exposures with improved background rejection will be required to attain sensitivity beyond the inverted mass ordering. A robust R\&D program is pursuing detector technologies with these capabilities, and we discuss the program briefly in \ref{subsec:beyondtonscale}

\subsection{Ton-scale experimental program}
\label{subsec:tonscale}

Experiments to observe \BBz\ decay are of paramount importance but are also challenging. If an experiment has evidence for the decay, a result that should earn a Nobel Prize, prompt confirmation will be necessary.  An observation in more than one isotope, each with significantly different detector uncertainties, will provide that confirmation.  The long time frame for construction and operation, however, demands that multiple experiments be pursued simultaneously. 

For the above reason, the U.S. program must include complementary experiments studying different isotopes with different detection techniques. From the view of experimental uncertainties, the biggest challenge is separating the \BBz\ signal from backgrounds, either inherent to the isotope or to the detector configuration.  
Furthermore, \BBz\ searches in multiple isotopes will mitigate the impact of theoretical uncertainties in the nuclear matrix elements, 
that may result in over-estimating the decay rate for a particular isotope. 
Finally, beyond 
providing robust evidence for 
the decay, the observation in multiple isotopes will help unravel the underlying physics that mediates \BBz\ decay.

In order to maximize the discovery potential for \BBz~decay at the ton-scale, the proposed US program consists of three experiments fielding very different experimental technologies, and using three different isotopes: CUPID, \Lthou\ and nEXO. These three experiments have undergone a DOE portfolio review, are ready to start construction, and are actively preparing for the Critical Decision process. The details of these experiments follow:

\textit{CUPID ---}The CUORE Upgrade with Particle Identification (CUPID)~\cite{CUPID-pCDR2019}, an upgrade to the currently-operating Cryogenic Observatory of Rare Events (CUORE) experiment at Gran Sasso National Laboratory (LNGS)\cite{CUORE-0Te}, is aimed at searching for $0\nu\beta\beta$ in $^{100}$Mo in the region of the inverted mass ordering. The proposed CUPID experiment leverages the extensive existing cryogenic and technical infrastructure built for CUORE. The baseline design for CUPID features an array of 1596 scintillating crystal bolometers and 1710 light detectors, each instrumented with germanium neutron transmutation doped (NTD) sensors, and organized into 57 towers. This technology provides exquisite energy resolution, 5\,keV FWHM, while the combination of the heat and scintillation light signal allows for efficient rejection of backgrounds due to alpha particles. The total isotopic mass of CUPID will be 240~kg of $^{100}$Mo.  With light and thermal readout, the estimated background index for CUPID is $<10^{-4}$~c/kg/keV/year. The experiment will have discovery potential in the entire inverted hierarchy region of neutrino masses. The collaboration estimates the half-life limit sensitivity (90\%) C.L. at $1.4\times10^{27}$ yr  and the half-life discovery sensitivity ($3\sigma$) of $1.0\times10^{27}$ yr. The re-use of the existing cryostat at LNGS allows for an economical deployment of CUPID and builds on the success of years of stable operation of the CUORE detector at base temperatures of 10~mK and a detailed understanding of the backgrounds from the cryostat. The light and thermal readout has been demonstrated by the CUPID-0\cite{CUPID-0Se}, CUPID-Mo\cite{CUPID-MOzero2022} and CROSS\cite{CROSS:2019xov} pathfinder experiments. Bolometric detectors are scalable, allowing gradual, phased deployment. In the case of a discovery, in principle, crystals based on different isotopes could be installed. The isotopic flexibility and scalability also make bolometers an interesting technology for beyond the ton-scale efforts\cite{CUPID-1T}, see Sec.~\ref{subsec:beyondtonscale}.

\textit{LEGEND ---} The Large Enriched Germanium Experiment for Neutrinoless \BB\ Decay (\Lthou) experiment~\cite{LEGEND-pCDR} utilizes the demonstrated low background and excellent energy performance of high-purity p-type, inverted coax, point contact (ICPC) Ge semiconductor detectors, enriched to more than 90\% in \nuc{76}{Ge}. The background rejection power of ICPC detectors begins with their superb energy resolution, demonstrated to have a full-width at half-maximum (FWHM) resolution of 0.12\% (0.05\% $\sigma$) at \qval. Pulse shape analysis of the signal distinguishes bulk \BBz\ decay energy depositions from surface events and backgrounds from $\gamma$ rays with multiple interaction sites. The granular nature of the Ge detector array allows rejection of background events that span multiple detectors. Finally, background interactions external to the Ge detectors are identified by LAr scintillation light. About 330 ICPC detectors with an average mass of 3 kg each are distributed among four 250-kg modules to allow independent operation and phased commissioning. In each module, the detectors are arranged into 14 vertical strings, supported by ultra-clean materials, and read out using ultra-low-background ASIC-based electronics. The detector strings are immersed in radiopure liquid Ar sourced underground and reduced in the \nuc{42}{Ar} isotope. The underground-sourced LAr is contained within an electroformed copper reentrant tube. Each of the four modules is surrounded by LAr sourced from atmospheric Ar, contained within a vacuum-insulated cryostat. The LAr volumes are instrumented with an active veto system comprised of optical fibers read out by Si photomultipliers. The cryostat is enveloped by a water tank providing additional shielding. \LEG\ reference designs are available for installation at either SNOLAB or LNGS. The \LEG\ collaboration aims to increase the sensitivity for the $^{76}$Ge \BBz\ decay half-life in a first phase (\Ltwo) to $10^{27}$~yr, and in a second phase (\Lthou) to $10^{28}$~yr, both for setting a 90\% C.L.~half-life limit and for finding evidence for $0\nu\beta\beta$ decay, defined as a 50\% chance for a signal at 3$\sigma$  significance. In \Ltwo\ about 200~kg of Ge detectors are operated in an upgrade of existing GERDA experiment infrastructure at the LNGS laboratory in Italy. \Ltwo\ is currently taking data.

\textit{nEXO ---} nEXO~\cite{nEXO-pCDR} is based on a Time Projection Chamber (TPC) and the use of five tonnes of liquid xenon (LXe) enriched to 90\% in $^{136}$Xe. The baseline location of the experiment is SNOLAB. The choice of LXe is directly derived from the success of EXO-200 and is motivated by the ability of large homogeneous detectors to identify and measure background and signal simultaneously. This approach takes maximum advantage of the large linear dimensions compared to the mean free path of $\gamma$-radiation. The nEXO TPC consists of a single cylindrical volume of LXe that is instrumented to read out both ionization and scintillation signals in the LXe to obtain $<1\%$ energy resolution~\cite{nEXO_energy_resolution} and strong background rejection. The ionization signal is readout using charge-collection tiles at the top of the TPC while scintillation light is collected with Silicon Photomultipliers (SiPMs) installed around the barrel of the cylinder. The TPC vessel is made from ultra-radiopure custom electroformed copper and is surrounded by a bath of HFE-7000~\cite{HFE}, which acts as a radiopure heat exchange fluid and an efficient $\gamma$-ray shield. The HFE-7000 cryostat is located in an instrumented water tank that serves as a muon veto and additional shielding layer. Information on particle interactions provided by the TPC includes several additional handles to reject backgrounds and improve confidence in a potential discovery. Energy reconstruction, event topology (single vs multi-site interactions), position reconstruction, and scintillation/ionization ratio, are combined using traditional and deep learning tools to effectively discriminate between signal and backgrounds. The nEXO background projections are grounded in existing radioassay data for most component materials and detailed particle tracking and event reconstruction simulations~\cite{Adhikari_2021nEXO}. This approach was validated by EXO-200, where the measured backgrounds closely matched the predictions~\cite{EXO-200_bkgd}. Based on these detailed evaluations, nEXO is projected to reach a 90\% CL sensitivity of $1.35\times10^{28}$ yrs, covering the entire inverted ordering parameter space, along with a significant portion of the normal ordering parameter space, for nearly all values of the nuclear matrix elements. The use of a liquid target has several unique advantages in the case of a discovery. nEXO could directly verify the discovery with a ``blank'' measurement by swapping the enriched xenon with natural/depleted xenon. The enriched target could be reused with a different detector technology, $e.g.,$ a discovery with nEXO may be followed by an investigation of energy and angular correlations in a gas TPC.

\subsection{Beyond ton-scale}
\label{subsec:beyondtonscale}
The field of neutrinoless double beta decay will continue beyond the current ton-scale experiments. If \BBz\ decay is discovered and confirmed by several such experiments, the next step will be to identify the mechanism behind LNV. Sensitivity to different models of LNV physics can be achieved by precision measurements of the half-lives of multiple isotopes, and by measuring the event topology---distributions in energy and opening angle of the decay electrons. 

If, on the other hand, \BBz\ decay is not observed, so that the inverted mass ordering is essentially ruled out,  increasing the scale and sensitivity of the experiments will be of paramount importance.
Better experiments will require even larger isotopic masses, at or above the ton scale, and very low (ideally negligible) backgrounds. Reconstructing the topology of the events will be important. 
Several concepts for experiments with sensitivity below the inverted ordering mass scale and well into the normal ordering region exist, as discussed above. It is important to support R\&D to identify the most promising technologies over the next decade, so that we can be ready to mount the ambitious next-next-generation experiments by the time the ton-scale experiments complete their operations. 

Among possible beyond-ton-scale  experiments are: NEXT, which will employ high pressure xenon gas time projection chambers with barium tagging;  \textsc{Theia}, a large-scale hybrid Cherenkov/scintillation detector that will be an outgrowth of the SNO+ and KamLAND-Zen experiments; 
Selena, which will employ high-resolution amorphous selenium / CMOS devices with electron imaging capabilities.  With novel techniques and sensor technologies, rich reconstruction of event topologies, and advanced particle identification, these experiments will be sensitive to half-lives in excess of $10^{28}$ years. The new detection capabilities of this future generation will also provide access to a wider physics program, including CPT and baryon-number-violation tests, precision low-energy solar neutrino measurements, and the possible study of supernova neutrinos.  

\subsection{Summary}

Let us recap the main points of this section:
If \BBz\ is observed, the implications will be profound. We will immediately know that neutrinos are their own antiparticles and that
lepton number is not conserved.  This will be direct evidence of physics beyond the Standard Model and point to
an explanation for the observed matter-antimatter asymmetry in the Universe.
Measurement of the rate would probe the neutrino mass scale, provide a terrestrial constraint on the standard
cosmological model, and yield insights into mass generation.  

Such extraordinary results require correspondingly convincing evidence.
\textit{No single experiment based on a particular isotope will be sufficient.}
What is required is independent observations in multiple isotopes, with different
experimental methods and systematics.
Since the 2015 LRP, the US nuclear physics community, in collaboration with our international partners, 
has developed a new generation of ton-scale experiments capable of probing the 
inverted-ordering parameter space and answering this challenge.
Three international experiments, CUPID, LEGEND, and nEXO, based on three different isotopes and technologies, 
all with significant US involvement, were deemed ready to
proceed following a comprehensive DOE Portfolio Review carried out during the summer of 2021.
Accordingly, and consistent with the 2015 Long Range Plan recommendation II,
the community is ready for {\it construction}  of  multiple ton-scale neutrinoless double beta decay experiments, each using a different isotope.
Mounting three experiments with three different isotopes can only be accomplished with both significant US involvement and support, and 
extensive collaboration with and contributions from international partners.   
These efforts must include support for a healthy nuclear theory program, which is vital for planning and intrepretation.   

The importance of discovering \BBz\ decay makes it essential that the
community continue to develop approaches for beyond-ton-scale experiments.  Such R\&D will be essential for learning about the source of LNV should the decay be observed at the ton scale, or for reaching beyond the inverted mass ordering, should that be needed.

\section{Recommendation II:  Targeted Program}

The second FSNN Town Hall recommendation is to vigorously pursue   a suite of experiments aimed at challenging the Standard Model and uncovering new phenomena.  The absence of new physics at the high energy frontier and of a consensus on what underlies the Standard Model are compelling  arguments for searching broadly for new phenomena,  either by increasing the precision of experiments on processes that the Standard Model allows or by exploring qualitatively new phenomena that it forbids or suppresses greatly. Nuclear science has 
the breadth and depth required to play an important role in this program, which has overlap with the fields of high energy physics and atomic and molecular physics.   Since the last Long Range Plan, the FSNN community has taken a leadership role in a number of complementary high-impact areas: precision tests of  the fundamental structure of the electroweak interactions, the search for new sources of time-reversal and CP symmetry breaking in multiple systems with unprecedented sensitivity, the search for baryon number violation, the study of neutrino masses and interactions, and the search for hypothetical  light new particles, such as sterile neutrinos.   Projects and initiatives in each of these areas vary in investment scale and degree of readiness.  The recommendation reflects the variety through three directives: 

\begin{itemize}
\item[(1)]  Capitalize on investments already made to complete larger scale projects (nEDM@SNS and MOLLER@JLAB);

\item[(2)]  Support mid- and small-scale programs that can have high impact. 
Among these, the highest priority is neutron and nuclear $\beta$ decay that can shed light on the possible non-unitarity in the quark-mixing matrix, and the search for  new sources of time reversal violation through EDMs. 

\item[(3)] Pursue emerging ideas, technologies and programs, which themselves vary in how ready they are.
The recommendation singles out  next-generation measurements of the absolute neutrino mass (Project 8), lepton flavor universality tests in the weak interactions (PIONEER), a search for new neutral current interactions (SoLID), and BSM searches enabled by FRIB and quantum sensing. 
\end{itemize}

As discussed in the supporting text of the recommendation, robust research support  is crucial to the success of this targeted program.     Erosion of research support  on one hand delays the achievement of scientific goals, which jeopardizes US leadership, 
and on the other reduces the ability of the FSNN community to develop new ideas and capabilities, such as  brighter UCN sources and isotope harvesting at FRIB, for future work at the precision frontier . 

In the following subsections, a targeted program of experiments and associated theory is presented. 
The presentation is organized by science topic; each topic includes experiments of different scale and degrees of readiness. 
Section \ref{sect:CPV} describes the searches for CP-violation through EDMs and other observables, section
\ref{sect:precision} describes precision tests of the Standard Model as probes of new physics, and section
\ref{sect:light} describes experiments to determine properties of neutrinos and search for new light particles. 

\subsection{CP-violation: Electric Dipole Moments and other observables}
\label{sect:CPV}
\subsubsection{Significance of Research}
The Standard Model fails to predict
the observed matter-antimatter asymmetry
in the Universe.  It does not break CP (simultaneous charge-conjugation and parity) symmetry strongly enough~\cite{Gavela:1993ts,Gavela:1994ds,Huet:1994jb}.
Permanent electric dipole moment (EDMs) of leptons, nucleons, atoms, and molecules violate both time-reveral and parity symmetries (and thus CP symmetry).  Because CP violation in the Standard Model is so weak, EDMs 
are extremely sensitive probes of new fundamental interactions that may violate CP strongly enough to induce the Universe's matter-antimatter imbalance. 

The current limits on the EDMs of the electron, neutron,
and $^{199}$Hg atom 
already probe BSM physics at energy scales that are much higher than those directly accessible to experiments at the energy frontier.
For example, BSM contributions to the EDMs
of elementary fermions scale (usually) as
\begin{equation*}
    d_f \approx \frac{m_f}{(4\pi)^n \Lambda_{\rm NP}^2} \sin \varphi_{\rm CP},
\end{equation*}
where $\Lambda_{NP}$ and $\varphi_{\rm CP}$ are the energy scale 
and CP-violating phase(s) in the BSM model, $m_f$ is the fermion mass and
$n$ is the number of loops at which the fermion EDMs are induced. 
The limit on the electron EDM $d_e < 4.1 \times 10^{-30}$ $e\cdot$cm (90\% C.L.), \cite{Roussy:2022cmp}, extracted from experiments with HfF
and ThO molecules, thus corresponds to $\Lambda_{\rm NP} \sim 50$ TeV if $n=1$, or $5$ TeV if $n=2$. 
 
\begin{figure}
\includegraphics[width=0.8\textwidth]{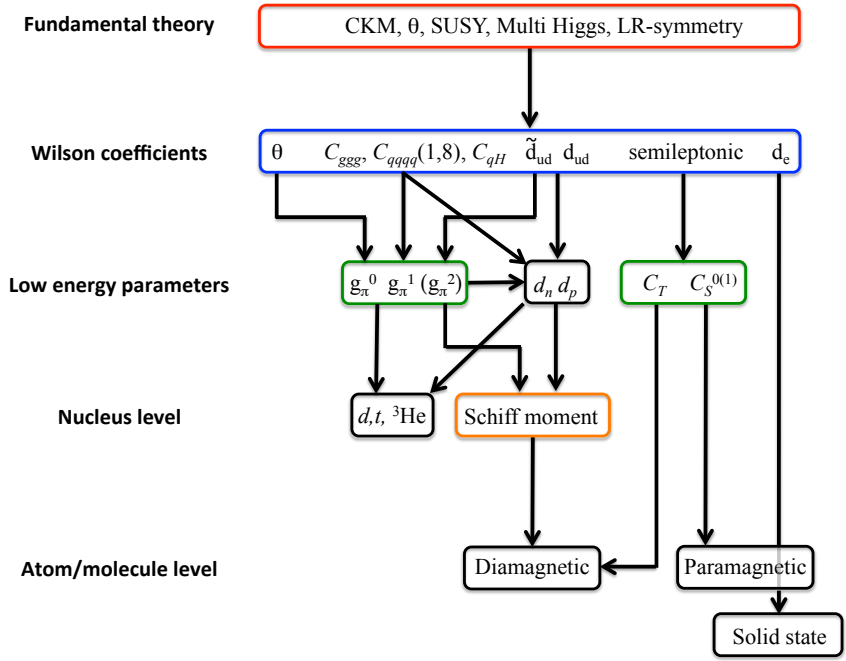}
\caption{
Broad brush illustration of the connections and complementarity between different EDMs. A broad EDM experimental program, coupled to advances in theory, is needed to constrain all possible manifestations of new sources of CP violation in models of physics beyond the Standard Model.
Figure taken from Ref. \cite{Chupp:2017rkp}. } 
\label{fig:metromap}
\end{figure}

As discussed below, the next generation of experiments will significantly improve existing bounds and explore EDMs in new systems. As new CP-violation can originate from very different high-energy mechanism (see Fig.~\ref{fig:metromap}), 
the existence of a broad program probing complementary systems is of paramount importance. Indeed,
while an observation of an EDM in any of these systems will be revolutionary, an understanding of the implications either of an observation or of tighter EDM bounds 
requires us, as with $0\nu\beta\beta$ decay and other FSNN processes, to connect low-energy observables first to the dynamics of quarks and gluons through nuclear-structure calculations, and then to physics at the electroweak and new-physics scales, through lattice QCD and EFT.  This chain of connections is necessary to understand the implications of EDM experiments by themselves, to quantify the complementarity EDM experiments probing different systems, and
also to understand the extent to which those experiments are related to others at high energy and with properties of other particles.  

\begin{figure}
\includegraphics[width=0.45\textwidth]{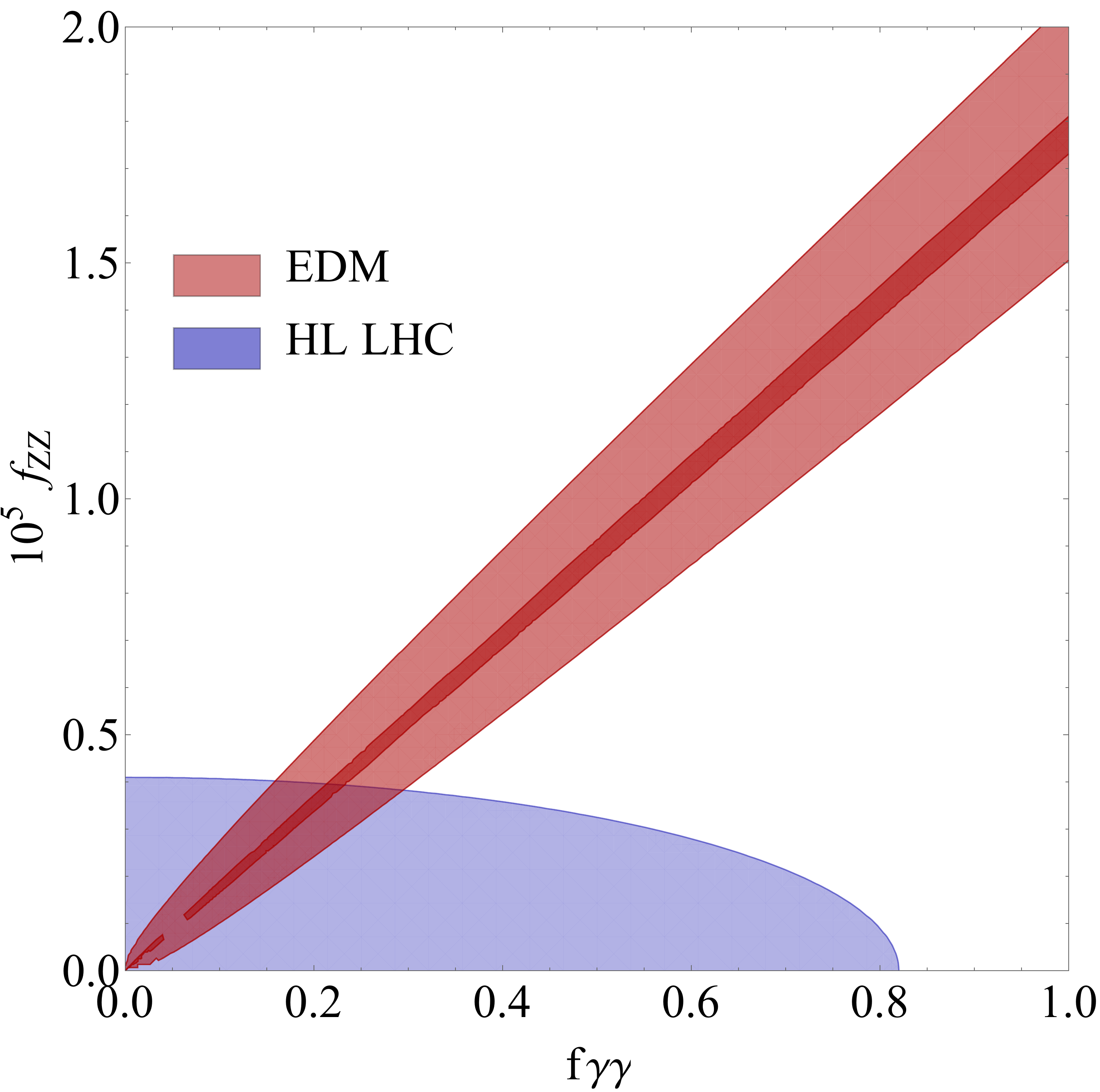}
\includegraphics[width=0.45\textwidth]{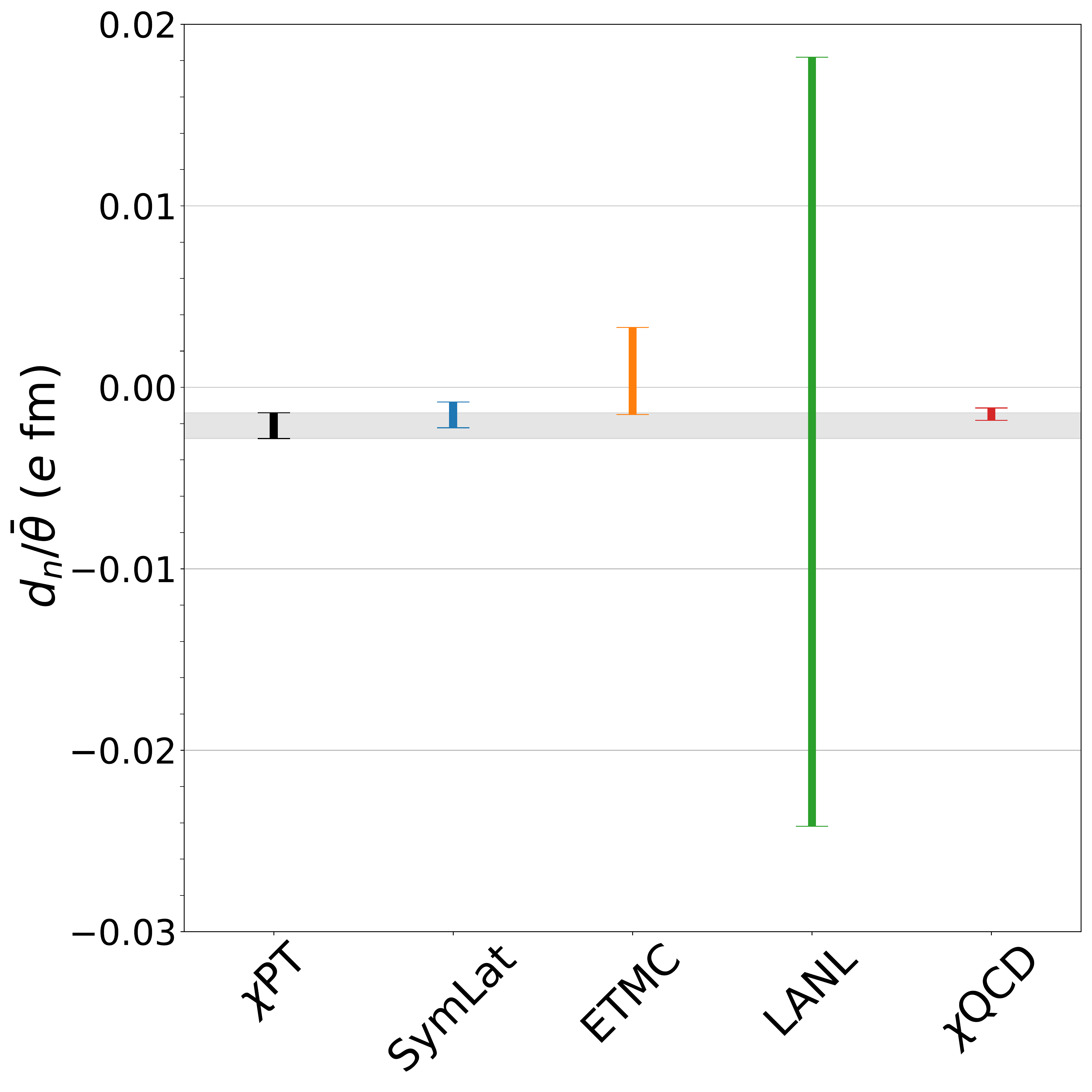}
\caption{
$(Left)$: EDM and LHC sensitivity to CP-violating couplings of the Higgs to two photons ($f_{\gamma\gamma}$) and two $Z$ bosons ($f_{ZZ}$). The allowed regions are obtained by marginalizing over the coupling to $\gamma Z$. The light  and dark red regions denotes 90\% CL limits from current EDM experiments
and assuming a factor of ten improvement on the electron and neutron EDM, respectively. The blue region denotes the projected bounds from the High Luminosity LHC from Ref. \cite{Gritsan:2022php}. Because EDM experiments are only sensitive to two linear combinations of Higgs couplings \cite{Cirigliano:2019vfc}, EDMs and LHC can constrain complementary regions in  parameter space. 
)
$(Right)$: Lattice QCD calculation of the neutron EDM induced by the QCD theta term. The black shaded area indicates the prediction from chiral perturbation theory \cite{Crewther:1979pi}, with the error band obtained by varying the renormalization scale in the chiral logarithm between $\mu = 600$ MeV and $\mu = 1200$ MeV. QCD sum rule estimates \cite{Pospelov:2005pr} find an effect of the same magnitude and with comparable error. The blue, orange, green and red lines denote the results of Refs. \cite{Dragos:2019oxn,Alexandrou:2020mds, Bhattacharya:2021lol,Liang:2023jfj}, extrapolated to the physics pion masses. These calculations have very different systematic errors. Details on systematic effects for each single calculation can be found in the original papers.}\label{fig:th_edm} 
\end{figure}

A high-impact example that illustrates the importance of multiple probes  is the study of BSM CP-violating interactions 
of the Higgs boson.  The left panel of Fig. \ref{fig:th_edm} illustrates the complementary sensitivities of present and future EDM experiments (red regions) and the High Luminosity LHC (blue region) to CP-violating couplings of the Higgs boson to photons and $Z$ bosons.
Achieving the full potential of EDM searches requires, for a given source of CP violation,  
first-principles theoretical calculations of the EDMs.  Lattice QCD is currently beginning to provide such calculations for the nucleon; 
see the right panel of Fig.~\ref{fig:th_edm}. Nuclear-structure theory is also important.
Details on theoretical developments and prospects appear in Section~\ref{sect:theory}.

Finally, new sources of T violation can appear in low-energy experiments that don't involve EDMs.  
Searches for T-odd correlations
in the transmission of polarized neutrons through polarized targets are sensitive to the P-odd T-odd nucleon-nucleon potential 
\cite{Bowman:2014fca}, which is quite a different quantity from the neutron EDM. These tests complement searches for nuclear Schiff moments.
Exploiting enhancements due to small energy splittings between states of opposite parity in heavy nuclei, the NOPTREX experiment, for example, can probe T-odd pion-nucleon couplings at a level that makes it competitive with measurements of atomic EDMs
\cite{Bowman:2014fca}.
Measurements or bounds on the T-odd $D$ coefficient in neutron $\beta$ decay can also compete with and in some cases do better than EDM 
experiments~\cite{RAMSEYMUSOLF2021136136,PhysRevD.87.116012}.  
A charge asymmetry in the  decay $\eta \to \pi^+ \pi^- \pi^0 $  is a C and CP violating observable~\cite{Gardner:2019nid} and is also independent of EDMs. 
Finally, a search for the T-odd transverse muon polarization in the K$_{\mu3}$ decay mode $(K^+\rightarrow\mu^+\pi^0\nu_\mu)$ has been proposed by the TREK collaboration~\cite{Shimizu:2019tar}.

In what follows, we summarize the status and opportunities in experimental searches of 
neutron and atomic / molecular EDMs that have the closest connection with nuclear science. 
Prospects for EDM searches in these and other systems, including the proton, are summarized in Ref.~\cite{Alarcon:2022ero}.

\subsubsection{Neutron EDM experiments}
The current limit on the neutron electric dipole moment (EDM), $|d_n| < 1.8 \times 10^{-26}$ $e$-cm (90\% C.L.), was reported in 2020 \cite{Abel2020} by an experiment at the Paul Scherrer Institut (PSI) in Switzerland, which employed an upgraded apparatus from that used for the previous best limit \cite{Baker2006, Baker2014, Pendlebury2015} of $|d_n| < 3.0 \times 10^{-26}$ $e$-cm (90\% C.L.) obtained at the Institut Laue Langevin (ILL) in France.  Notably, the $1\sigma$ systematic error on the PSI result, $\pm 0.2 \times 10^{-26}$ $e$-cm, was reduced by a factor of five over that in the previous best limit.  Thus, the PSI result demonstrated the feasibility, with improvements in the statistical reach of new experiments under development, for further improvements in the sensitivity to the level of at least $\sim 10^{-27}$ $e$-cm.

At present, there are at least eight neutron EDM experiments under development worldwide, for which a summary is given in Table \ref{tab:neutronEDM} (based on \cite{NeutronWhitePaper}).  There are two experiments underway in the U.S., both of which utilize ultracold neutrons (UCN) and are discussed below, the nEDM@SNS experiment \cite{Ahmed2019}, which projects an ultimate sensitivity of $\sim 3 \times 10^{-28}$ $e$-cm through the deployment of several novel features for both the statistical reach and control of systematic errors, and the LANL nEDM experiment, which projects an intermediate-step sensitivity of $3 \times 10^{-27}$ $e$-cm and has been designed to be complementary to the nEDM@SNS experiment.  Of the six other worldwide experiments, two experiments in an R\&D stage use beams of cold neutrons, and the remaining four experiments use UCN.  The non-U.S. UCN experiments include: (1) the PNPI experiment at the ILL, (2) the n2EDM experiment at PSI, (3) the PanEDM experiment at the ILL, and (4) the TUCAN experiment at TRIUMF.  All four of the non-U.S. UCN experiments, along with the LANL nEDM experiment, are based on Ramsey’s method of separated oscillatory fields, with two measurement cells at room temperature.  Their sensitivity reach is projected to be on the order of $\sim$several $\times 10^{-27}$ $e$-cm.  The nEDM@SNS experiment, based on a novel concept first discussed by Golub and Lamoreaux \cite{Golub1994}, will use two different measurement techniques and projects a sensitivity on the order of $\sim 3 \times 10^{-28}$ $e$-cm.

\begin{table}[t]
    \centering
    \resizebox{\textwidth}{!}{%
    \begin{tabular}{|c|c|c|c|c|c|} \hline\hline
    Experiment:& Neutron& Measurement& Measurement& 90\% C.L. ($10^{-28}$ $e$-cm)& Year 90\% C.L. \\
    Facility& Source& Cell& Techniques& With 300 Live Days&
    Data Acquired \\ \hline\hline
    Crystal: JPARC& Cold Neutron Beam& Solid& Crystal Diffraction (High Internal $\vec{E}$)& $< 100$& Development \\ \hline
    Beam: ESS& Cold Neutron Beam& Vacuum& Pulsed Beam&
    $<50$& $\sim 2030$ \\ \hline
    PNPI: ILL& ILL Turbine (UCN)& Vacuum& Ramsey Technique,&
    Phase 1 $<100$& Development \\
    & PNPI/LHe (UCN)& & $\vec{E}=0$ Cell for Magnetometry&  $<10$& Development \\ \hline
    n2EDM: PSI& Solid D$_2$ (UCN)& Vacuum& Ramsey Technique, External Cs& $<15$& $\sim 2026$ \\
    & & & Magnetometers, Hg Co-Magnetometer& & \\ \hline
    PanEDM& Superfluid $^4$He (UCN),& Vacuum& Ramsey Technique, Hg Co-& $<30$& $\sim 2026$ \\
    ILL/Munich& Solid D$_2$ (UCN)& & External $^3$He and Cs Magnetometers& & \\ \hline
    TUCAN:& Superfluid $^4$He (UCN)& Vacuum& Ramsey Technique, Hg Co-& $<20$& $\sim 2027$ \\
    TRIUMF& & & Magnetometer, External& & \\
    & & & Cs Magnetometers& & \\ \hline
    nEDM:& Solid D$_2$ (UCN)& Vacuum& Ramsey Technique, Hg Co-& $<30$& $\sim 2026$ \\
    LANL& & & Magnetometer, Hg External& & \\
    & & & Magnetometer, OPM& & \\ \hline
    nEDM@SNS:& Superfluid $^4$He (UCN)& $^4$He& Cryogenic High Voltage, $^3$He& $<20$& $\sim 2029$ \\
    ORNL& & & Capture for $\omega$, $^3$He Co-Magnetometer& $<3$& $\sim 2031$ \\
    & & & with SQUIDs, Dressed Spins,& & \\
    & & & Superconducting Magnetic Shield& & \\
    \hline\hline
    \end{tabular}
    }
    \caption{Summary of neutron EDM experiments under development worldwide, with projected 90\% C.L. sensitivity (in units of $10^{-28}$ $e$-cm, and the projected date by which data will be acquired to achieve the projected sensitivity.}
    \label{tab:neutronEDM}
\end{table}

The nEDM@SNS experiment \cite{Ahmed2019} is the most ambitious of all worldwide neutron EDM experiments, with a projected sensitivity approximately two orders of magnitude below the current limit.  The experiment will be mounted on the Fundamental Neutron Physics Beamline (FNPB) at the Spallation Neutron Source (SNS) at Oak Ridge National Laboratory (ORNL), where a nearly mono-energetic cold neutron beam will scatter from phonons in superfluid $^4$He, thus producing UCN in the measurement cells. This allows for a relatively high density of UCN to be produced, as there are no source-to-experiment transport losses of UCN. The superfluid $^4$He also acts as an electrical insulator permitting electric fields of at least 75 kV/cm, as demonstrated in small-scale and medium-scale prototype systems~\cite{Ito:2015hwa, Phan:2020nhz}.  Polarized $^3$He is then used as both a co-magnetometer and monitor of the UCN precession frequency. Magnetometry is possible via SQUID sensors that measure the time-dependent magnetization of the polarized $^3$He, while the UCN frequency is monitored via the spin-dependent neutron-$^3$He capture reaction that produces scintillation light from the reaction products. The polarized $^3$He not only allows for two independent techniques to be used for the EDM search (monitoring the frequency of the free precession and using critical spin dressing, see \cite{Golub1994,Ahmed2019}), it provides direct access to characterize one of the largest systematic effects in neutron-EDM experiments -- the so-called geometric phase false EDM effect. A small change in the operating temperature of the experiment of $\sim 0.1$ K can greatly increase the size of this false EDM effect in $^3$He, which thus permits measurement of the magnitude of this systematic effect in a small fraction of the overall experiment running time.

At the time of the last Long Range Plan, the nEDM@SNS experiment was beginning an intense R\&D program (i.e., the Critical Component Demonstration phase) whereby high-fidelity prototypes of the most challenging components were constructed. In some cases, these were the full-scale components to be used in the experiment, while others demonstrated the feasibility of the techniques.  At present, the Magnetic Field System is being reassembled and commissioned at the SNS, while the Central Detector System and Polarized $^3$He System are under construction. Construction of the new building at the SNS to house the experiment and installation of cold neutron guides, followed by commissioning and data-taking, is planned on a $\sim 5$ year timescale.

The LANL nEDM experiment is based on the proven Ramsey’s method of separated oscillatory fields at room temperature, featuring a double precession chamber geometry. The LANL nEDM experiment is complementary to the nEDM@SNS experiment. It takes advantage of the LANL UCN source, one of the strongest UCN sources in the world and the only UCN source currently operating in North America, providing the U.S. neutron EDM community with an opportunity to perform a neutron EDM experiment and obtain competitive physics results on a shorter time scale, while the development and construction of the nEDM@SNS experiment continues.

Soon after the last Long Range Plan, the LANL UCN source went through a major upgrade, increasing the output by a factor of four \cite{Ito2018}.  A dedicated UCN beamline was constructed for the LANL nEDM experiment. A sufficient UCN density for a neutron EDM experiment with a $1\sigma$ sensitivity of $3 \times 10^{-27}$ $e$-cm was demonstrated under conditions relevant for a neutron EDM experiment \cite{Ito2018,Wong2022}.  A large, high-shielding factor magnetically shielded room (MSR) has been installed in the experimental area, and the $B_0$ coil system which provides the uniform and stable magnetic field has been fabricated and installed inside the MSR and its performance has been characterized. Various magnetometers (including a $^{199}$Hg based co-magnetometer and a $^{199}$Hg based external magnetometer as well as optically pumped magnetometers) are being developed. The precession chambers, electrodes, and UCN valves are currently being assembled. The instruments are being commissioned for imminent data-taking with a Ramsey precession measurement in 2023.

As mentioned above, the LANL nEDM experiment is complementary to the nEDM@SNS experiment. Various capabilities and expertise developed for the LANL nEDM experiment are benefitting the nEDM@SNS experiment. These include mutual technical interests, such as the development of a system to scan for magnetic impurities and the development of methods to fabricate components with minimal magnetic contamination, and also the development of the next generation workforce. The training of early career scientists on the LANL nEDM experiment is important for realizing the workforce that will subsequently operate and analyze the data from the nEDM@SNS experiment.

\subsubsection{Atomic and molecular EDM experiments}
\newcommand{\iso}[2]{$^{#1}${#2}}

Experiments using methods from the atomic/molecular/optical (AMO) physics community are already making a major impact on the search for BSM CP violation.  They are now poised to not only continue making this impact, but also to expand both its breadth and depth~\cite{Alarcon2022Snowmass,Safronova2018Review,Chupp2019Review}. In the search for hadronic CPV, the nucleus in a given atom or molecule is sensitive to underlying CP-violating interactions complementary to those probed by the neutron 
EDM~\cite{Abel2020nEDM}, which vary across different nuclei. One current experiment of this type, using \iso{199}{Hg} atoms~\cite{Graner2016EDM}, has already reached a sensitivity to new BSM physics that is comparable to that of current neutron-EDM searches.  Such experiments leverage AMO techniques that offer advanced quantum control for sensitivity and robustness.  Despite their small scale, these experiments already probe BSM physics at scales as high as $\sim$100~TeV, and are set to reach over 1~PeV in the next decade~\cite{Alarcon2022Snowmass,Safronova2018Review,Chupp2019Review}. 

There are two main CP-violating AMO observables that arise from hadronic effects: nuclear Schiff moments and nuclear magnetic quadrupole moments. Both are broadly sensitive to hadronic CPV, including $\theta_\mathrm{QCD}$, quark EDMs and chromo-EDMs, CP-violating pion exchange effects, and more, thus offering a powerful complement to electron-EDM searches.  These moments are discussed further in a number of comprehensive reviews~\cite{Alarcon2022Snowmass,Safronova2018Review,Chupp2019Review,Ginges2004EDMs}.  Schiff moments arise from the slight difference between the nuclear charge and mass distributions. The charged constituents of the nucleus must feel zero static force on average in equilibrium; however, not all of the constituents are charged, leading to the Schiff moment, a residual electrostatic moment.  The Schiff moment interacts with electrons and mixes opposite parity electronic states, resulting in CPV energy shifts.   Magnetic quadrupole moments arise from CP-violating magnetic effects in the nucleus, for example from a valence nucleon with an EDM from its spin.  Like Schiff moments, magnetic quadrupole moments interact with electrons to create CP-violating atomic/molecular energy shifts.
Hadronic CPV effects ($\theta_\mathrm{QCD}$, quark chromo-EDMs, ...) also enter AMO EDM observables though the semi-leptonic CP-odd 
interaction $C_S \bar{e}\gamma_5 e\bar{N}N$~\cite{Flambaum:2019ejc}. This has some advantages over the Schiff and magnetic quadrupole moments, in that it contributes in the same systems used to target $d_e$, and that calculations of $C_S$ are under relatively good 
control. The downside is that the sensitivity is somewhat weaker, but already at an interesting level~\cite{Flambaum:2019ejc}. 

Similar to AMO searches for the electron EDM, both the Schiff moment and magnetic quadrupole moment have a roughly $Z^{2-3}$ scaling of ``field enhancement'' -- the electromagnetic environment inside atoms and molecules results in amplified shifts due to these CP-violating moments -- and are enhanced by a factor of $\sim10^{3-4}$ in molecules, which have a higher polarizability compared to atoms.  However, a critical difference is that the Schiff and magnetic quadrupole moments are properties of a given nucleus, and can therefore have their own intrinsic amplifications as well; in other words, the Schiff or magnetic quadrupole moment of a particular nucleus can be large or small, and the sensitivity of an atom or molecule to this nuclear moment can be large or small as well.

Several AMO experiments to search for Schiff and magnetic quadrupole moments are in development; they promise to complement, or for some sources of hadronic CPV even exceed, the reach of the \iso{199}{Hg} and nEDM~\cite{Abel2020nEDM} experiments in the next 3--5 years.  The methods and species are wide-ranging, and include \iso{223/225}{Ra} atoms~\cite{Parker2015RaEDM}, \iso{129}{Xe} atoms~\cite{Allmendinger2019EDM,Sachdeva2019EDM},  \iso{171}{Yb} atoms~\cite{Zheng2022YbEDM}, \iso{223/225}Rn atoms~\cite{Tardiff2008Rn}, \iso{173}{YbOH} molecules~\cite{Kozyryev2017PolyEDM,Pilgram2021YbOH}, and \iso{205}{TlF} molecules~\cite{Grasdijk2021TlF}.  The use of different species provides critical complementary information on many different underlying mechanisms for hadronic CPV, and the use of very different methods gives robustness against systematic errors. In the longer term ($\sim$10 years), emerging new approaches hold  clear promise for many orders of magnitude in increased sensitivity~\cite{Alarcon2022Snowmass}. 

Several major developments are enabling this new generation of experiments with rapidly improving sensitivity. One is the improvement in the ability to control the quantum states of molecules, with methods similar to those developed over the last few decades for atomic systems~\cite{Safronova2018Review,Hutzler2020Review,Fitch2021Review}. This in turn is making it possible to take advantage of the 3--4 order of magnitude amplification of 
CP-violating signals in molecules as compared to atoms. Molecules can also offer new controls over systematic errors that are unavailable in atoms.  Molecular methods have led to orders of magnitude improved sensitivity to the electron EDM, and similar experiments are underway to use the same amplification to study hadronic CPV.  

Another development is the recognition that CPV effects within the nuclear medium can be greatly enhanced---by factors of 10--1,000, and in some cases even more---in heavy nuclei with strong quadrupole and octupole deformation. Experiments with \iso{173}{YbOH} and \iso{173}{YbF} molecules, now underway, will take advantage of the amplified magnetic quadrupole moment (magnetic quadrupole moment) in quadrupole-deformed \iso{173}{Yb} nuclei~\cite{Kozyryev2017PolyEDM,Pilgram2021YbOH,Ho2023YbFMQM,Flambaum2014MQM}. Deformed nuclei with the greatest CPV amplification, typically between a factor of 100 to 1,000,  via Schiff moments~\cite{Haxton1983CPV,Auerbach1996OctDev,Dobaczewski2005RaNSM,Sushkov1985NucCPV,Flambaum2002NSMAtoms} are pear-shaped (octupole-deformed) radioactive actinides, including \iso{225}{Ra}, \iso{223}{Fr}, \iso{229}{Pa}, and others.  An experiment searching for the Schiff moments of pear-shaped \iso{223/225}{Ra} nuclei in laser-cooled and optically-trapped atoms is underway at Argonne National Laboratory, and has already produced limits on the Schiff moment of the $^{225}$Ra nucleus~\cite{Parker2015RaEDM}.  This experiment continues to be upgraded, and has recently demonstrated a number of improvements~\cite{Booth2020,Ready2021} to reach higher fields and improve laser cooling; these are ready to be used in making an improved measurement. The Radium EDM experiment at Argonne National Laboratory is unique among hadronic CPV experiments with published upper limits in that it has the potential to improve upon its established limit by several orders of magnitude in the near future. Isotope harvesting at FRIB~\cite{Abel2019FRIB}, will result in the increased availability of many of the necessary short-lived isotopes listed above, thereby increasing experimental sensitivity through improved count rates.

Very recently, it has become increasingly realistic to combine both molecular and nuclear amplification factors to potentially achieve many orders of magnitude of improved sensitivity to hadronic CPV effects~\cite{Alarcon2022Snowmass}.  One approach is to use quantum-controlled, radioactive molecules that contain heavy pear-shaped nuclei~\cite{Kozyryev2017PolyEDM,Isaev2017RaOH,Yu2021RaOCH3,Fan2021RaPoly,RadMolWhitepaper2023}, which combine both nuclear and molecular enhancements to achieve such large intrinsic sensitivity that the ability to trap and manipulate even a single molecule at a time would enable probes at the frontier of hadronic CPV~\cite{Yu2021RaOCH3}.  The past few years have seen critical demonstrations, including spectroscopy in neutral radioactive molecules~\cite{GarciaRuiz2020} and the trapping, cooling, and control of radioactive molecular ions~\cite{Fan2021RaPoly}.  In parallel, new capacities for isotope harvesting~\cite{Abel2019FRIB} and radiochemistry at the FRIB will provide offline access to certain ``enhancer'' isotopes such \iso{225}{Ra} and \iso{229}{Pa}, in practical quantities for ultrasensitive nuclear CPV searches in atoms and molecules.

Several next-generation CP-violation search schemes with molecules containing pear-shaped nuclei are currently being developed~\cite{RadMolWhitepaper2023} and have new physics sensitivities projected to be several orders of magnitude beyond the state-of-the-art \iso{199}{Hg} EDM experiment.   Experiments that leverage the extreme sensitivity of these exotic nuclei are made possible by broad advances in working with Short-Lived Atoms and Molecules (SLAMs), which brings together AMO science, nuclear physics, physical chemistry, radiochemistry, radioactive beam facilities, and more~\cite{RadMolWhitepaper2023}.  A community to support and coordinate SLAM physics is being developed, with one of the major goals being the construction of a center for SLAM science at FRIB called the Pear Factory.  FRIB will offer access to unstable nuclei with extremely large sensitivity enhancements, from very short-lived species ($\lesssim$~1~day) needing on-line capability, to longer-lived harvested isotopes for off-line experiments.  The Pear Factory will combine and coordinate all of the necessary facilities and expertise; beam scientists and radiochemists will take the nuclei from ``beam to bottle''; physical chemists, spectroscopists, and AMO scientists will take the nuclei from ``bottle to experiment''; nuclear, AMO, and high-energy theorists will take the work from ``experiment to impact.''

This effort would facilitate the coordinated and collaborative development of a variety of techniques and precursory measurements and calculations needed to set the stage for ultrasensitive CPV searches in the hadronic sector, primarily using radioactive molecules. We estimate that this stage would be performed in the first half of the LRP period. 
The outcome would include clarity about which isotopes, molecules, and techniques are the most promising for an ultrasensitive next generation search for CPV in the hadronic sector. At the second stage, the community would coalesce around at least two different CPV search experiments with different techniques and, importantly, different systematics. We estimate that this stage would require 
5 years to build the experiments and deliver first science results.

Because of the complexity of nuclei and the multiple sources of CP-violating effects, it is critical to support a broad platform of approaches both within AMO and outside of it, such as direct studies with nuclei, neutrons, and protons.  A single result in a single system -- even a non-zero one -- cannot be used to robustly relate the measurement to the underlying fundamental physics sources.

\subsection{Precision tests of the Standard Model as probes of new physics}
\label{sect:precision}

 \subsubsection{Parity Violating Electron Scattering}
An important strategy to determine the full extent of validity of the electroweak (EW) theory and search for new dynamics from MeV to multi-TeV scales involves indirect probes, where ultra-precise measurements of
EW observables at energy scales well below the EW symmetry breaking are compared to accurate theoretical predictions. The MOLLER~\cite{MOLLER:2014iki} and SoLID~\cite{JeffersonLabSoLID:2022iod} experiments at Jefferson Lab (JLab), like other low energy measurements that have been proposed in the Fundamental Symmetries area, pursue such a strategy.  The energy, luminosity, and stability of the electron beam at JLab are uniquely suited to carry out these parity-violating electron scattering (PVES) measurements.
At energies much below the mass of the $Z^0$ boson (the ``$Z$-pole''), $Q^2\ll M_Z^2$, the Lagrangian of the EW neutral-current (NC) interaction relevant to electron-electron (M\o ller) scattering or electron deep inelastic scattering (DIS) off quarks inside the nucleon is given by~\cite{Zyla:2020zbs}:
\begin{eqnarray}
L_{NC} &=& \frac{G_F}{\sqrt{2}} \left[ g_{AV}^{ee}\bar e\gamma_\mu\gamma^5 e\bar e\gamma^\mu e+ 
g_{AV}^{eq} \, \bar e\gamma^\mu\gamma_5 e \bar q\gamma_\mu q 
+ g_{VA}^{eq} \, \bar e\gamma^\mu e \bar q\gamma_\mu \gamma_5 q %
\right], \label{eq:L}
\end{eqnarray}
where $G_F=1.166\times 10^{-5}$~GeV$^{-2}$ is the Fermi constant, while the $g^{ee}_{AV}$, $g^{eq}_{AV}$, and $g^{eq}_{VA}$
are effective four-fermion couplings, where $q$ is the quark flavor. 
The $AV$ and $VA$ couplings are both parity-violating (PV) and thus can be isolated by measuring PV observables, such as the cross section asymmetry between a right-handed and left-handed electron beam scattering from the target: 
\begin{eqnarray}
  A_{PV} &=& \frac{\sigma_R-\sigma_L}{\sigma_R+\sigma_L}~. 
\end{eqnarray}
The coupling $g_{AV}^{ee}$, (related to the weak charge of the electron $Q^e_W\equiv -2g_{AV}^{ee}$) has been measured by SLAC E158~\cite{Anthony:2005pm} and will be measured with five times improved precision in the planned MOLLER experiment~\cite{MOLLER:2014iki} at JLab. The $g_{AV}^{eq}$ coupling is best determined by combining atomic parity violation~\cite{Wood:1997zq,Guena:2005uj,Toh:2019iro} and elastic PVES experiments such as Qweak~\cite{Androic:2013rhu,Androic:2018kni} and the planned P2 experiment~\cite{Becker:2018ggl} at the MESA facility in Mainz.
The $g_{VA}^{eq}$ coupling requires spin-flip of quarks and can only be accessed in DIS, 
for which only two experiments exist to date from SLAC~\cite{Prescott:1978tm,Prescott:1979dh} and JLab 6 GeV~\cite{PVDIS:2014cmd}.

\begin{figure}[t]
\includegraphics[width=0.49\textwidth]{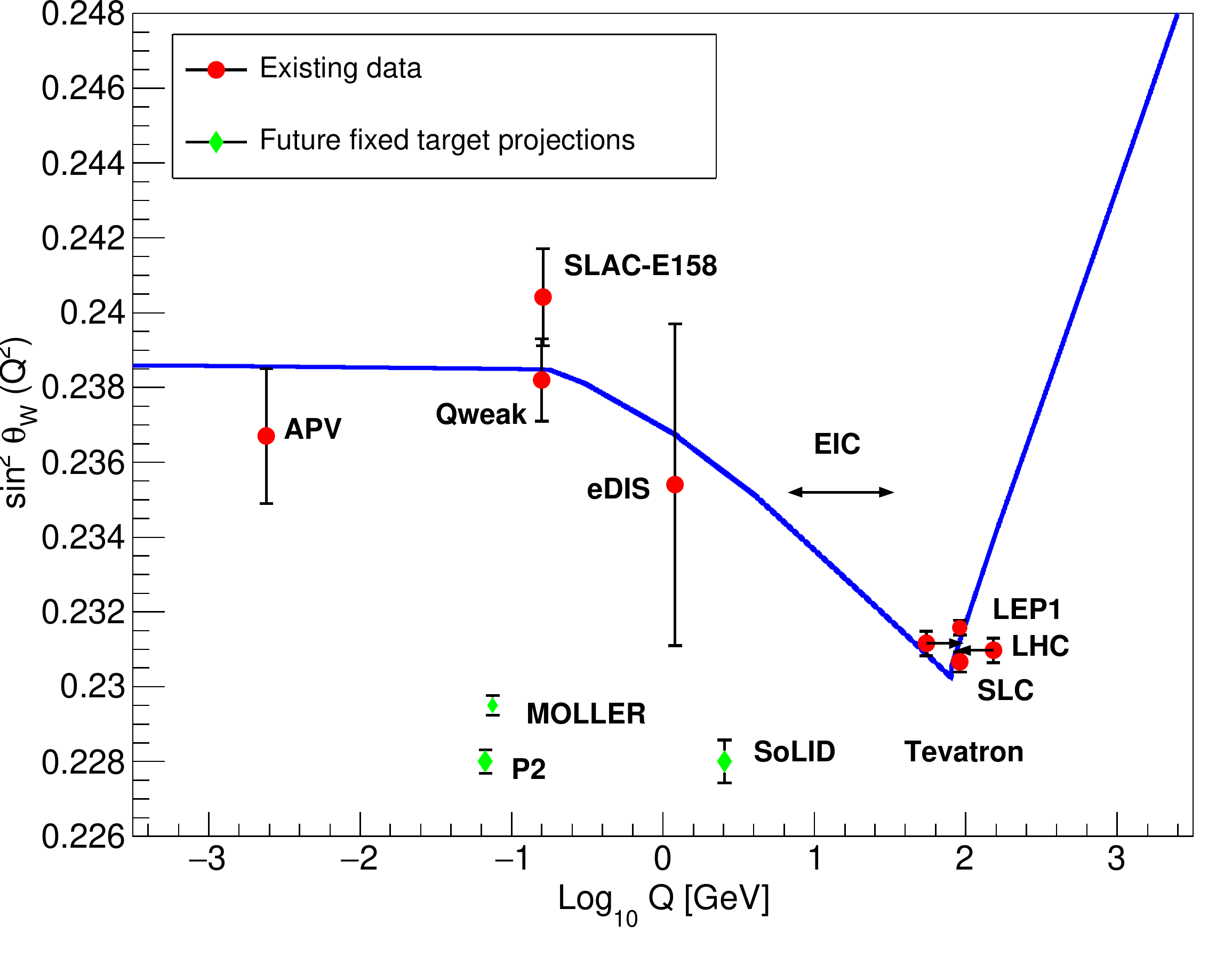}
\includegraphics[width=0.4\textwidth]{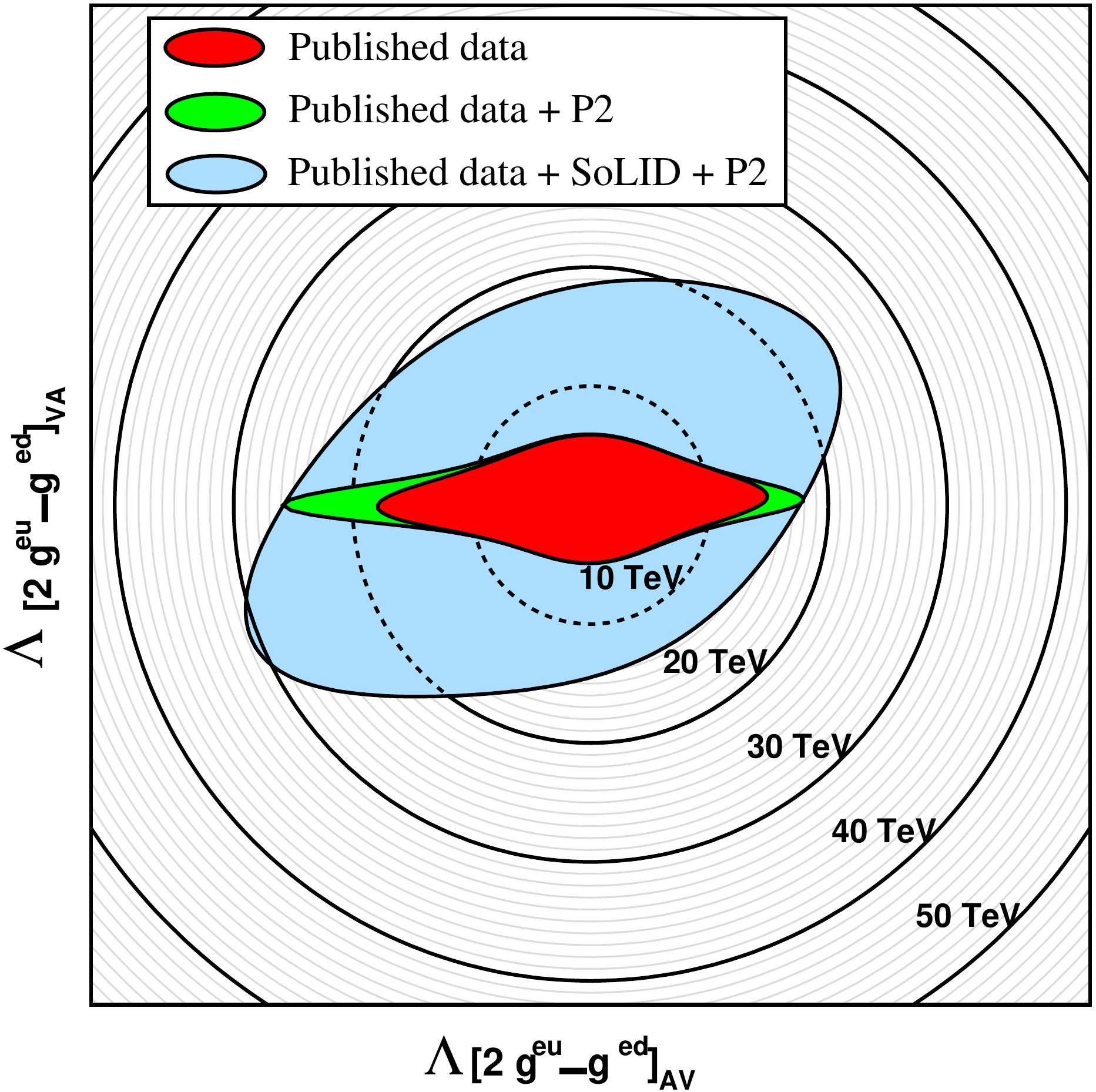}
\caption{ Figure from~\cite{JeffersonLabSoLID:2022iod}. {\it Left:} 
existing and planned experimental determinations of the weak mixing angle $\sin^2\theta_W$. Data points for Tevatron and LHC are shifted horizontally for clarity. 
The solid line shows the scale dependence of the weak mixing angle defined in the $\overline{MS}$ scheme~\cite{Erler:2017knj}. {\it Right:} Projection on the mass scale limits (colored regions are excluded) of BSM $eq$ contact interaction with $VA$ vs. $AV$ parity structure for the $2u-d$ quark flavor combination, obtained using all world data (red), with adding P2 (green), and with adding both P2 and SoLID deuteron PVDIS (light blue). See text for details and Eq.~(\ref{eq:ciqmodified}) for the definition of mass scales. }
\label{fig:c2}
\end{figure}

The weak mixing angle $\sin^2\theta_W$ has played a central role in the development and validation of the EW theory, especially testing it at the quantum loop level. 
A key feature of the experiments discussed here is that the $A_{PV}$ measurements will be carried out at $Q^2\ll M_Z^2$.
Since  
$\sin^2\theta_W$ ``runs" as a function of $Q^2$ due to EW radiative corrections, one can use $\sin^2\theta_W$ as a bookkeeping parameter to compare the consistency of the full $Q^2$ range
of weak NC measurements, as shown on the left of 
Fig.~\ref{fig:c2}. The theory uncertainty in the low energy extrapolation is comparable to the width of the line in the 
figure~\cite{Erler:2004in,Erler:2017knj}. The upcoming MOLLER~\cite{MOLLER:2014iki} and the SoLID~\cite{JeffersonLabSoLID:2022iod} deuteron PVDIS measurement, along with the P2 experiment~\cite{Becker:2018ggl} at Mainz, 
will provide three new cornerstone measurements of the weak mixing angle $\sin^2\theta_W$ at energy scales between atomic PV and high energy colliders.  The projected improvements over the existing measurements are a factor of 5 for MOLLER and SoLID and a factor of 3 for P2. 
Moreover, the proposed MOLLER $A_{PV}$ measurement would achieve a sensitivity of $\delta(\sin^2\theta_W) = \pm 0.00028$, would be the first low $Q^2$ measurement to match the precision of the single best high energy measurement at the $Z^0$ resonance, and the most precise anticipated weak mixing angle measurement currently proposed over the next decade at low energy.

These next generation PVES experiments have precision goals that result in a very sensitive discovery reach for flavor- and CP- conserving scattering amplitudes in the next decade; see recent reviews that situate these measurements in broader contexts~\cite{Kumar:2013yoa, Cirigliano:2013lpa, Erler:2013xha}.  They are complementary to other precision low energy experiments and the energy frontier efforts at the Large Hadron Collider (LHC)~\cite{Erler:2019hds}. If the LHC continues to agree with the Standard Model with high luminosity running at the full 14 TeV energy, then these measurements will be a significant component of a global strategy to discover signatures of a variety of physics that could escape LHC detection. Examples include hidden weak scale scenarios such as compressed supersymmetry~\cite{Ramsey-Musolf:2006evg}, lepton number violating amplitudes such as those mediated by doubly charged scalars~\cite{Cirigliano:2004mv}, and light MeV-scale dark matter mediators such as the ``dark'' $Z$~\cite{Davoudiasl:2012ag,Davoudiasl:2012qa}.

A fairly general and model-independent way of characterizing the BSM physics search potential of an experiment is to express BSM physics in terms of contact interactions that perturb the SM Lagrangian~(\ref{eq:L}), {\it i.e.}, by replacements of the form~\cite{Erler:2014fqa},
\begin{eqnarray}
\frac{G_F}{\sqrt{2}} g_{ij} \rightarrow \frac{G_F}{\sqrt{2}} g_{ij} + \eta_{ij}\frac{g^2}{(\Lambda_{ij})^2}\ ,
\label{eq:ciqmodified}
\end{eqnarray}
where $ij=AV,VA,AA$ and can be for either $ee$ or $eq$ interaction, $g$ is the coupling and $\Lambda$ is the mass scale of BSM physics, {\it i.e.} the coupling and the mass or interaction scale of the hypothetical BSM particle being exchanged. 
If the new physics is strongly coupled, $g^2 = 4\pi$, then the 90\% C.L. mass limits reached or to be reached by MOLLER, P2, and SoLID PVDIS on $g_{VA}^{ee}$, $g_{AV}^{eq}$ and $g_{VA}^{eq}$ are, respectively~\cite{JeffersonLabSoLID:2022iod}:
\begin{eqnarray}
 \Lambda_{VA, \mathrm{MOLLER}}^{ee}&=& g \sqrt{\frac{\sqrt{2}}{G_F 1.96\Delta g_{VA}^{ee}}}=39~\mathrm{TeV}, \\
 \Lambda_{AV, \mathrm{P2}}^{eq}&=& g \sqrt{\frac{\sqrt{2}\sqrt{5}}{G_F 1.96\Delta \left(2g_{AV}^{eu}+g_{AV}^{ed}\right)}}=49~\mathrm{TeV},\\ 
 \Lambda_{VA, \mathrm{SoLID+world}}^{eq} &=& g \sqrt{\frac{\sqrt{2}\sqrt{5}}{G_F 1.96\Delta \left(2g_{VA}^{eu}-g_{VA}^{ed}\right)}}=16~\mathrm{TeV}; 
\end{eqnarray}
where the $\sqrt{5}$ for P2 and PVDIS cases are to represent the ``best case scenario'' where BSM physics affects maximally the quark flavor combination being measured. 
The expected uncertainty $\Delta\left(2g_{VA}^{eu}-g_{VA}^{ed}\right)$ 
is obtained by combining SoLID PVDIS with existing world data. If one instead looked for the maximal mass limit expected from PVDIS for that observable's combination of $g_{AV}^{eq}$ and $g_{VA}^{eq}$ then it would be 22~TeV from SoLID alone~\cite{Erler:2014fqa}. Furthermore, the mass limit on $(2g_{VA}^{eu}-g_{VA}^{ed})$ vs. $(2g_{AV}^{eu}-g_{AV}^{ed})$ is shown in the right panel of Fig.~\ref{fig:c2}, illustrating the different exploratory potential on the parity-violating electron-quark BSM contact interactions.

Another model-independent way of expressing the new physics potential of PVES is provided in the framework of Standard Model Effective Field Theory (SMEFT), which showed that low energy data 
are complementary to those from high energy facilities. More specifically, BSM parameter constraints obtained from the LHC Drell-Yan cross section data were found to have a flat direction~\cite{Boughezal:2021kla}, and adding P2 and PVDIS projections helps remove such ambiguity in the determination of BSM parameters. 

For the specific case of MOLLER, the remarkable feature of its sensitivity to four-lepton flavor-conserving contact interactions has been emphasized~\cite{Falkowski:2015krw}. Not only does the contact interaction scale reach exceed the scales probed at LEP-200, the highest energy electron-positron collider to collect data, but there is unique sensitivity to a specific linear combination of left- and right-handed four electron operators to which all other collider measurements happen to be insensitive. Indeed, in the current global analysis, the E158 result~\cite{SLACE158:2005uay} is used to break the degeneracy. The MOLLER measurement will allow the extension of the current limits for these operators from $\Lambda/g \sim$ 2~TeV to more than 7~TeV. 

Improving this sensitivity over the entire multi-dimensional space of new operators is particularly important if higher sensitivity searches at the LHC yield no new discoveries. 
For example, the MOLLER measurement is one of the rare low $Q^2$ observables with sensitivity to  lepton number
violating amplitudes mediated by doubly-charged scalars, which naturally arise in extended Higgs sector models containing complex triplet 
representations of SU(2).  In the context of a left-right symmetric model, the proposed MOLLER measurement would lead to the
most stringent probe of the left-handed charged scalar and
its coupling to electrons, with a reach of 5.3 TeV,
significantly above the LEP 2 constraint of about 3~TeV. Moreover, such sensitivity is complementary to other 
sensitive probes such as lepton-flavor violation and neutrinoless double-beta decay searches~\cite{Cirigliano:2004mv, Dev:2018sel, Gardner:2019mcl}.

If an anomaly is observed at the LHC, then the next generation PVES measurements will have the sensitivity to be part of a few select measurements that will provide important constraints to choose among possible BSM scenarios to explain the anomaly.  Examples of such BSM scenarios that have been explicitly considered for these measurements include: new particles predicted by the Minimal Supersymmetric Standard Model observed through radiative loop effects (R-parity conserving) or tree-level interactions (R-parity violating)~\cite{Kurylov:2003zh,Ramsey-Musolf:2006evg} and TeV-scale $Z^\prime$s~\cite{Erler:2011iw} which 
arise in many BSM theories.

Both the MOLLER and SoLID PVDIS experiments are sensitive to the interesting possibility of a light MeV-scale dark matter mediator known as the ``dark'' $Z$~\cite{Davoudiasl:2012ag,Davoudiasl:2012qa}.  It is denoted as 
$Z_d$ with mass $m_{Z_d}$,  and it stems from a spontaneously broken $U(1)_d$ gauge symmetry associated with
a secluded ``dark" particle sector. The $Z_d$ boson can couple to SM particles through
a combination of kinetic and mass mixing with the photon and 
the $Z^0$-boson, with couplings $\varepsilon$ and $\varepsilon_Z = \frac{m_{Z_d}}{m_Z}\delta$ respectively. 
In the presence of mass mixing ($\delta \neq 0$),  a new source of ``dark'' parity violation arises~\cite{Davoudiasl:2012ag} such that it has negligible effect on other precision EW observables at high energy, 
but is quite discernible at low $Q^2$ through a shift in the weak mixing angle~\cite{Davoudiasl:2012qa}.

For the SoLID PVDIS experiment, the $g_{VA}^{eq}$'s in the SM are small, so they could be particularly sensitive to BSM physics. The $g_{VA}^{eq}$'s are sensitive to models that involve leptophobic $Z'$s~\cite{Buckley:2012tc, Gonzalez-Alonso:2012aib}. These are additional neutral gauge bosons ($Z'$) with negligible couplings to leptons, and thus would cause only sizable modification to the quark axial coupling (thus $g_{VA}^{eq}$) while leaving the $g_{AV}^{eq}$ unaffected. 

Looking forward, in addition to MESA and a possible energy upgrade of JLab, a series of upgrades for the LHC are being discussed~\cite{LHeCStudyGroup:2012zhm,LHeC:2020van,FCC:2018byv,FCC:2018evy} that will venture into the unexplored energy range much beyond the $Z$-pole~\cite{Britzger:2020kgg}. The EIC, coming online within the next 1-2 decades, likely will have decent sensitivity to EW parameters in between JLab and high-energy colliders as well~\cite{Boughezal:2022pmb}.  Among all existing and planned facilities, JLab 
 is one of the few that can provide direct access and high precision measurements of the SM effective couplings owing to both its high luminosity fixed-target settings and the relatively low beam energies, and thus holds a unique place in the test of the SM across all energy scales.

The MOLLER experiment was approved by JLab PAC34 in January 2009, received an ``A'' scientific rating and was allocated 344 days of beamtime by PAC37 in January 2011, and had both its rating and beam time allocation reaffirmed by PAC49 in July 2021.  The experiment received strong endorsement from a Department of Energy (DOE) Science Review in 2014, received CD-0 approval from DOE in December 2016 and CD-1 approval from DOE in December 2020.  The MOLLER project team is currently preparing for CD-3A and CD-2 reviews planned for calendar 2023. The project recently received Inflation Reduction Act funding that will allow the experiment to remain on a technically driven schedule and to obtain early physics results in the same timeframe as complementary results from the LHC and the Mainz Microtron.  The MOLLER collaboration currently consists of $\sim$ 180 members representing 34 institutions from 4 countries. The current schedule for the experiment calls for construction of the apparatus to be completed in late 2024 followed by installation in Hall A and production running in the 2026 - 2028 timeframe.

The Solenoidal Large Intensity Device (SoLID)~\cite{JeffersonLabSoLID:2022iod} is a spectrometer planned for Hall A of JLab, that will combine large angular and momentum acceptance with the capability to handle very high data rates at high luminosity.  A very rich physics program can be realized with the detector, including the tomography of the nucleon in 3-D momentum space from Semi-Inclusive Deep Inelastic Scattering (SIDIS), expanding the phase space in the search for new physics and novel hadronic effects in PVDIS, and a precision measurement of J/$\psi$ production at threshold that probes the gluon field and its contribution to the proton mass.  Here, the focus is on the PVDIS deuteron measurement which will use a 50 ${\mu}$A longitudinally polarized electron beam incident on a 40-cm long liquid deuterium target.  Sub-percent level asymmetry measurements will be made over a wide ($x$,$Q^{2}$) range to allow for measurement of and correction for higher-twist and charge symmetry violation hadronic effects to allow for extraction of the fundamental electron-quark couplings of interest.

In addition to the PVDIS measurement, the SoLID 3-D nucleon tomography (or SIDIS) program \cite{JeffersonLabSoLID:2022iod} has a connection to fundamental symmetries through  tensor charge, a basic QCD quantity related to the quark/nucleon spin that is defined by the tensor current matrix element, and  
the connection is through nucleon and quark EDMs. 
Nucleon EDMs receive contributions from quark EDMs with  the flavor-dependent nucleon tensor charge being the corresponding coefficient in front of each quark EDM \cite{Ellis:1996dg,Bhattacharya:2012bf,Pitschmann:2014jxa,Xu:2015kta}. 
The SoLID SIDIS program with three approved 
experiments \cite{JLabPR:E12-10-006,JLabPR:E12-11-007,JLabPR:E12-11-108} 
will allow us to obtain essential information on TMDs (transverse-momentum-dependent parton distribution functions) from the neutron/proton in the valence quark region, and determine the flavor separation of TMDs for up and down quarks with unprecedented precision.
With the high-precision measurement of the transversity TMD, SoLID will improve the precision of the neutron and proton tensor charge  determination by one order of magnitude 
compared to existing global analyses 
and allow for their quark flavor separation. 
This will bring the phenomenological precision on the tensor charges in the same ballpark of the lattice calculations, 
as such providing a benchmark test for lattice QCD predictions \cite{FlavourLatticeAveragingGroup:2019iem,PNDME18,Gupta:2018lvp,ETMC20}.
The combination  of  lattice QCD and the SoLID SIDIS  and EIC program~\cite{GAMBERG2021136255,AbdulKhalek:2021gbh} 
input on the tensor charges will provide a very robust  framework to 
interpret future nucleon EDM experiments with much improved sensitivity (see Section~\ref{sect:CPV}), 
probing new physics scales in the tens of TeV range~\cite{Liu:2017olr,Ye:2016prn,Gupta:2018lvp}.

The PVDIS program of SoLID was approved by JLab PAC35 in January 2010, received an ``A'' scientific rating and was allocated 169 days of beamtime by PAC37 in January 2011, and had both its rating and beam time allocation reaffirmed by PAC50 in July 2022. The SoLID collaboration currently consists of $\sim$ 270 members representing 70 institutions from 13 countries. 
A Pre-Conceptual Design Report (pCDR) had gone through multiple reviews and an MIE was submitted to DOE in 2020.
The experiment received a successful Science Review from DOE in March 2021.
Currently, pre-R\&D activities for the detector and DAQ system of SoLID are ongoing and a cold test of the CLEO-II magnet is underway.  It is anticipated that SoLID would run after MOLLER and physics results would be expected in the 2030's.

In summary, the MOLLER and the SoLID PVDIS experiments 
employ the technique of parity-violating electron scattering to unprecedented precision using the 11 GeV electron beam at Jefferson Lab. 
These experiments are made feasible in part by the progress in experimental techniques developed for the recently completed nuclear weak form factor measurements PREX~\cite{PREX:2021umo}  and CREX~\cite{CREX:2022kgg}  (Section~\ref{sect:pvesh}) such as the control and correction of the polarization-induced asymmetry in the electron beam at the part per billion level and the accuracy in monitoring the electron beam polarization at the 0.5\% level.
Both experiments 
represent a special opportunity to probe physics beyond the Standard Model, with a unique window to new physics at MeV and multi-TeV scales.  They will significantly improve our knowledge of flavor- and CP- conserving interactions at low energies.  The results will be complementary to the anticipated results from the high luminosity running at the LHC.

 \subsubsection{Precision beta decays}
 \label{sect:beta}
Neutron and nuclear $\beta$-decay have historically played a key role in  defining  the Standard Model (SM) 
and now provide a powerful probe for new physics \cite{Gonzalez2019}. The US fundamental symmetries research program made major strides 
since the last long range plan, helping lead breakthroughs in the theoretical analysis of electroweak radiative corrections and Beyond Standard Model (BSM) scenarios.  The experimental program included the development of the most precise measurement of the neutron lifetime, the highest precision angular correlation measurements for polarized and unpolarized nuclear $\beta$-decay to date, and exciting new technologies such as Cyclotron Radiation Emission Spectroscopy for $\beta$-spectroscopy and Superconducting Tunnel-Junction sensors for nuclear recoil measurements. This intensive effort is motivated by the possibility for ``clean'' predictions from the SM for these decay processes, enabling high precision tests for new physics. The activity and detailed plans for this community are described in detail in two white papers~\cite{nucl-bet-wp2022,NeutronWhitePaper}, from which we draw heavily for this summary. The goals for $\beta$-decay research, explained in the following text, include:
\begin{itemize}[noitemsep,nolistsep]
  \item Firmly establishing the $V_{ud}$ input to the Cabibbo-Kobayashi-Maskawa (CKM) unitarity test:
  \begin{itemize}[noitemsep,nolistsep]
  \item Reducing theoretical uncertainties of the superallowed data sets, targeting uncertainties in the unitarity test for $V_{ud}$ similar to those for $V_{us}$.
  \item Bringing the unitarity test from the neutron decay to the current precision level of the superallowed $0^+ \rightarrow 0^+$ decays.
  \end{itemize}
\item Probing for a Fierz interference term at the $10^{-3}$ level (LHC limits) and developing a strategy to further improve the limit.
\item Exploring new techniques to search for BSM physics uniquely suited to $\beta$-decay, such as sterile $\nu$ branches and tests of the SM helicity structure.
\end{itemize}

Precision studies of neutron and nuclear $\beta$-decay are well motivated  ``broad-band" probes of  Beyond Standard Model (BSM), thanks to the sub-permille precision that can be achieved both 
in theory and experiment. These decays not only probe the structure of the charged-current electroweak interactions to unprecedented levels  ($V-A$,  scalar, and tensor currents), 
but also provide a sensitive tool to search for sterile neutrinos.  We briefly discuss below the main physics drivers before going into details of past accomplishments and future opportunities.

\begin{wrapfigure}{r}{0.5\textwidth}
	\begin{centering}
   \includegraphics[width=0.9\linewidth]{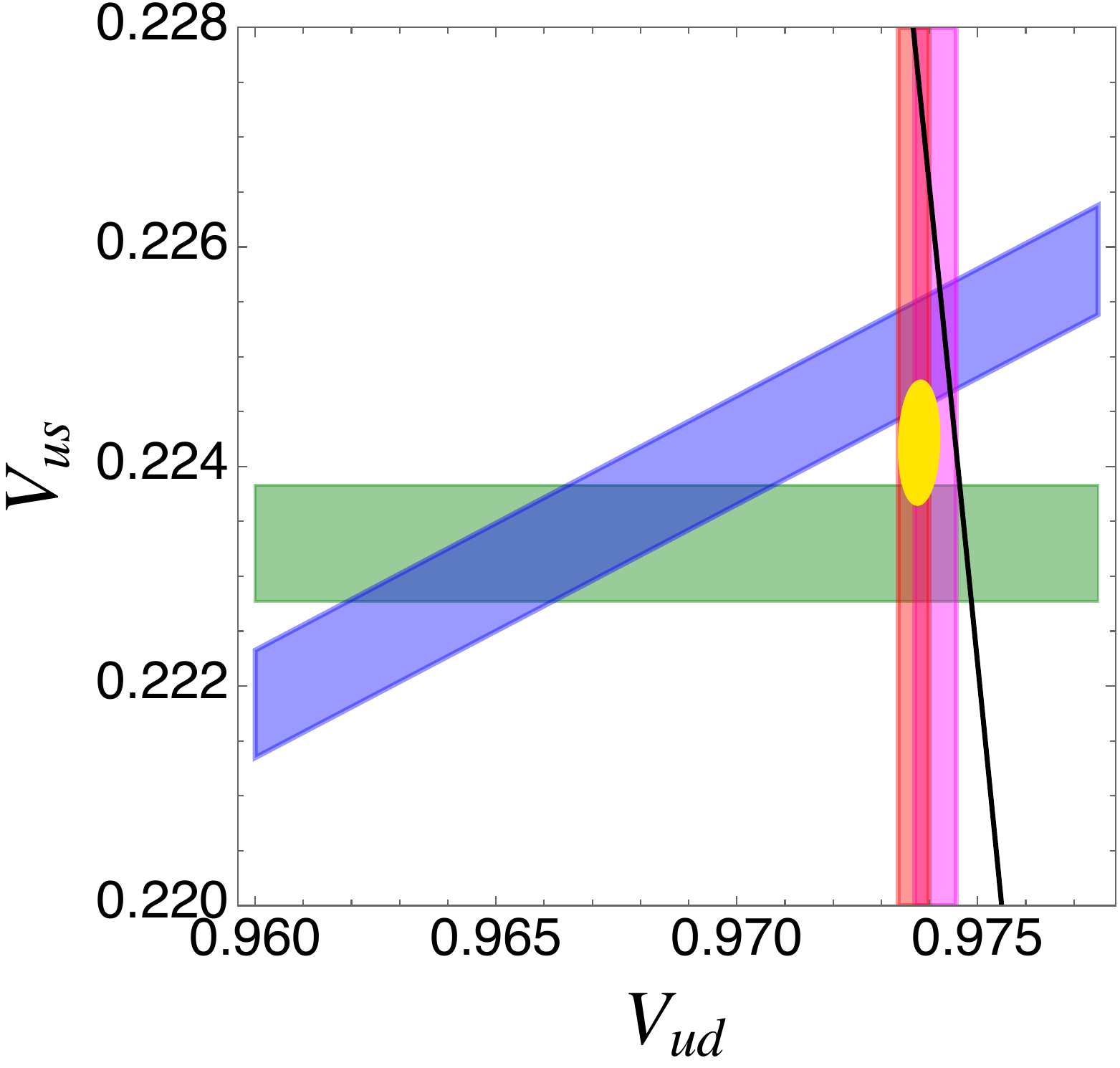}
		\par\end{centering}
	\caption{Values of $|V_{ud}|$ obtained from superallowed $0^+\rightarrow 0^+$ nuclear $\beta$-decays (red) and neutron $\beta$-decay (violet), $|V_{us}|$ from semileptonic kaon decays ($K_{\ell 3}$, green), and $|V_{us}/V_{ud}|$ from leptonic kaon/pion decays ($K_{\mu 2}/\pi_{\mu 2}$, blue). The yellow ellipse represents a global fit of the two matrix elements, and the black line assumes the first-row CKM unitarity. Figure taken with permission from Ref.~\cite{Cirigliano:2022yyo}.
 }
\label{fig:VudVus}
\end{wrapfigure}

Renewed interest arose in first-row CKM unitarity following 
an upward shift in the $K \to \pi$ vector form factor relevant for the extraction of $V_{us}$ from $K \to \pi \ell \bar{\nu}_\ell$ decays~\cite{FlavourLatticeAveragingGroupFLAG:2021npn} and 
 the 2018 analysis of Seng \textit{et al.} \cite{Seng:2018yzq} of the inner radiative correction to neutron and nuclear beta-decay.  Their work, based on a novel dispersion relation analysis~\cite{Seng:2018yzq}, resulted in a reduction in the uncertainty for these corrections and a shift of the previous, state-the-art $|V_{ud}|$ central value~\cite{Marciano:2005ec}. This finding was confirmed by several independent studies~\cite{Czarnecki:2019mwq,Seng:2020wjq,Shiells:2020fqp,Hayen:2020cxh}.
Fig.\ref{fig:VudVus} summarizes the current status of $|V_{ud}|$, $|V_{us}|$, $|V_{us}|/|V_{ud}|$ and displays several anomalies. For instance, the combination $|V_{ud}|_{0^+}^2+|V_{us}|_{K_{\ell 3}}^2-1=-0.0021(7)$ exhibits a deficit from unitarity at the level of 3$\sigma$, and the two different determinations of $|V_{us}|$ from semileptonic and leptonic kaon decays also show a $\sim 3\sigma$ disagreement. They are now known collectively as the ``Cabibbo Anomaly''. 
The current precision is such that the CKM unitarity test is in the league of `precision electroweak observables'. 
Moreover, the Cabibbo Anomaly has  been shown to play an important role in global fits to BSM physics  
and in the interpretation of the $W$ boson mass anomaly~\cite{Cirigliano:2022qdm,Blennow:2022yfm,Bagnaschi:2022whn}.
The current hints of new physics demand a very strong theoretical and experimental program to assess the uncertainties and study the implications for new physics.

The most precise determination of $|V_{ud}|$ presently comes from superallowed $0^+\rightarrow 0^+$ nuclear $\beta$ decays, $|V_{ud}|_{0^+}=0.97367(30)_{\text{th}}(11)_{\text{exp}}$~\cite{Cirigliano:2022yyo,Hardy:2020qwl}, 
where currently the uncertainty is dominated by nuclear structure dependent radiative corrections~\cite{Seng:2018qru,Gorchtein:2018fxl}.
As outlined below, the synergy of ab-initio theory and improved measurements will be the key to reduce these uncertainties. 

On the other hand, free neutron decay is theoretically cleaner, but is limited by experimental uncertainties of the neutron lifetime $\tau_n$ and the axial-to-vector coupling ratio $\lambda = C_A/C_V$. Using the PDG averages, one obtains $|V_{ud}|_n^{\text{PDG}}=0.97441(13)_{\text{th}}(87)_{\text{exp}}$; however, adopting the single best measurement of $\tau_n$ from UCN$\tau$~\cite{UCNt:2021pcg} and $\lambda$ from PERKEO III~\cite{Markisch:2018ndu} respectively returns $|V_{ud}|_n^{\text{best}}=0.97413(13)_{\text{th}}(40)_{\text{exp}}$, with the total uncertainty already comparable to that from superallowed $\beta$ decays. With future improvements in the radiative corrections from lattice QCD and in the experimental precision of $\tau_n$ and $\lambda$, neutron $\beta$ decay could eventually surpass $0^+\rightarrow 0^+$ as the best avenue to extract $|V_{ud}|$.
{\it Addressing the potential  breakdown of unitarity and the Cabibbo Anomaly is the highest priority for the $\beta$-decay community}.

Another sensitive probe of new physics arises from 
a comparison of the experimental  axial-to-vector coupling ratio, $\lambda = C_A/C_V$  and direct computation
from lattice QCD. 
This test  probes the existence of right-handed currents \cite{Alioli:2017ces} and was made possible by a significant increase in precision from lattice QCD determinations of $g_A$ over the last 4 years \cite{Aoki2021}, with individual calculations achieving sub-percent precision \cite{Chang2018, Walker-Loud2020}. Furthermore, attractive solutions to the Cabibbo Anomaly propose the existence of right-handed currents~\cite{Cirigliano:2022yyo}, which position neutron $\beta$ decay to provide independent constraining power on both CKM unitarity and BSM physics scenarios.

Finally, scalar / pseudoscalar and tensor currents can be generated in a variety of BSM scenarios, such as those 
with charged scalars or leptoquarks. 
Scalar and tensor interactions can be probed quite competitively in $\beta$ decays
and in high transverse mass tail of the charged-current Drell-Yan process at the LHC \cite{Cirigliano:2012ab,Alioli:2018ljm,Boughezal:2021tih,Allwicher:2022gkm,Allwicher:2022mcg}. 
By carrying out a comprehensive analysis of $\beta$-decay data, Ref.~\cite{Falkowski:2020pma}
found bounds on the scalar and tensor couplings normalized to the Fermi constant to be at the level of $10^{-3}$, 
which is quite close and complementary to the  Drell-Yan processes at the LHC. 
These bounds on $\epsilon _T$ can be converted to a neutron Fierz interference term $|b| \lesssim 1.3 \cdot 10^{-3}$ (95\,\% CL), which is within reach of the next generation of experiments. High precision $\beta$-spectroscopy is under development to push $\beta$-decay limits well above those from the LHC through Cyclotron Radiation Emission Spectroscopy (CRES), with pilot studies on $^{6}$He and $^{19}$Ne now complete. Limits for T-couplings with right-handed neutrinos have also been improved through $\beta-\nu$ correlation measurements in $^{8}$Li and $^{8}$B decays, with further work expected on the mass 8 system. 
Recent progress~\cite{Hayen2018,Glick-Magid2022formalism,King2022,Glick-Magid2022,Sargsyan2022,Zelevinsky2017} both in theory and experiments make the sub-permille precision an achievable goal.  

Theoretical developments and prospects are discussed in Section~\ref{sect:theory}. Below, we discuss experimental progress and prospects.

\subsubsection*{Nuclear Decays}
The nuclear $\beta$-decay experimental program involves a diverse set of measurements, focusing primarily on (1) input for the CKM Unitarity test and the Cabibbo Anomaly, (2) high precision spectroscopy and angular correlation measurements to probe for exotic Scalar (S) and Tensor (T) couplings, and (3) experiments that constrain sterile neutrino branches (which overlaps with the neutrino physics community).  For each area of activity, we first briefly review progress during the current LRP and then outline prospects for the new LRP.  A brief summary of a few additional recognized R\&D targets is included as well.

\textit{Unitarity Progress: }
During the current LRP period, Hardy and Towner reported numerous refinements to the $0^+ \rightarrow 0^+$ superallowed data set \cite{Hardy:2020qwl}, in particular developing high precision measurements for mirror-superallowed transitions in mass 26, 34 and 38 nuclei to help test the accuracy of corrections due to isospin violating state admixtures. 
While the  impact on the $0^+ \rightarrow 0^+$ superallowed data set was modest, the framework developed to test theory 
will have lasting impact 
for the upcoming LRP, where there is a significant shift of focus to the lowest mass cases to provide the most effective tests of theory.

Superallowed mixed transitions between $T=1/2$ states, also referred to as mirror nuclear decays, have also been proposed as a means to extract $V_{ud}$ \cite{Naviliat2009}.  For these decays, both Fermi and Gamow-Teller decays are possible, with the ratio between Fermi and Gamow-Teller matrix elements for nuclear decays specified by a ``mixing ratio'', $\rho = M _F /M _{GT}$.  While this requires an additional experimental input through an angular correlation, because $\rho$ must be known to high precision, substantial enhancements in sensitivity are available through near-cancellation of the observable.  Cases of interest include a factor of 4 for $^{17}$F up to a factor 13 for $^{19}$Ne~\cite{Hayen2020}, as well as a number of additional cases in the important mass region $A \le 20$.
There has been a surge of experimental activity in the past years around mirror transitions at various institutions world-wide including: half-life ($^{37}$K~\cite{Shidling2014}, $^{21}$Na~\cite{Shidling2018} and $^{29}$P~\cite{Iacob-InPrep}) and branching ratio ($^{37}$K~\cite{Ozmetin-InPrep}) measurements at Texas A\&M University; half-life measurements of $^{11}$C~\cite{Valverde2018},
$^{13}$N~\cite{Long2022}, $^{15}$O~\cite{Burdette2020}, $^{25}$Al~\cite{Long2017} and $^{29}$P~\cite{Long2020} at the University of Notre Dame; $Q_{EC}$-value measurements of $^{11}$C~\cite{Gulyuz2016}, $^{21}$Na and $^{29}$P using LEBIT at NSCL~\cite{Eibach2015}; and with significant development of $99.13(9)\%$ nuclear polarization via optical pumping~\cite{fenkerNJP}, a precise $\beta$-asymmetry measurement of $^{37}$K using TRINAT at TRIUMF improved the value of $V_{ud}$ for this isotope by a factor of 4~\cite{Fenker2018}.

\textit{ Unitarity Prospects:} the focus is on a coordinated effort to systematically evaluate nuclear-structure related corrections and reduce the $0^+ \rightarrow 0^+$ superallowed data st's uncertainties to levels comparable to those from $V_{us}$ in the unitarity sum. Given the anticipated emphasis, for theory, on the mass region $6\le A \le 20$ where/ multiple high precision nuclear structure predictions are possible, there is a strong need for more precise measurements of the branching ratio of the $0^+\rightarrow 0^+$ transitions of $^{10}$C and $^{14}$O. These species play an important role in the superallowed data set for $V_{ud}$ and to constrain Fierz terms.
Recently, the use of superconducting tunnel junctions has shown tremendous promise for precision spectroscopy of recoiling ions following nuclear $\beta$-decay with vastly different systematic corrections to traditional approaches \cite{Fretwell2020, Friedrich2021}. Unlike other quantum sensors, the microsecond(s) response time enables both high precision and high count rate spectroscopy. 
Through the planned development of these sensors at radioactive ion beam facilities (the SALER experiment), measurements of these branching ratios could be performed through recoil spectroscopy as a way of avoiding common systematic effects. Additional information on recoil-order and isospin-breaking corrections may be obtained using precise electroweak nuclear radii measurements in several isotriplet systems \cite{Seng:2022epj,Seng:2022inj}. 

The $\beta$-delayed proton decays of $^{20}$Mg, $^{24}$Si, $^{28}$S, $^{32}$Ar and $^{36}$Ca, to be studied at TAMUTRAP~\cite{shidling2021}, will provide alternate $0^+\rightarrow0^+$ cases once the $^{3}$He-LIG system at the Cyclotron Institute is fully commissioned.  This program  connects mass $A=20$ to a set of near-proton-dripline cases and extends the high precision data-set. These measurements have vastly different experimental systematic uncertainties to the standard super-allowed cases and provide a demanding test of isospin-symmetry-breaking corrections.

For the nuclear mirror decays, there are multiple on-going analyses and planned half-life, branching ratio, and $Q_{EC}$-value measurements.  There are also several efforts to measure correlation parameters in these systems, including more precise angular correlation measurements ($\beta -\nu$, $\beta$-asymmetry, $\nu$-asymmetry, \ldots) of K and Rb isotopes with TRINAT.  These experiments will continue to define the cutting edge for polarized nuclear decays, with optical probes of the trapped samples already establishing 0.3\% precision for the $\beta$-asymmetry in $^{37}$K decay \cite{Fenker2018}, with roughly a factor of three improvement envisioned for runs planned early in the LRP period. This is particularly relevant because with this level of precision for the polarization, the recoil asymmetry in $^{37}$K can provide the input for $V_{ud}$ from this nucleus with comparable accuracy to the superallowed decay species. The focus on high precision data for nuclear systems with $A\le 20$ is well aligned with plans for $\beta-\nu$ measurements in multiple mirror transitions (including the very sensitive $^{17}$F and other isotopes with $A\le 20$) with  the St.\ Benedict ion trapping system~\cite{Brodeur2016, OMalley2020}.  Precision recoil spectroscopy opens an alternate technology for these measurements as well, with SALER targeting measurements of mirror isotopes such as $^{11}$C, where their long lifetimes make them less suitable for ion or optical trap measurements.  

The envisioned experimental campaign should permit a coordinated benchmarking of theory with systems like $^{10}$C, $^{11}$C, $^{14}$O and other nuclei with $A\le 20$ to establish more rigorous uncertainty budgets for nuclear structure dependent corrections. This then sets the stage for a systematic revision of the uncertainties for $V_{ud}$ over the full range of nuclear decays.

\textit{Exotic Couplings Progress}: For left-handed neutrinos, the strongest limits on exotic S-couplings are derived from an endpoint analysis of the $0^+ \rightarrow 0^+$ superallowed decays \cite{Hardy:2020qwl}.  The limits for T-couplings with left-handed neutrinos are constrained through global fits to neutron and nuclear decays (which also constrain S-couplings at comparable levels to the endpoint analysis) \cite{Falkowski2021}.  In addition to these constraints, atom-trap and ion-trap techniques have been used to collect and suspend samples of $\beta$-emitting isotopes, permitting angular correlation measurements.  Experiments with the BPT at the ATLAS facility at ANL, have achieved increasingly precise results for the $\beta -\nu$ angular correlations in $^{8}$Li~\cite{sternberg2015,burkey2022} and $^{8}$B~\cite{gallant2022}. Atom traps have been used to determine this correlation in $^{6}$He~\cite{mueller2022} and to polarize $^{37}$K atoms to measure the $\beta$ asymmetry~\cite{fenkerNJP,Fenker2018}. These experiments have achieved precision as good as 0.3\%, placing limits on the possible existence of tensor interactions and currents with right-handed neutrinos. A recent global fit of nuclear and neutron $\beta$-decay data shows a hint of BSM tensor coupling to right-handed neutrinos at the 3$\sigma$-level \cite{Falkowski2021}. This effect could be generated by various BSM effects such as a TeV-range leptoquark coupling to light quarks, positrons, and right-handed neutrinos, and can be directly confirmed with the inclusion of correlation measurement data of mirror transitions (with well-defined paths to further improve the precision in these cases)\cite{Falkowski2021}.

\textit{Exotic couplings prospects:} The priority for the LRP period is to establish direct sensitivity for exotic couplings at the $10^{-3}$ level for both S- and T-couplings and to define a concrete strategy to push those limits down an order of magnitude.  For Fierz terms, the most promising strategy is based on  Cyclotron Radiation Emission Spectroscopy (CRES) measurements in multiple nuclei,  including $^{6}$He, $^{14}$O and $^{19}$Ne at CENPA.  
The CRES approach, first demonstrated with low-energy electrons \cite{Asner:2015a} has recently been applied to the study of higher-energy $\beta$ particles from the decays of $^{6}$He and $^{19}$Ne~\cite{CRES-arXiv}. This ``non-destructive'' energy measurement has a radically different error budget from conventional, detector based experiments, with a long term goal of sensitivities at the $10^{-4}$ for Fierz terms, probing mass scales near 20 TeV.
By studying both $\beta^{-}$ and $\beta^{+}$ decays, the sign of the Fierz interference term changes, and therefore measuring both significantly reduces most systematic effects. In addition, by confining $\beta$-emitters in a specially-designed Penning trap for CRES measurements, the approach can be extended to study any isotope.

For couplings to right-handed neutrinos, angular correlation measurements offer a path forward with uncertainties at or below 0.1\% planned.  Further increase in sensitivity using the mass-8 system is being pursued and will require access to high-intensity beams of $^{8}$Li and $^{8}$B. New trap-structure designs to minimize $\beta$-particle scattering and efforts to better characterize the detector-array performance will further reduce uncertainties. In addition, a better understanding of the low-lying continuum level structure of $^8$Be, including resolving the question of the existence of low-lying intruder states, and the associated recoil-order contributions will be needed. Additional devices like TAMUTRAP and St. Benedict are poised to further extend the reach of precision angular-correlation measurements as needed, and the HUNTER collaboration \cite{Martoff:2021vxp} is proposing a precision EC spin asymmetry measurement for which linear dependence on tensor couplings offers strong potential for advances. 

\textit{$\beta$-decays for neutrino physics progress and prospects:} Energy and momentum conservation in nuclear $\beta$ decay allows  model-independent searches for  
massive neutrinos 
coupled to the electron flavor, and is a uniquely powerful method for BSM physics searches in this area.  This includes the absolute neutrino mass measurements of the light mass states via $\beta$ decay endpoint measurements as well as the search for new, heavy (mostly sterile) mass states as an expansion to the $3\times3$ PMNS matrix. 
Current and future efforts   towards a sensitivity of $m_\beta \leq 0.04$ eV/c$^2$ at 90\% C.L.~\cite{Project8:2022wqh}  in tritium $\beta$-spectrum measurements   are described in Section~\ref{sect:nu-mass}.
To go beyond $m_\beta \leq 0.04$ eV/c$^2$, however, new experimental paradigms must be considered.  Of growing interest are ultra-low Q value $\beta$-decays that would occur from the ground state of the parent isotope to an excited nuclear state in the daughter with $Q_{ES} = Q_{GS} - E^{*} \lesssim 1$ keV. Such decays could provide new candidates for direct neutrino mass determination experiments~\cite{Cattadori2007_115In} and further insight into atomic interference effects in $\beta$-decay at low energies~\cite{Mustonen2010_ULQCalcs}.  A number of isotopes have been found that could have an ultra-low $Q$ value transition~\cite{Mustonen2010_ULQs,Mustonen2011_135Cs,Haaranen2013_115Cd,Suhonen2014_ULQs,Gamage2019_ULQs,Keblbeck2022_ULQs}, but more precise $Q$ value (from Penning traps), and in some cases energy level data is needed~\cite{Horana2022_75As,Ramalho2022_75As,Ge2022_111In,Eronen2022_131I,Ge2021_159Dy,deRoubin2020_135Cs}.  Experimentally, these ultra-low $Q$ values are challenging to implement, however recent work with trapped nanoscale objects may permit a variety of isotopes to be characterized while reaching sensitivities that are sufficient to resolve the requisite momenta in a single nuclear decay~\cite{Carney:2022pku}.

The search for sub-MeV sterile neutrinos via precision nuclear decay measurements is among the most powerful methods for BSM massive-neutrino searches since it relies only on the existence of a heavy neutrino admixture to the active neutrinos, and not on the model-dependent details of their interactions.  Sub-MeV sterile neutrinos are well motivated, natural extensions to the Standard Model (SM) that have been extensively studied over the past 25 years~\cite{Dod94,Adh17,Boy19}.  
In these experiments, the neutrino ``missing mass" is reconstructed using precise momentum measurement of all other products (including the recoil nucleus) from decay of a nucleus at rest.  The experimental situation is simplified dramatically in neutron-deficient nuclei where the 3-body $\beta$ decay mode is energetically forbidden, and thus the parent nucleus \textit{only} undergoes nuclear electron capture (EC) decay.  Precision measurement of the low-energy nuclear
recoil and all the other (low energy) decay products allows the neutrino four-momentum and mass to be directly probed.  Measurements of this type are currently being performed using $^7$Be decay by the BeEST experiment~\cite{Leach:2021bvh,Fretwell2020,Friedrich2021} and planned for $^{131}$Cs by the HUNTER experiment~\cite{Martoff:2021vxp}.  
The BeEST experiment currently sets the best laboratory limits in the 100 - 850~keV mass range~\cite{Friedrich2021} 
and the projected sensitivities for this class of experiments are impressive (Fig.~\ref{fig:betadecay-sterile-excl}).

Exotic neutrino physics searches also feed into the Reactor-Antineutrino Anomaly, where $\beta$-shape functions are needed to percent-level $\Bar{\nu}$ flux predictions~\cite{fallot2012,hayes2014,sonzogni2015,rasco2016,fijalkowska2017,estienne2019,hayen2019} and provide a method to directly measure dominant backgrounds for dark matter searches and for neutrino-less-double $\beta$ decays ($0\nu\beta\beta$)~\cite{stuckel2022}.

The concrete goals for this community are to realize the current precision targets for the BeEST and the HUNTER experiments. These will place the most stringent constraints on sub-MeV sterile neutrino branches.  $R\& D$ towards more sensitive probes for absolute neutrino mass and opportunities to contribute to related neutrino physics problems will also be pursued.

\textit{Rare decays and other R\&D topics:}  $\beta$-decay research offers other opportunities to constrain BSM physics, motivating ongoing research and development.  For example, it was suggested that the neutron lifetime anomaly could be caused by a dark decay branch~\cite{Fornal2018}, motivating a number of neutron decay experiments (see the ``Neutron Decays" subsection) and papers exploring the implications for neutron stars and astrophysical constraints~\cite{Berryman2022}.  This process could have as a consequence that very loosely bound neutrons in exotic nuclei could decay in a similar way, and the residual nucleus would be the signature of such a decay~\cite{Pfutzner2018}.  A promising candidate is the decay of $^{11}$Be where the dark decay would produce $^{10}$Be as residue~\cite{Ayyad2019,Ayyad2022,Lopez2022}. The question if there is, or if there is not, a signature for a dark decay is not yet solved because of scattered results for the $^{10}$Be production ratio~\cite{Riisager2020}, motivating ongoing measurements on $^{10}$Be and other species such as $^6$He~\cite{Savajols2022}.  Another focus for R\&D in the neutron and nuclear community is the development of improved tests for T non-invariance in $\beta$-decay. For example, the MORA experiment~\cite{Delahaye:2019a,Benali:2020a} and a measurement with TRINAT~\cite{Behr2022BAPS} are being developed for nuclear decays and R\&D for the BRAND experiment~\cite{Bodek:2019a,Dhan:2022a} and a next generation emiT~\cite{Mumm:2011a,Scott:2022a} are underway with neutrons.

\subsubsection*{Neutron Decays}

\begin{figure}[t]
	\begin{centering}
   \includegraphics[width=0.45\linewidth]{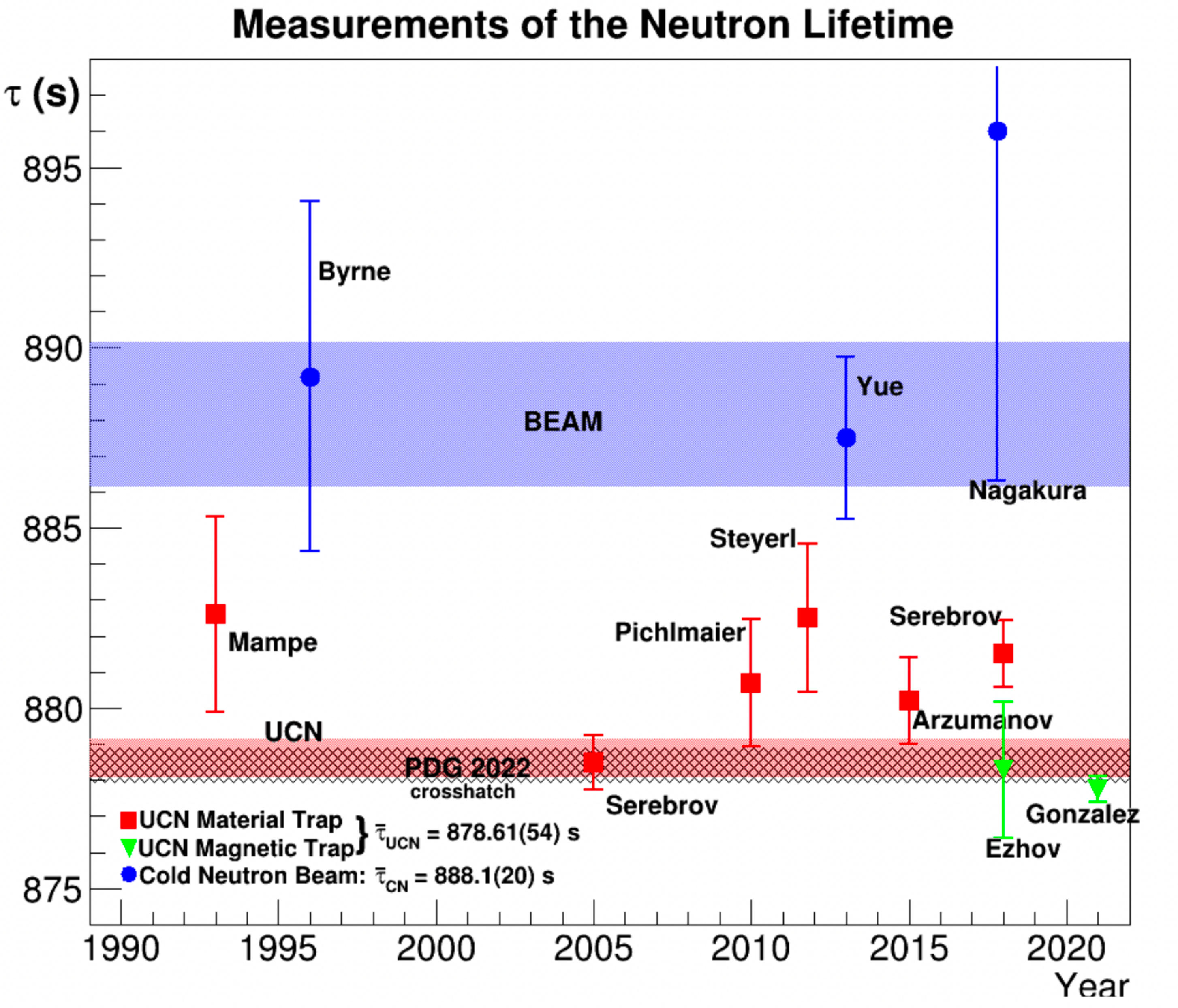}
\hspace{0.75cm}
\includegraphics[width=0.45\linewidth]{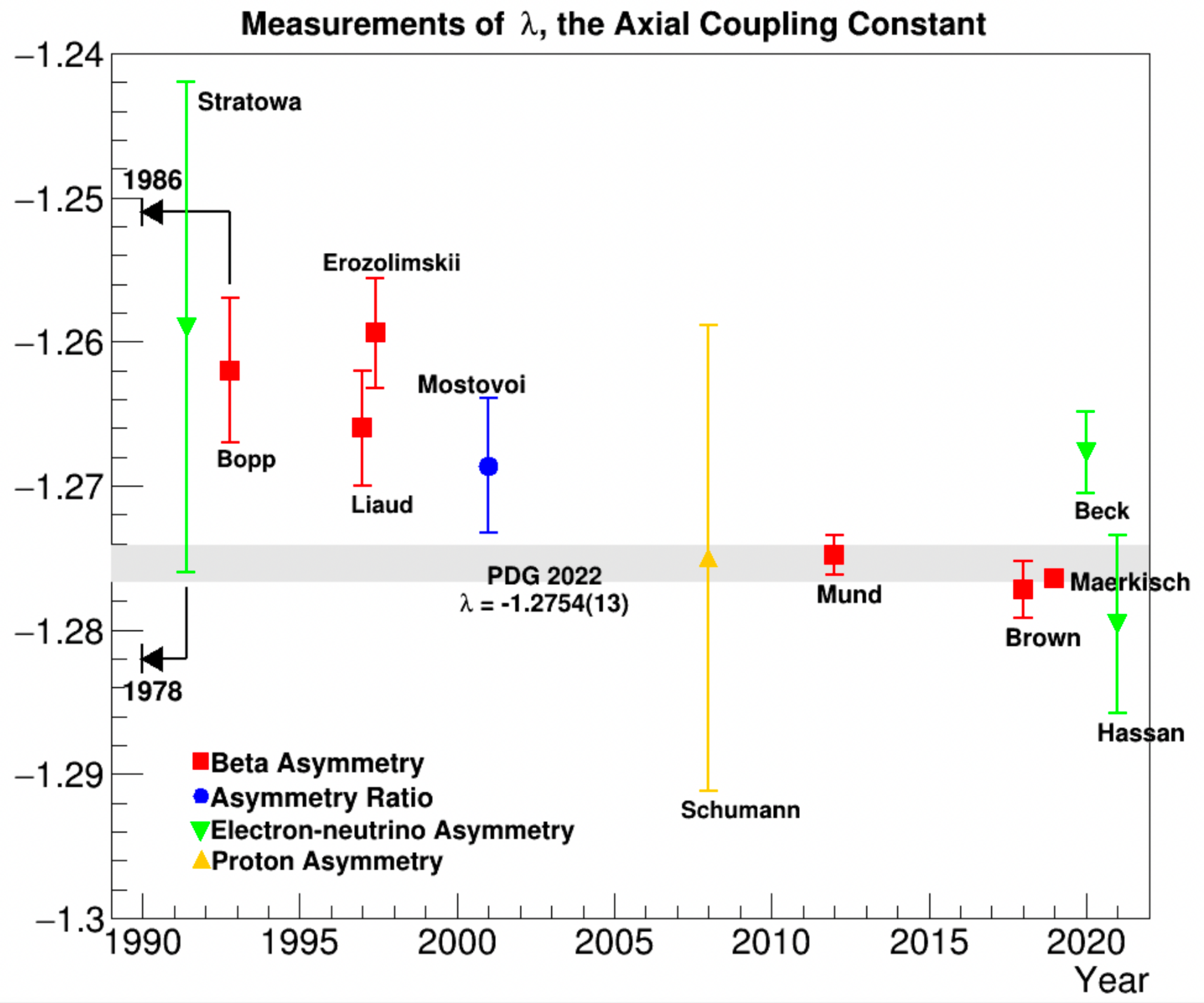}
		\par\end{centering}
  \caption{Left -- most recent and/or precise lifetime results from the global neutron $\beta$-decay experimental program, including measurements using UCN (squares, with ${\bar \tau}_{UCN}=878.6(5)$~s in pink band with uncertainty scaled by 1.9)~\cite{Mampe:1993a,Serebrov2005a,Pichlmaier:2010zz,Steyerl:2012a,Arzumanov2015,Serebrov2018,Ezhov2018,Gonzalez2021}
  and cold neutron beams (circles, ${\bar \tau}_{UCN}=888.1(2.0)$~s in blue
  band)~\cite{Byrne:1996a,Yue:2013a,Nagakura:2017a} and the crosshatch band indicating the PDG 2022
  average~\cite{PDG2022}, $\bar \tau _{PDG} = 878.4(5)$~s with scale factor 1.8.  Right -- most recent and/or precise measurements of $\lambda=g_A/g_V$, including measurements using the $\beta$-asymmetry (squares)~\cite{bopp86,yerozolimsky97,liaud97,Mund:2012a,Brown2018,Maerkisch:2018a}, ratios
  of electron-proton coincidence asymmetries (circles)~\cite{Mostovoi2001}, the proton-asymmetry
  (triangles)~\cite{Schumann:2008b} and the electron-antineutrino asymmetry (inverted
  triangles)~\cite{Stratowa1978,Byrne:1996a,Beck:2020a,Hassan:2021}.
  Also shown is the PDG 2022 average~\cite{PDG2022} $\lambda = -1.2754(13)$ (scale factor $2.7$). 
}
\label{fig:betadecay_world}
\end{figure}

 \textit{Progress:} The potential impact for neutron data to the Cabibbo Anomaly is the primary driver for the neutron decay experimental program, stemming from the fact that $\left | V_{ud} \right |$ determined from the neutron does not rely on nuclear structure dependent corrections.
 Over the last LRP period there has been a burst of productivity from the neutron beta-decay experimental community (see Fig. \ref{fig:betadecay_world}), with four new experimental results for the neutron lifetime and four new (or updated) results for angular correlations. 
 
Within the recently completed or updated lifetime experiments \cite{Arzumanov2015,Ezhov2018,Serebrov2018,Gonzalez2021},  
 one of the most significant steps in the field was an improvement of a factor of two over previous lifetime measurements by the LANL-based UCN$\tau$ experiment, which reported a first physics result in 2018~\cite{Pattie:2018a} and a value of $877.75(22)_{stat}(+22/-16)_{sys}$~s~\cite{Gonzalez2021} in 2021. The UCN$\tau$ experiment uses an asymmetrical, bowl-shaped magnetic trap to store neutrons. UCN are introduced from the bottom of the trap, and the spectra  are such that detected UCN have insufficient energy to overcome the gravitational potential barrier required to exit through the top surface of the bowl.  The stored populations are monitored using an {\it in situ} detector lowered into the trap.  The combination of extremely low UCN loss from the trap, strong control of systematic uncertainties through the {\it in situ} detector, and the large number of stored UCN possible in this high volume trap coupled to the LANL UCN source~\cite{Saunders:2013RSI,Ito2018}, have established this as the highest precision experiment to date.  The UCN$\tau$ result reinforces the ``lifetime puzzle'' --- the noticeably different lifetimes obtained in bottle and beam experiments --- and pulls the global average for the lifetime to $\tau_n = 878.4(0.5)$~s, with a scale factor of 1.8~\cite{PDG2022}, suggesting underestimated systematic uncertainties even within the UCN experiments.  The UCN$\tau$ collaboration also published the strongest direct limits to date for neutron decay to a ``dark" particle with the emission of a $\gamma$-ray~\cite{Tang2018}.

The beam-based BL2 experiment is ongoing at the National Institute of Standards and Technology (NIST), and is designed to probe the systematic uncertainty budget of the BL1 experiment and provide an improved value for the neutron lifetime.  The BL1 experiment was last updated by Yue {\it et al.}~\cite{Yue:2013a}, which provided the driving motivation for the current ``lifetime-puzzle".  Experimental running of BL2 should be complete in 2023, with publication following on a few year time scale and a targeted uncertainty of better than 2~s.  The NIST team also published a high precision measurement of the radiative decay branch in neutron decay in 2016~\cite{Bales:2016a} which stands as the definitive measurement for that process.
Furthermore, a determination of the lifetime from the dependence of the number of thermal neutrons as a function of altitude above the Moon's surface, 
with that surface being the source of thermalized neutrons, has reached a 15~s precision~\cite{wilson_ref}.
The long-standing discrepancy between the reported value for the neutron lifetime in storage and beam experiments remains unresolved, and has led to a flurry of proposals and activity to find additional neutron $\beta$-decay channels \cite{Fornal2018,Dubbers2019,Berezhiani2009,Broussard2022}, with no positive results so far.

There were four new angular correlation measurements: the $\beta -{\bar \nu}$ correlations reported by the aCORN experiment~\cite{Hassan:2021} and the aSPECT collaboration~\cite{Beck:2020a} and the $\beta$-asymmetry measurements reported by the UCNA collaboration\cite{Brown2018} and the PERKEO III collaboration~\cite{Maerkisch2019}.  As can be seen from Fig. \ref{fig:betadecay_world}, the results from beta asymmetry measurements are consistent, but differ by $3\sigma$ from the determination from $\beta -{\bar \nu}$ correlation in aSPECT.

 The UCNA experiment published a ``final" analysis of the $\beta$-asymmetry in 2018~\cite{Brown2018} with a combined result (all UCNA measurements) for the $\beta$-Asymmetry parameter of $A_0= -0.12015(34)_{stat}(63)_{syst}$, which yielded for the axial coupling constant, $\lambda = -1.2772(20)$.  UCNA is the only angular correlation experiment which has used UCN, exploiting the ability to produce and store very highly polarized populations of UCN with negligible neutron-generated backgrounds.  UCNA remains the highest precision, independent cross-check of the cold neutron beam measurements PERKEO II and PERKEO III which define the state-of-the-art determinations of $\lambda$.  The UCNA collaboration also published the most precise limits for neutron decay to dark particles with the emission of $e^+ -e^-$ pairs~\cite{Sun2018}, and the first direct limits on Fierz terms in neutron decay~\cite{Hickerson2017,Sun2020}.
 
 Rapid progress has also been made on the $\beta$-$\bar \nu$ asymmetry.  The aCORN experiment produced the first increment in precision for the $\beta$-$\bar \nu$ parameter $a$ in 15 years with their publication in 2017~\cite{Darius:2017a} and a final result in 2021 of $a = -0.10782(124)_{stat}(133)_{sys}$~\cite{Hassan:2021} and $\lambda = -1.2796(62)$. As mentioned above, the current precision for $a$ is defined by the results of the aSPECT experiment.

\textit{Prospects:} The primary goal for the US neutron $\beta$-decay community during the next LRP is to determine the value of $V_{ud}$ from neutron decay with a precision competitive with the $0^{+}\rightarrow 0^+$ decays, namely $\delta V_{ud} \sim 3 \times 10^{-4}$ or better.  Using the Particle Data Group global averages as the standard, this will require less than a factor of two improvement in the uncertainty for the lifetime, and a factor of 3 for $\lambda$ ($\Delta\tau_n \sim 0.3$~s and $\Delta\lambda/|\lambda|\sim 0.03\,\%$ is needed, which includes understanding of potential discrepancies between methods at this level).  These improvements are within reach.   

There are several US-based lifetime experiments planned or proposed for the next LRP period: UCN$\tau +$ and UCNPro$\beta$e at the LANL UCN source, Space-based lifetime at APL, and BL3 at NIST.  The strategy for UCN$\tau +$ is to improve the statistical uncertainty using an adiabatic transfer technique to load the existing magnetic trap. Because a number of the constraints for key systematic uncertainties (including contributions from quasi-bound UCN and phase-space evolution) are limited by the statistical sensitivity of the experiment, uncertainties below 0.15~s appear feasible. Commissioning and a start for running of UCN$\tau +$ is planned for 2024.  UCNPro$\beta$e is designed to measure the branching ratio for $\beta$-decay relative to all decay modes (the total disappearance rate) for neutron decay.  The sensitivity target for the branching ratio is 1.2~s, 
giving UCNPro$\beta$e the potential to play a critical role if the discrepancy between the NIST beam experiments and UCN storage experiments persists. Commissioning of UCNPro$\beta$e is planned for 2025, with final data taking in 2027. To improve the statistical uncertainty of the lifetime measured with the space-based approach to 3~s, a dedicated mission with orbital measurements or landed lunar experiments has been proposed~\cite{Lawrence:2020tln}. The BL3 experiment builds on the strategies developed in Yue {\it et al.}~\cite{Yue:2013a} for high precision determination of the density of the neutron beam, with a scaled-up trap volume and increased neutron flux at the NG-C beamline.  The ability to achieve 1~s precision in a day of running will ensure that extensive characterization of the systematic error budget will be possible. The initial run at NIST is planned to begin in 2026, with a precision goal $< 0.3$~s.

There are also three US-based angular correlation experiments which are already underway or could be mounted during the next LRP period which can also provide a precision for $\lambda$ comparable to the most precise measurement to date, PERKEO III.  The only experiment currently in commissioning is Nab~\cite{Baes2014,Fry19}, with an expected sensitivity to $\lambda$ of about 0.04\,\%.
First decay data for Nab is possible before the coming shut-down at the SNS in the fall of 2023, and data-taking is planned until roughly 2025. 
This experiment is the first to use the combined electron and proton energy spectra to reach the ultimate sensitivity to the $\beta$-$\bar \nu$ parameter.  This experiment has the potential to resolve the current tension between recent $\beta$-asymmetry measurements and the aSPECT result. It will also provide a critical contribution to the high precision data set, with the measurement subject to a distinctly different set of systematic uncertainties than the previous and on-going $\beta$-asymmetry measurements. 

A natural extension of the Nab experiment can make use of the Nab spectrometer and a polarized neutron beam to perform simultaneous measurements of the $\beta$-asymmetry and angular correlations involving polarized protons.  This experiment, called pNab, will require almost no modification of the existing Nab apparatus, since the capability for highly polarized neutron beams and spin analysis is now incorporated into the baseline capability for Nab. Although this experiment is not yet approved for the FNPB beamline, it would provide a new measurement of $\lambda$ with a goal of $\Delta\lambda/\lambda=0.02\,\%$ and new methods to control sources of systematic uncertainties through coincident detection of electrons and protons and ratios of spin-dependent observables.  
Research and development towards an upgrade of the UCNA experiment is currently underway at the Los Alamos UCN source. This experiment, called UCNA+, would utilize the high UCN densities available in the LANL UCN source to reduce statistical uncertainties and an improved detector package to minimize scattering corrections. These improvements push the projected sensitivity for the $\beta$-asymmetry below 0.2\,\%, making it comparable in sensitivity to Nab and PERKEO III. Given the current uncertainty in the schedule for PERC and the control of key systematic uncertainties through the use of UCN, UCNA+ could have a significant impact.

Overall, the potential impact of the US program on the CKM unitarity test is very high.  The planned neutron lifetime measurements provide a robust basis to establish the lifetime at the required 0.3~s level and a path to clarify the current discrepancy between beam and storage experiments. 
The US is in a leadership position with these measurements.  
  In contrast, there is only one new angular correlation measurement, Nab, mounted at an operational beamline and being commissioned.  A successful measurement at a sufficient precision to effectively achieve the 0.03\,\% goal in $\lambda$, when taken together with PERKEO III is possible.  
  The availability of measurements with significantly different methodology and sources of systematic error has historically been extremely important in this subfield. Given this, 
  although the PERC collaboration pursues similar goals (with somewhat more optimistic precision targets for the $\lambda$ parameter), 
  there is a clear benefit to implementing pNab and/or UCNA+: this will ensure the US program plays a decisive role in the understanding of the Cabibbo Anomaly and maintains worldwide leadership.

\subsubsection{Precision muon and meson experiments}
\newcommand{\abs}[1]{\ensuremath \left|#1\right|}
\newcommand{\pie}{$\pi\to e$}

US Nuclear physicists have been involved in an ambitious program in low-energy muon and pion physics. The experimental campaign in the US, Europe, and Japan represents a targeted set of scientific investigations using high precision tests of the Standard Model or great sensitivity to rare or forbidden processes\footnote{The Fermilab Muon Campus serves the $g-2$ and Mu2e experiments. The multiple pion and muon beam lines at PSI serve, or have served the MuCap, MuLan, MuSun, MUSE, PiBeta, PEN, MEG, Mu3e, and PIONEER experiments. PSI is investing in a next-generation, High-Intensity Muon Beam (HIMB) for the future.  J-PARC is developing a facility for the Museum and Muon g-2/EDM experimental programs.}. These include: the most precise measurement of the muon's anomalous magnetic moment and various charged-lepton-flavor-violating processes to sensitively test the completeness of the Standard Model; determinations of fundamental quantities including the magnetic moment ratio $\mu_\mu / \mu_p$, the lepton mass ratio $m_{\mu} / m_e$, the Fermi constant $G_F$, the muon's electric dipole moment, the proton charge radius $r_p$; and, a program of muon capture to explore elusive features of weak interactions involving nucleons and nuclei (see Ref.~\cite{Gorringe:2015cma}). 
A new rare pion decay program has recently been started~\cite{PIONEER:2022yag}, which will perform the world's most sensitive test of lepton flavor universality, determine $V_{ud}$ through pion beta decay, and set sensitive limits on possible exotic decay modes and the existence of heavy neutrinos.
Here we will briefly highlight the current efforts in which US Nuclear physicists are playing leading roles.

\begin{figure}
\begin{center}
\hfil\null
\vspace{0.1in}
\includegraphics[width=0.8\textwidth]{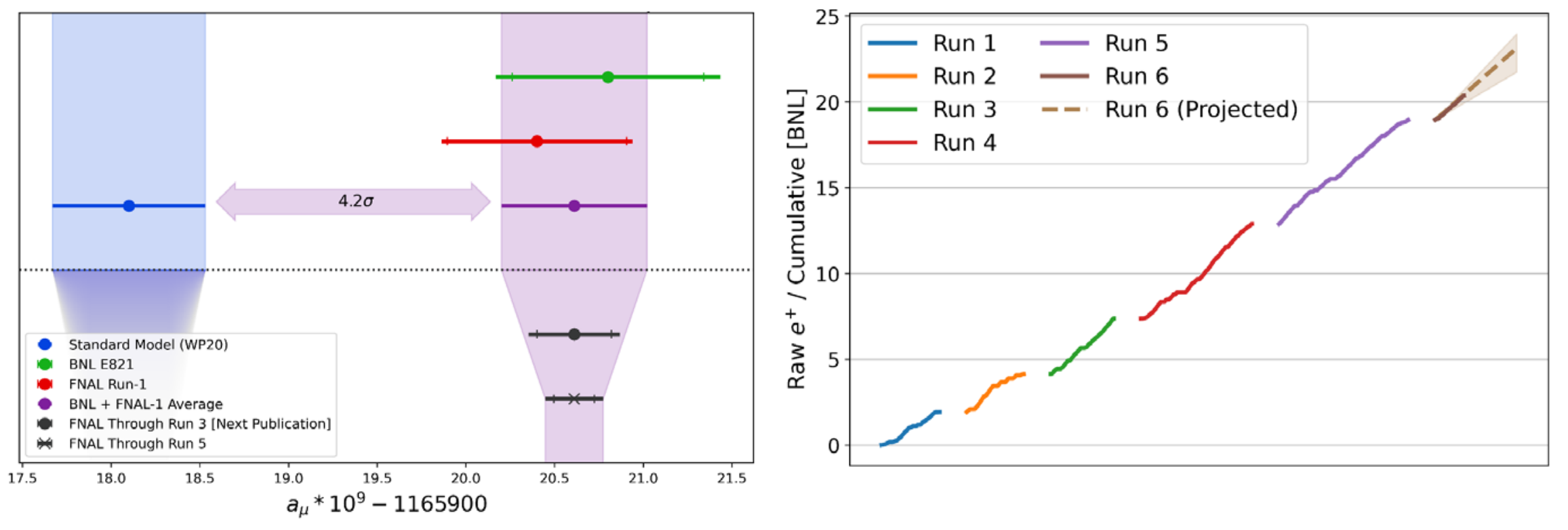}
\caption{Left: The SM prediction for $a_\mu$ is compared to the previous BNL result, the FNAL Run-1 result, and the world average.  The SM uncertainty will reduce in the coming years as will our experimental uncertainty based on already existing data. The placement of the future data points is purely arbitrary.  Right:  The online summary of data accumulation for the existing Runs 1-5, and the anticipated data taking Run-6}
\label{fig:g2result}
\end{center}
\end{figure}

\textit{Muon g-2:} A major highlight of the last Long Range Plan period is the success and first results from the Fermilab muon $g-2$ experiment (E989). It is presently taking data in its 6th and final campaign aiming to measure the muon's anomaly, or $g-2$, to a precision of better than 140\,ppb.  The data acquired to date are sufficient to meet this goal and represent more than 20 times that obtained by the BNL E821 experiment in the early 2000s. Motivating this experiment is the long standing result~\cite{Muong-2:2006rrc} from BNL that exceeds by $\sim 3$ sigma the recent 2020 community White Paper SM prediction~\cite{Aoyama:2020ynm}.   Fermilab E989 -- founded by nuclear physicists in 2009 -- published its first results in April 2021~\cite{Muong-2:2021ojo,Muong-2:2021ovs,Muong-2:2021vma,Muong-2:2021xzz}.  The measurement confirmed the result from BNL and when combined with it, the world average experimental value deviates from the SM prediction by $4.2\,\sigma$, see Fig.~\ref{fig:g2result}. This result has had tremendous impact. The BNL result has more than 3000 citations and the first Fermilab result already has more than 1100. 

Currently, an intense campaign to address the important hadronic vacuum polarization contribution using the tools from lattice is in progress, as well as our continued work by the Muon $g-2$ Collaboration to analyze the large volume of obtained data. 
New results from the Runs-2/3 periods are being planned for release by summer, 2023. Processing of Runs-4/5 is nearly complete, and pre-processing of ongoing Run-6 data is taking place in real time. The many analysis efforts fall into broad categories of muon precession, precision magnetic field, and beam dynamics.  Within each, several groups independently analyze data and evaluate systematics. The outlook is that uncertainties with all previously known categories will be smaller than the Proposal targets and several newly discovered effects are well under control.  These are often addressed using dedicated systematic measurements for beam dynamics related issues, and summer periods then the accelerator is off to study magnetic field transients and calibration procedures.  A chart of data accumulation vs. time is shown in Fig.~\ref{fig:g2result}.  The $g-2$ effort will continue for at least 3 more years of analysis,  documentation, and selected systematic measurements. At present, there are no new plans for additional runs.  

\textit{MUSE:} At the Paul Scherrer Institute in Switzerland, the MUSE Collaboration is poised to take data on low-energy $\mu-p$ and $e-p$ scattering in an effort to contribute to the long-standing quest to understand the charged proton radius. While originally arising from the high-precision muonic-hydrogen spectroscopy Lamb-shift experiments, there has been an engaged world effort to revisit both hydrogen spectroscopy experiments and low $Q^2$ electron scattering data on the proton.  MUSE has a unique approach to the latter. This experimental campaign is ongoing and has been presented in the 2022 Town Hall Meeting on Hot \& Cold QCD.

\textit{PIONEER} is a new opportunity that is well aligned with this Long Range Plan timeline.  It is a recently approved next-generation, rare-pion decay experimental program~\cite{PIONEER:2022yag} that will take place at the Paul Scherrer Institute (PSI) in Switzerland. The Program Advisory Committee stated that they ``enthusiastically support this proposal with high priority.”\footnote{PAC report: R-22-01.1: Studies of rare pion decays (PIONEER) (D. Bryman, D. Hertzog, T. Mori et al.) The PAC also stated that ``The theoretical motivations for these measurements as tests of the SM are very strong and have become more urgent in recent years in connection with possible violations of lepton universality and CKM unitarity. This is an ambitious, long-term project with a strong collaboration."}

\begin{figure}[b]
\centering
\includegraphics[width=0.8\textwidth]{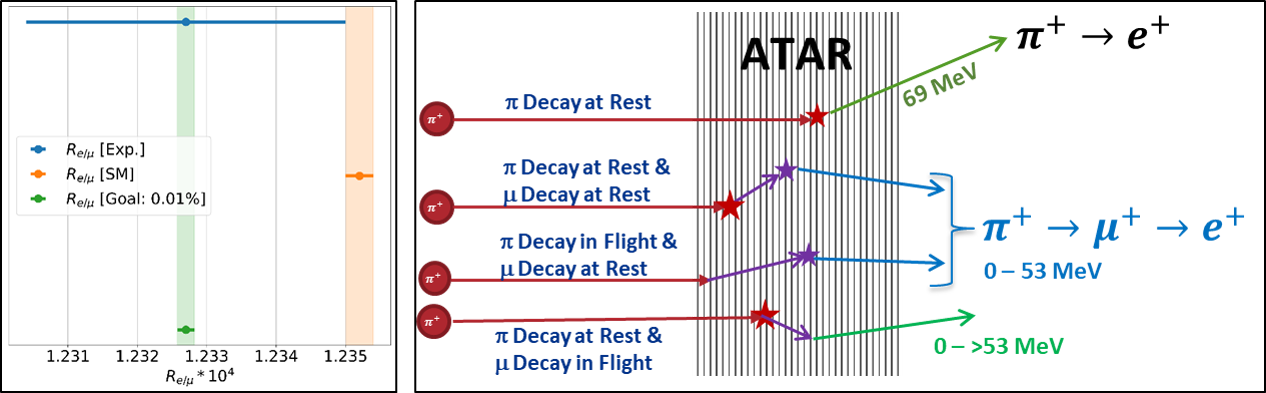}
\caption{Left: Current experimental measurement of $R_{e/\mu}$, the SM theoretical prediction, and the proposed precision goal of the PIONEER experiment. Right: The key event types involved in the measurement with their topologies as one might observe using the highly segmented active target (ATAR). 
}
\label{fig:Theory-Experiment}
\end{figure}

Phase I of PIONEER will focus on a  measurement of the charged-pion branching ratio to electrons vs. muons -- $R_{e/\mu}$.  This is a test of lepton flavor universality (LFU) to be performed at an order of magnitude greater sensitivity than previous experiments.  It is strongly motivated by a variety of anomalies in flavor physics\footnote{See Marciano, Pich, Dror, Hoferichter, and Crivellin presentations at the \href{https://indico.cern.ch/event/1175216/}{Rare Pion Decay Workshop}.} 
including the above mentioned Muon $g-2$ results, $B$-quark meson LFU violation results, and even the current tension with CKM unitarity~\cite{Coutinho:2019aiy,Crivellin:2020lzu}. 
The program will probe non-SM explanations of these anomalies through sensitivity to quantum effects of new particles across a wide range in their masses. 
The $R_{e/\mu}$ ratio is predicted (Cirigliano and Rosell~\cite{Cirigliano:2007ga}) to 1 part in $10^{4}$, 15 times more precisely than the current experimental result (Fig.~\ref{fig:Theory-Experiment}, Left). PIONEER is being designed to distinguish the different primary pion decay modes using both topology and energy information in the target (Fig.~\ref{fig:Theory-Experiment}, Right) and high-resolution calorimetry. The goal is to match the precision of theory.   

Later Phases of PIONEER will measure the ultra-rare ($\sim 10^{-8}$ BR) pion beta decay process $\pi^+ \to  \pi^0 e^+ \nu$ to determine $\abs{V_{ud}}$ in the most theoretically pristine manner. A clean determination will contribute to the evaluation of CKM unitarity, which is very important in light of the recently emerged tensions. 
Throughout all Phases, PIONEER will improve the sensitivity to a host of exotic decays, including probes for the effects of heavy neutrinos~\cite{Shrock:1980vy,Shrock:1980ct,Abela:1981nf,Minehart:1981fv,Bryman:1983cja,Azuelos:1986eg,Britton:1992pg, PiENu:2015seu,PIENU:2017wbj,PIENU:2019usb, Bryman:2019ssi,Bryman:2019bjg, PIENU:2021clt}, unique capabilities to search for pion decays to various light dark sector 
particles~\cite{Altmannshofer:2019yji,Dror:2020fbh,Batell:2017cmf,PIENU:2021clt},
and lepton-flavor-violating decays of the muon into light new physics
particles $\mu \to e X$.

\begin{figure}[htb]
\centering
\includegraphics[width=0.8\textwidth]{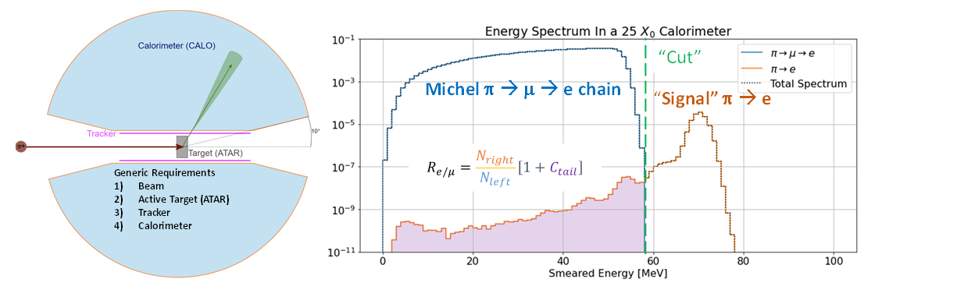}
\caption{Left: Generic components of the PIONEER experimental setup.  The intense positive pion beam enters from the left and is brought to rest in a highly segmented active target (ATAR).   Decay positron trajectories are measured from the ATAR to an outer electromagnetic calorimeter (CALO) through a tracker.  The CALO records the positron energy, time and location. Right: Representative positron energy spectra from  muon decays (blue) and from \pie~ decays (orange) for a high-resolution calorimeter. 
The shaded region below the 58\,MeV dashed cut line indicates the hidden $\pi \rightarrow e$ tail fraction that must be determined to accurately extract the branching ratio.  ATAR-based event reconstruction is designed to reduce the Michel spectrum to a level below the shaded tail. 
}
\label{fig:GenericDetector}
\end{figure}

The experimental design is benefitting heavily from the collective experiences of the previous generation rare-pion decay experiments PIENU (TRIUMF) \cite{PiENu:2015seu} and PEN/PiBeta (PSI) \cite{PEN:2018kgj, Pocanic:2014jka}), and from advanced, state-of-the-art detector and electronics technologies that were not available in the past. The generic experiment is sketched in the left panel of Fig.~\ref{fig:GenericDetector}, highlighting four core requirements:  1) A high-intensity, low-momentum positive pion beam having good $\Delta P/P$, a small spot size, and low muon and positron contamination; 2) a highly segmented, low-mass, stopping target with excellent energy, time, and spatial resolutions; 3) a high-resolution, large-acceptance electromagnetic calorimeter with 25 radiation length depth to fully contain positron showers; and, 4) a fast, thin, cylindrical tracker surrounding the target to connect trajectories from the target to the calorimeter.  These core systems require calibration and monitoring instrumentation and state-of-the-art high-speed digitization and a triggerable data acquisition system.
In each case, active development and R\&D efforts are ongoing, which require initial funding support. Brief comments follow.

The appropriately modified PSI PiE5 beamline will be used following its current commitment to the MEG\,II and Mu3e cLFV experiments.   The PIONEER Collaboration did carry out initial beam tuning measurements in June 2022; in the future, both prototype detector tests and beam-development runs will take place at PSI. 
The active target (ATAR) is the key advance for PIONEER. 
Its 5000 Low Gain Avalanche Detector (LGAD) strips will enable tracking, timing and energy information of the stopping pion, its subsequent decay trajectory to either to a 4.2\,MeV muon (99.99\% BR) or its rare decay to a positron ($1.23\times 10^{-4}$ BR). It is designed to distinguish these events in a subset of data. This will enable that the true calorimeter response of the $\pi$-to-$e$ channel below the Michel electron end point to be observed directly.
The calorimeter must resolve the $\pi$-to-$e$ and $\mu$-to-$e$ channels with resolution at or better than 2\% for the $\pi$-to-$e$ monoenergetic positron energy of 69 MeV.  The lead 
 technological solution follows from the MEG-II experience with LXe, which is intrinsically uniform in an open spherical geometry. 
As an alternative tapered LYSO crystals, creating an segmented ball much like the PEN calorimeter are also being explored in bench tests. 
The right panel of Fig.~\ref{fig:GenericDetector} illustrates the anticipated calorimeter energy spectrum from both Michel decays and the rare $\pi \rightarrow e$ channel. 
A fast, segmented tracker is envisioned to surround the ATAR, providing a link between the positron tracks emerging from ATAR and the calorimeter.  Resistive Micro Well ($\mu$-RWELL) technology appears to be a good candidate.
Finally, the experiment will require custom high-speed, electronics, digitization, and data acquisition efforts and a complete simulation framework. These efforts are already in progress.

\textit{The Collaboration} The PIONEER collaboration consists of participants from PIENU, PEN/PiBeta, and MEG II, as well as international experts in rare kaon decays, low-energy stopped muon experiments, the Muon $g-2$ experimental campaign, high-energy collider physics, neutrino physics, and other areas. Groups from the US are currently funded from DOE HEP and NP offices and from NSF Nuclear Physics.  Groups from Canada, Japan, Switzerland, and Germany are supported by their respective funding sources.

\subsubsection{Hadronic Parity and Time-Reversal Violation}
\def\lsim{\mathrel{\rlap{\lower4pt\hbox{\hskip1pt$\sim$}}
    \raise1pt\hbox{$<$}}}         
\def\gsim{\mathrel{\rlap{\lower4pt\hbox{\hskip1pt$\sim$}}
    \raise1pt\hbox{$>$}}}         
    
    \newcommand{\nc}{\newcommand}  

\nc{\ra}{\rightarrow}  
\nc{\slsh}{\slash\hspace*{-0.22cm}}
\def\ie{{\it i.e.}}
\def\eg{{\it e.g.}}
\def\etc{{\it etc}}
\def\etal{{\it et al.}}
\def\ibid{{\it ibid}.}
\def\to{\rightarrow}
\def\Re{{\cal R \mskip-4mu \lower.1ex \hbox{\it e}\,}}
\def\Im{{\cal I \mskip-5mu \lower.1ex \hbox{\it m}\,}}
\def\be{\begin{equation}}
\def\ee{\end{equation}}
\def\bea{\begin{eqnarray}}
\def\eea{\end{eqnarray}}
\def\bit{\begin{itemize}}
\def\eit{\end{itemize}}
\nc{\eref}[1]{(\ref{#1})}
\nc{\Eref}[1]{Eq.~(\ref{#1})}

\nc{\vev}[1]{ \left\langle {#1} \right\rangle }
\nc{\bra}[1]{ \langle {#1} | }
\nc{\ket}[1]{ | {#1} \rangle }
\nc{\fb}{\,{\rm fb}^{-1}}
\nc{\ev}{{\rm eV}}
\nc{\kev}{{\rm keV}}
\nc{\Mev}{{\rm MeV}}
\nc{\gev}{{\rm GeV}}
\nc{\tev}{{\rm TeV}}
\nc{\mev}{{\rm MeV}}

\def\D{{\cal D}}
\def\L{{\cal L}}
\def\M{{\cal M}}
\def\O{{\cal O}}
\def\W{{\cal W}}

\def\One{\ensuremath{\bf 1}\xspace}
\def\hc{\ensuremath{\mbox{\rm h.c.}}\xspace}
\def\tr{\ensuremath{\mbox{\rm tr}}\xspace}
\def\half{\ensuremath{\frac{1}{2}}\xspace}
\def\thalf{\ensuremath{\frac{3}{2}}\xspace}
\def\third{\ensuremath{\frac{1}{3}}\xspace}
\def\tthird{\ensuremath{\frac{2}{3}}\xspace}

\def\Pl{{\mbox{\scriptsize Pl}}}
\def\lum{{\cal L}}
\def\eff{{\mbox{\scriptsize eff}}}
\def\CM{{\mbox{\scriptsize CM}}}
\def\BR{\mbox{\rm BR}}
\def\ee{e^+e^-}
\def\ppb{{\mbox{p\bar p}}}
\def\qqb{{\mbox{q\bar q}}}
\def\inpb{{\mbox{{\rm pb}^{-1}}}}
\def\infb{{\mbox{{\rm fb}^{-1}}}}
\def\sstw{\sin^2\theta_w}
\def\cstw{\cos^2\theta_w}
\def\mz{m_Z}
\def\gz{\Gamma_Z}
\def\mw{m_W}
\def\mt{m_t}
\def\gt{\Gamma_t}
\def\mh{m_h}
\def\gmu{G_\mu}
\def\GF{G_F}
\def\alphas{\alpha_s}
\def\msb{{\bar{\ssstyle M \kern -1pt S}}}
\def\lmsb{\Lambda_{\msb}}
\def\ELER{e^-_Le^+_R}
\def\EREL{e^-_Re^+_L}
\def\ELEL{e^-_Le^+_L}
\def\ERER{e^-_Re^+_R}
\def\eps{\epsilon}

\newcommand{\NIST}{National Institute of Standards and Technology (NIST) \renewcommand{\NIST}{NIST}}
\newcommand{\LHC}{Large Hadron Collider (LHC) \renewcommand{\LHC}{LHC}}
\newcommand{\LRP}{Long Range Plan (LRP) \renewcommand{\LRP}{LRP}}
\newcommand{\BSM}{Beyond-Standard-Model (BSM) \renewcommand{\BSM}{BSM}}
\newcommand{\CP}{charge conjugation and parity (CP) \renewcommand{\CP}{CP}}
\newcommand{\FNPB}{undamental Neutron Physics Beamline) (FNPB) \renewcommand{\FNPB}{FNPB}}
\newcommand{\EDM}{Electric Dipole Moment (EDM) \renewcommand{\EDM}{EDM}}
\newcommand{\LANL}{Los Alamos National Laboratory (LANL) \renewcommand{\LANL}{LANL}}
\newcommand{\SM}{Standard Model (SM) \renewcommand{\SM}{SM}}
\newcommand{\CKM}{Cabibbo-Kobayashi-Maskawa (CKM) \renewcommand{\CKM}{CKM}}

Searches for P-odd and/or T-odd effects in nuclei can test the Standard Model (SM) and uncover BSM interactions. NN weak interaction amplitudes probe one of the most poorly-understood sectors of the SM. The relative sizes of different quark-quark weak interaction amplitudes, which in turn induce NN weak interactions, are very sensitive to quark-quark correlations in the nucleon and to low energy nonperturbative NN strong interaction dynamics. The measurement of NN weak amplitudes therefore offers a unique, dynamically-rich regime in which to test the standard  model.  The NN weak interaction is also a test case for our ability to trace symmetry-violating effects of a known, short range quark-quark interaction across many nonperturbative strong interaction scales. This is an exercise that also must be performed for many other searches for symmetry-violating low energy nuclear observables beyond the SM such as electric dipole moments and neutrinoless double beta decay. Interpreting such experiments requires calculation of matrix elements in heavy nuclei, which cannot be directly measured and where theoretical methods give a wide range of results.

The theory of NN weak interactions has undergone a qualitative change. The well-known DDH model used to guide theoretical and experimental work has been supplemented with improved input~\cite{Gardner:2022mxf} and surpassed by theory approaches with a more direct connection to QCD, such as lattice gauge theory~\cite{Wasem:2011tp}, pionless and chiral effective field theories ~\cite{Haxton:2013aca,Schindler:2013yua, deVries2013,Gardner:2017xyl}, and related ``hybrid" approaches involving combinations of lattice and EFT calculations ~\cite{Feng2018, Sen:2021dcb} with dynamical approximations using the $1/N_{c}$ expansion~\cite{Zhu2009,Phillips2015, Schindler2016} and the factorization approximation for nucleon-meson matrix elements~\cite{Gardner:2022dwi}. This work has mapped out a path toward the determination of the 5 low-energy constants in the pionless EFT NN weak interaction and has enabled specific predictions for NN weak processes under different dynamical assumptions. Parallel improvements in the theoretical treatment of strong interactions has led to more reliable predictions for the relative contributions of different NN weak amplitudes in few body systems~\cite{Viviani2014,Hyun2017,Lazauskas2019}.

An additional major goal is to use the NN weak amplitudes determined from few-nucleon measurements to calculate parity-odd observables in mid-mass and heavy nuclei. NN weak interactions induce parity-odd nuclear anapole moments~\cite{Zeldovich1957,Flambaum:1980sb}, whose effect can be measured in experiments using methods from atomic/molecular/optical physics and quantum sensing~\cite{Wood:1997zq}.  Calculations of atomic/molecular structure needed to determine anapole moments from such measurements routinely achieve uncertainties of $<10\%$~~\cite{Hao:2020zlz, Hao2018PRA}, and in some atoms much lower~\cite{Derevianko2007Proceedings}. Improved calculations of anapole moments in light nuclei ($Z \lesssim 10-20$) are believed possible~\cite{Hao:2020zlz}, and promising new approaches for heavy nuclei are being pursued. A new technique to measure nuclear anapole moments of heavy nuclei~\cite{Demille2008} has been demonstrated~\cite{Altuntas2018} which takes advantage of systematically small energy differences between opposite-parity levels in molecules~\cite{Flambaum1985} that can be tuned experimentally to near-degeneracy using external magnetic fields to greatly enhance the P-odd asymmetry compared to experiments with atoms~\cite{Wood:1997zq}.

More experimental and theoretical work in atomic, molecular, and nuclear systems is needed to (over)determine the low energy NN weak interaction amplitudes. Continued extension of the  NN weak EFT calculations to more few body systems is essential for the interpretation of measurements. The initial goal for lattice gauge theory is to calculate the $\Delta I=2$ NN weak amplitude, which is computationally easier to access than the other NN weak amplitudes due to the absence of disconnected diagrams. Additional work on dynamical models which can help develop insight into the physics behind the relative size of different NN weak interaction amplitudes is also needed.   
Opportunities for sensitive experiments exist in few nucleon systems. Parity-odd neutron spin rotation in n-$^{4}$He can reach a sensitivity of $10^{-8}$ rad/meter for the P-odd rotary power in 1/2 year of running on the NG-C beam at NIST to provide a strong constraint on a known linear combination of NN weak amplitudes and can distinguish between recent predictions based on $1/N_{c}$ arguments and a combined renormalization group $+$ lattice-constrained factorization calculation. A pulsed slow neutron beamline proposed for the European Spallation Source~\cite{ANNI2019} could search for (1) neutron-proton parity-odd spin rotation, which is one of the few experimentally-accessible observables with sensitivity to the $\Delta I=2$ NN weak amplitude, (2) the parity-odd gamma asymmetry in $\vec{n}+D \to T+\gamma$, which is a sufficiently simple system to be treatable in terms of two-body NN weak amplitudes~\cite{Song2012}. An upgraded HiGS facility~\cite{Howell:2020nob} could search for parity-odd photodisintegration in $\vec{\gamma}+D \to n+p$ very near threshold, another observable sensitive to the $\Delta I=2$ NN weak amplitude. 

The ZOMBIES experiment projects to build on its recent proof-of-principle to measure anapole moments of several nuclei in the range $Z \gtrsim 40$~\cite{Demille2008, Altuntas2018}, initially $^{137}$Ba in the molecule BaF.  New approaches are being explored~\cite{Norrgard2019, Hutzler2020, Udrescu2022BAPS} to enable measuring anapole moments of very light nuclei, where accurate  nuclear-structure calculations are already being performed.  These rely on the same principle as ZOMBIES, but use recently-developed methods for increased quantum control of molecules---such as direct laser cooling \cite{Shuman2010Nature, Vilas2022Nature}, and quantum state readout of trapped molecular ions \cite{Cairncross2017PRL}---to achieve better energy resolution and enable measurements even of radioactive nuclei~\cite{GarciaRuiz2020Nature}. Several experiments aiming to measure anapole moments in atoms are also in development~\cite{Tsigutkin:2009zz, Gwinner2022QST}. 

Searches of  T-odd correlations
in the transmission of polarized neutrons through polarized targets are sensitive to nucleon-nucleon P-odd/T-odd and P-even/T-odd potentials 
\cite{Bowman:2014fca} and can exploit enhancements due to small energy splitting between resonances of opposite parity in heavy nuclei to reach interesting sensitivity. T-odd interactions from new sources beyond the SM can generate two types of terms in the neutron forward scattering amplitude: a P-odd/T-odd term of the form $\vec{s}_{n} \cdot (\vec{k}_n \times \vec{I})$, where $\vec{s}_{n}$ is the spin of the neutron, $\vec{k}_{n}$ is the neutron momentum, and $\vec{I}$ is the polarization of the nucleus, and a P-even/T-odd term of the form $(\vec{k}_n \cdot \vec{I})((\vec{s}_{n} \cdot (\vec{k}_n \times \vec{I}))$. These two flavors of T violation come from very different types of BSM interactions. A positive signal could reveal the new CPV sources that are needed for baryogenesis but are difficult to probe directly in high energy experiments  \cite{Cirigliano:2016nyn,Alioli:2017ces,Cirigliano:2019vfc,Gritsan:2022php}.  In forward transmission experiments one can realize a null test for T which, like electric dipole moment searches, is in principle free from the effects of final state interactions~\cite{Gudkov:1990tb,Gudkov:2013dp,Bowman:2014fca}. Amplifications of $P$-odd neutron amplitudes in compound nuclear resonances by factors of $~10^6$ above the $~10^{-7}$ effects expected for weak NN amplitudes compared to strong NN amplitudes have already been observed~\cite{Mitchell:1999zz} in several heavy nuclei. 
It is important that these experiments measure the ratio of TRIV effects to PV effects at the same p-wave resonance which leads to almost complete cancellation of nuclear reaction effects.  This is a big advantage comparing to absolute measurements of symmetry violations (like PV effects).
A similar resonance mechanism can amplify a $P$-even and $T$-odd amplitude by a factor of $10^{3}$~\cite{Barabanov:1986sz, Bunakov:1988eb, Gudkov:1991qc}. Direct experimental upper bounds on $P$-even and $T$-odd NN amplitudes~\cite{Huffman:1996ix} are only ~1\,\% of strong NN amplitudes.  

The NOPTREX collaboration is actively pursuing this opportunity through neutron p-wave resonance spectroscopy measurements on several relevant nuclei at LANSCE, FRM, JPARC, and CSNS. A proposal to conduct a P-odd/T-odd $\vec{s}_{n} \cdot (\vec{k}_n \times \vec{I})$ search in $^{139}$La has been submitted to JPARC. NOPTREX can also search for amplified P-odd effects on p-wave resonances in so-far-unmeasured nuclei and help test the statistical theory of symmetry violation in neutron-nucleus resonances~\cite{Tomsovic2000} against the extensive data set from the TRIPLE collaboration, in combination with new information on NN weak amplitudes. More theoretical work is needed to better understand the statistical theory of P-odd/T-odd and P-even/T-odd interactions. 
This includes the development of techniques for the calculation of PV and TRIV nuclear matrix elements using DDH-like  and EFT nucleon  interactions to understand the sensitivity of the TRIV effects to the details of CP-odd nucleon interactions (and as a consequence, to different sources of CP-violations),  and  a statistical spectroscopy approach for the further analysis of the experimental results in terms
of constraints on different models of CP-violation.

\begin{table}[h]
    \centering
\resizebox{\textwidth}{!}{%
    \begin{tabular}{|c|l|r|r|r|r|}
    \hline    
         & Experiment & Observable & Result/Goal & Implications & Facility  \\
         \hline\hline
         & NPD$\gamma$ & $A^{np}_{\gamma}$ & $[-3.0 \pm 1.4(stat) \pm 0.2(sys)] \times 10^{-8}$~\cite{NPDGamma:2018vhh} & n-p weak interaction & ORNL SNS \\
         & n-$^{3}$He PV  & $A_{PV}$ & $[1.58 \pm 0.97 (stat) \pm 0.24 (sys)] \times 10^{-8}$~\cite{n3He:2020zwd} & smallest P-odd NN asymmetry & ORNL SNS  \\
         & n-$^{4}$He & $d\phi_{PV}/dz$ (rad/m) & $[+2.1 \pm 8.3(stat.) \pm 2.9(sys.)] \times 10^{-7}$~\cite{Swanson:2019cld}  & bounds $\mu$eV-eV Z$^{`}$ bosons~\cite{Yan13} & NIST \\
         & ZOMBIES  &  &  & &  \\ 
         & n-$^{4}$He & $d\phi_{PV}/dz$ (rad/m) & 10$^{-8}$ & NN weak theory test & NIST NG-C  \\
         & NDT$\gamma$ & $A_{\gamma}$ & 10$^{-7}$  & NN weak theory test &  ESS \\
         & $\gamma$DNP & $A_{n}$ & 10$^{-8}$ & $\Delta I=2$ NN weak & HIGS2  \\
         & NOPTREX &   ${<V_{PT}>} \over {<V_{P}>}$ & 10$^{-5}$ & P-odd/T-odd NN null search & JPARC \\
         & NOPTREX & ${<V_{T}>} \over {<V_{strong}>}$ & 10$^{-5}$ & P-even/T-odd NN null search  & CSNS/LANSCE  \\ 
        \hline
    \end{tabular}
    }
    \caption{Accomplishments since last LRP and future opportunities in hadronic parity and time reversal violation}
    \label{tab:IUbudgetEquip}
\end{table}

The NPDGamma collaboration reported~\cite{NPDGamma:2018vhh} the parity-odd asymmetry $A^{np}_{\gamma}=[-3.0 \pm 1.4(stat) \pm 0.2(sys)] \times 10^{-8}$ in $\vec{n}+p \to D + \gamma$ to determine the $\Delta I=1$, $^{3}S_{1} \to ^{3}P_{1}$ component of the weak nucleon-nucleon interaction. The n3He Collaboration reported~\cite{n3He:2020zwd} the smallest asymmetry of any parity-odd asymmetry in NN interactions measured so far: $A_{PV}=[1.58 \pm 0.97 (stat) \pm 0.24 (sys)] \times 10^{-8}$ in the emission direction of the proton in polarized neutron capture on $^{3}$He,  $\vec{n}+^{3}$He $\to ^{3}$H $+ p$. Both of these measurements were completed at the FnPB beam at SNS. The final analysis of an upper bound on parity-odd neutron rotary power in n$+^{4}$He measured at NIST of $d\phi/dz=[+2.1 \pm 8.3(stat.) \pm 2.9(sys.)] \times 10^{-7}$ rad/m~\cite{Swanson:2019cld} was published. This null measurement was used to place the most stringent constraints on exotic parity-odd interactions of neutrons with matter with ranges between millimeter and atomic scales.

\subsubsection{Baryon Number Violation: neutron oscillations}
\label{sect:BNV}
A unique opportunity has emerged during the past decade to search for baryon number violation (BNV) through the transformation of neutrons to antineutrons (a $\Delta B = 2$ process) using neutron beams. Such an experiment, sited at the European Spallation Source (ESS) in Lund, Sweden, can improve the experimental sensitivity to antineutron appearance by three orders of magnitude over previous beam measurements, thanks to key design elements incorporated into ESS facility planning: an unprecedented, Large Beam Port coupling cryogenic moderators to a neutron beamline and the availability of a roughly 300~m long footprint for the required drift of free neutrons through an evacuated beam tube to an annihilation target. The development of NNBAR relies on strong US NP scientific leadership as part of an international and multidisciplinary collaboration, and the earliest stage of the program at ESS is already approved. During the upcoming LRP, small-scale experiments at ORNL targeting complementary science provide early R\&D opportunities in this LRP period as a staged program towards the future large scale, high sensitivity search beyond this decade.

Violation of baryon number B is one of Sakharov's required conditions to explain baryogenesis. Both BNV and LNV are predicted in the SM due to non-perturbative electroweak effects but are too small to be observed experimentally. The need for BSM BNV and LNV has motivated construction of large detectors in the past decades to search for proton decay, a $\Delta B=1$ process which is now highly constrained~\cite{Dev:2022jbf} (and as the studied modes conserve $\Delta (B-L)$ these are unable to explain baryogenesis due to the sphaleron mechanism), and searches for $\Delta L=2$ neutrinoless double beta decay to support leptogenesis~\cite{Davidson:2008bu}, which is the highest priority recommendation from this field. Neutron-antineutron oscillations ($\Delta B=2$) offer another, viable path for the BNV required for baryogenesis\footnote{Searches for the $\Delta B=2$ processes $e^- p\rightarrow e^+ \bar{p}$, $e^- p\rightarrow \bar{n}\bar{\nu}$, and $e^- n\rightarrow e^-\bar{n}$ have also been proposed at ARIEL~\cite{ARIEL-wp2022}.}. $\Delta B=2$ processes not only provide a direct test of baryogenesis mechanisms, but also have a symbiotic relationship with the issue of neutrino masses and feature in a number of scenarios of physics beyond the Standard Model (recently summarized:~\cite{Proceedings:2020nzz, Barrow:2022gsu, Dev:2022jbf}).  Current limits correspond to new physics scales of $\sim$100 TeV.

The most compelling approach for an improved search for $n\rightarrow\bar{n}$ uses free neutrons, which is background-free and theoretically cleanest. The last free neutron search at the ILL three decades ago was background-free, with the limit $\tau_{n\bar{n}} > 0.86\times10^8$\,s (90\,\% C.L.)~\cite{BaldoCeolin:1994jz}. 
Since then, $n\rightarrow\bar{n}$ searches have been performed using a complementary approach in large volume detectors, with the best limit from SuperKamiokande of $\tau_{n\bar{n}} > 4.7\times10^8$\,s (90\,\% C.L.)~\cite{Super-Kamiokande:2020bov}. DUNE can reach  $\tau_{n\bar{n}} > 5.53\times10^8$\,s (90\,\% C.L.)~\cite{DUNE:2020ypp} or beyond depending on how well backgrounds and other systematics are controlled~\cite{Barrow:2021odz}.  Leveraging the significant advances in neutronics and detection technology in the past few decades, an improved search using free neutrons in NNBAR can reach $\tau_{n\bar{n}}\sim10^{9-10}\,$s~\cite{Addazi:2020nlz}.

The ESS, which will be the world's most powerful spallation source, has recognized particle physics as a priority~\cite{CapGapESS2018}.  In addition to the world-unique Large Beam Port and space for a  300\,m beamline earmarked for NNBAR, an ESS team is also engaged in an optimization of the lower moderator for fundamental physics including NNBAR~\cite{Abele:2022iml}. The NNBAR Conceptual Design Report is being prepared for late 2023 by institutes in Sweden, Denmark, Germany, France and the US, and should include an evaluation of the scope of experiment construction; progress was recently reported~\cite{Backman:2022szk}. While the time horizon of NNBAR is beyond this LRP period and will require international collaboration and support, modest support for US-based R\&D efforts in this decade to develop techniques and technology, and to understand effects such as neutron phase preserving collisions~\cite{Kerbikov:2018mct,Nesvizhevsky:2018tft} which can reduce the project scale, is needed. These efforts will ensure ongoing ESS construction is optimized for NNBAR, enable preparation for the next LRP period, and develop US NP workforce continuity (with key expertise in tracking calorimeter construction, neutron transport and neutron source technology) and leadership in the project.

A program of smaller scale searches for related $\Delta B=1$ processes of sterile or ``mirror'' neutron oscillations $n\rightarrow n'$~\cite{Berezhiani:2005hv} are underway through a collaborative effort between US universities, ORNL and ESS, addressing questions including BNV, the nature of dark matter, and anomalies such as the neutron lifetime discrepancy. Constraints here are weaker and were obtained using UCN disappearance~\cite{Serebrov:2008her, Altarev:2009tg, nEDM:2020ekj, Mohanmurthy:2022dbt} with some reported anomalous signals~\cite{Berezhiani:2012rq,Berezhiani:2017jkn} and using fast neutrons from reactors~\cite{Almazan:2021fvo, Stasser:2020jct}. At ORNL, cold neutron regeneration techniques~\cite{Berezhiani:2017azg} were used to exclude $n\rightarrow n'$ as an explanation for the neutron lifetime anomaly~\cite{Broussard:2021eyr}.  
Further studies are ongoing, supported by DOE NP and Swedish Research Council + ESS, which leverage the technical capabilities at ORNL including the Small Angle Neutron Scattering instrument at the High Flux Isotope Reactor through the General User program as well as expertise in neutron source, transport, and characterization~\cite{Broussard:2017yev, Broussard:2019tgw}. These early measurements inform a proposed high sensitivity search for $n\rightarrow n'$ in the dedicated HIBEAM experiment at ESS~\cite{Addazi:2020nlz}, where a low-sensitivity stage-0 $n\rightarrow n'$ project is approved and funded by ESS, and planned for 2027--2028. The later stages of HIBEAM are being designed with key US NP participation and are timely opportunities for this LRP period. The end goal of HIBEAM includes a first search for $n\rightarrow n'\rightarrow \bar{n}$~\cite{Berezhiani:2020vbe} and a low sensitivity search for $n\rightarrow \bar{n}$ at similar sensitivity to ILL. These developments in the next decade will lay the groundwork for a exceptional opportunity to improve experimental sensitivity to $n\rightarrow \bar{n}$ by $\times$1000 in the NNBAR experiment to address the critically important question of baryogenesis.

\subsection{Properties of neutrinos and hypothetical light particles} 
\label{sect:light}

\subsubsection{Absolute neutrino mass measurements and sterile neutrinos}
\label{sect:nu-mass}
The existence of non-zero neutrino mass, established by a suite of neutrino flavor-oscillation experiments~\cite{Super-Kamiokande:1998kpq, SNO:2001kpb, SNO:2002tuh, KamLAND:2002uet, SNO:2008gqy, T2K:2011ypd, DayaBay:2012fng, RENO:2012mkc}, is the first direct evidence of beyond-Standard-Model physics in a laboratory setting. Knowledge of the absolute neutrino-mass scale -- not just the splittings that oscillation experiments measure between the neutrino-mass values $\{m_1, m_2, m_3\}$ -- is essential information to understand how neutrino masses should be incorporated into the Standard Model~\cite{Formaggio:2021nfz}. The neutrino-mass scale is also crucial for other cutting-edge scientific investigations. 

Since neutrinos comprised a significant fraction of the energy density in the early universe, their mass, energy spectrum, and associated collision-less damping scale shaped the formation of large-scale structure. The cosmic microwave background, including lensing, is imprinted with the signature of the neutrino mass. Recently, the Dark Energy Survey has inferred a limit of $\Sigma m_i < 0.13$~eV (95\% C.L.) based on a combination of their own gravitational-lensing and galaxy-clustering data with data sets from baryon acoustic oscillations, Type Ia supernovae, redshift-space distortions, and the Planck cosmic-microwave-background measurement~\cite{DES:2021wwk}; next-generation observations target a sensitivity allowing $>3\sigma$ detection of $\Sigma m_i$, even at the normal-ordering floor of $\Sigma m_i = 0.06$~eV~\cite{Adhikari:2022sve}. Of course, these tight bounds must be interpreted within the $\Lambda CDM$ concordance model. As explored in, e.g., Ref.~\cite{Abazajian:2022ofy}, external input on the mass scale will test and sharpen the cosmological standard model.

If neutrinos have a Majorana nature, the rate of neutrinoless double beta decay (Sec.~\ref{sect:LNV}) depends on the effective neutrino-mass observable $\langle m_{\beta \beta} \rangle = \left| \sum U^2_{ei} m_i \right|$ -- presuming that the process is dominated by light-neutrino exchange. (Here, $U_{ei}$ denotes the element of the PMNS matrix that couples the electron flavor to the $i^{th}$ neutrino-mass eigenvalue.) 
The next generation of experiments aims to probe the entire region allowed by the inverted mass ordering, and may even have sensitivity in the normal mass ordering case when the lightest neutrino mass eigenvalue is $> 10\,{\rm meV}$. However, if neutrinos are \textit{not} Majorana fermions, then the neutrinoless double-beta decay rate will be zero and these experiments will have no sensitivity to the neutrino-mass scale. Even in the case of Majorana fermions, it must also be noted that, if the lightest neutrino mass value is less than about 0.002~eV, $\langle m_{\beta \beta} \rangle$ is relatively insensitive to the mass scale in both the inverted and normal orderings.

The most sensitive direct (i.e., model-independent) way to measure neutrino mass is via high-precision spectroscopy of $\beta$ decay near its endpoint~\cite{Formaggio:2021nfz}. The spectral shape reveals the neutrino-mass scale $m_\beta$:
\begin{equation}\label{eqn:m_beta}
 \ m_{\beta}^2 = \sum_{i=1}^3 \left| U_{ei} \right|^2 m_i^2 = m_1^2 + \left| U_{e2} \right|^2 \Delta m_{21}^2 + \left| U_{e3} \right|^2 \Delta m_{31}^2 
\end{equation}
The mass-squared splittings $\Delta m_{ij}^2$, measured in oscillation experiments, give a lower limit on $m_\beta$:  $m_{\beta} > 0.048$\,eV/$c^2$ if $m_3$ is the lightest state (inverted ordering), and  $m_{\beta} > 0.0085$\,eV/$c^2$ if $m_1$ is lightest (normal ordering) (95\% confidence level)~\cite{ParticleDataGroup:2020ssz}. This sensitivity arises directly from conservation of energy in $\beta$ decay: the mass 
of the neutrino represents a packet of energy that the $\beta$ cannot carry away.

Assuming CPT symmetry, a spectral measurement for any type of $\beta$ decay -- $\beta^+$, $\beta^-$, or electron capture -- can reveal $m_\beta$. In practice, a relatively low $Q$ value is favored because it increases the fraction of decays that carry useful information about $m_\beta$. Significant effort has been invested in measuring  spectra from ${}^{163}$Ho electron-capture decays ($Q= 2.83$~keV), primarily by the ECHo~\cite{Mantegazzini:2021yed} and HOLMES~\cite{Alpert:2014lfa} collaborations; current efforts target a sensitivity on the order of $m_\beta < 20$~eV. However, direct $m_\beta$ limits have historically been driven by the sensitivity achievable with tritium $\beta$ decay, ${}^3\mathrm{H} \rightarrow {}^3\mathrm{He}^+ + \beta^- + \bar{\nu_e}$, which is super-allowed with a $Q$ value of 18.6~keV and a half-life of 12.3~years. 

\begin{figure}[htb]
    \centering
    \includegraphics[width=0.9\textwidth]{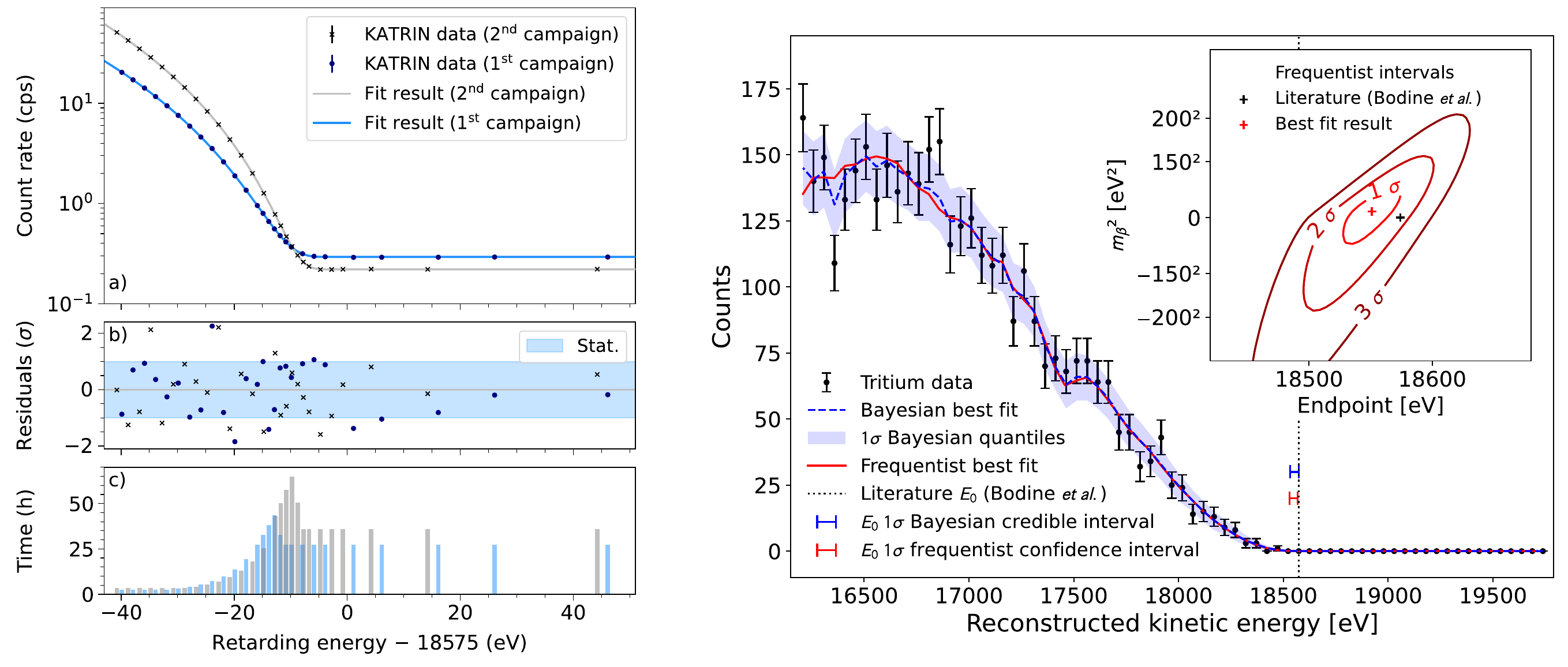}
    \caption{ \textbf{Left}: Top: Measured KATRIN endpoint tritium integrated spectra from the first and second neutrino-mass measurement campaigns, along with the results of a joint fit with a common neutrino-mass-squared parameter. Center: Residuals from the two campaigns, normalized to the uncertainties. Bottom: Measurement time distribution. The ``retarding energy'' is the energy threshold set by the MAC-E filter. Adapted from Ref.~\cite{KATRIN:2021uub}. \textbf{Right}: First tritium spectrum measured by Project~8 using CRES, along with Frequentist and Bayesian fits incorporating knowledge of the detector response. Figure from Ref.~\cite{Project8:2022hun}.}
    \label{fig:beta_spectrum}
\end{figure}

The Karlsruhe Tritium Neutrino (KATRIN) experiment has set the current world-best limit, $m_{\beta} < 0.8~$eV (90\% C.L.)~\cite{KATRIN:2021uub}, using integral spectroscopy of $\beta$s from T$_2$ decay. These results, based on the two measurement campaigns in 2019 (Fig.~\ref{fig:beta_spectrum}), are statistics-limited; KATRIN operations have continued and the next neutrino-mass result, anticipated this year, is expected to achieve a sensitivity of about 0.5~eV. The collaboration is investigating several possible improvements to the experimental setup in order to further reduce systematic uncertainties, discriminate background events, and improve statistical power by transitioning to a differential beta-spectrum measurement. Dedicated neutrino-mass operations are planned to continue through 2025, after which the apparatus will also be able to search for keV-scale sterile neutrinos with the novel TRISTAN detector (see below). In 2027 and beyond, the KATRIN apparatus will be used as an R\&D testbed for further tritium-based neutrino-mass work.

Meanwhile, the Project~8 collaboration is pursuing a new, frequency-based technique for measuring the energy spectrum of tritium beta decay. Cyclotron radiation emission spectroscopy (CRES) uses the microwave-frequency cyclotron emission from electrons in a magnetic field $B$ to precisely measure their kinetic energy $K$ via $f = eBc^2/(m_e c^2 + K)$~\cite{Monreal:2009za}. Due to the fW power levels, electron trapping for readout is essential.  
The technique was introduced in 2009~\cite{Monreal:2009za}, with first verification and proof-of-concept measurement performed in 2014 using a gaseous~$^{83{\rm m}}$Kr source of mono\-energetic electrons~\cite{Project8:2014ivu}. Project~8 has now demonstrated this technology on tritium, extracting the world's first CRES neutrino-mass limit -- $m_\beta < 180$~eV (90\% C.L.) -- after 82~live days (Fig.~\ref{fig:beta_spectrum})~\cite{Project8:2022hun}.

To reach the targeted sensitivity of 0.04~eV, Project~8 must solve two primary R\&D challenges. First, the imaged volume of tritium must be scaled up from the current 1~mm$^2$ to the m$^3$ scale, likely using a microwave cavity at relatively low magnetic field. Second, once the $m_\beta$ sensitivity improves to about 0.1~eV, the dominant systematic will become spectral broadening from rotational and vibrational molecular states populated by the $\beta$ decay of T$_2$. Project~8 therefore plans to use an atomic T source, which will require cracking molecules; cooling atoms in stages; and then trapping cooled atoms in a magnetogravitational trap in the cavity. Project~8 has defined a sequence of milestones in addressing both of these challenges, leading by 2029 to the planned operation of a pilot-scale cavity-based T$_2$ neutrino-mass experiment (Phase III) with 1-eV sensitivity. This experiment is a crucial milestone for Phase IV, a full-scale atomic tritium experiment with 0.04-eV sensitivity, envisioned to be built and operated after this long-range planning period.

Additional candidates for ultra-low $Q$-value ground-state-to-excited-state decays ($Q < 1$~keV) have been proposed based on literature searches~\cite{Kopp:2009yp,Gamage:2019xvx,Keblbeck:2022twm}, but a program of precision measurements of the parent and daughter atomic masses, and of specific excitation energy levels, is required in order to establish whether these candidate transitions are energetically possible. For shortlisted candidates, the next steps would be to observe each specific decay experimentally, and then to perform R\&D for neutrino-mass experiments. 

\begin{figure}[htb]
    \centering
    \includegraphics[width=0.6\textwidth]{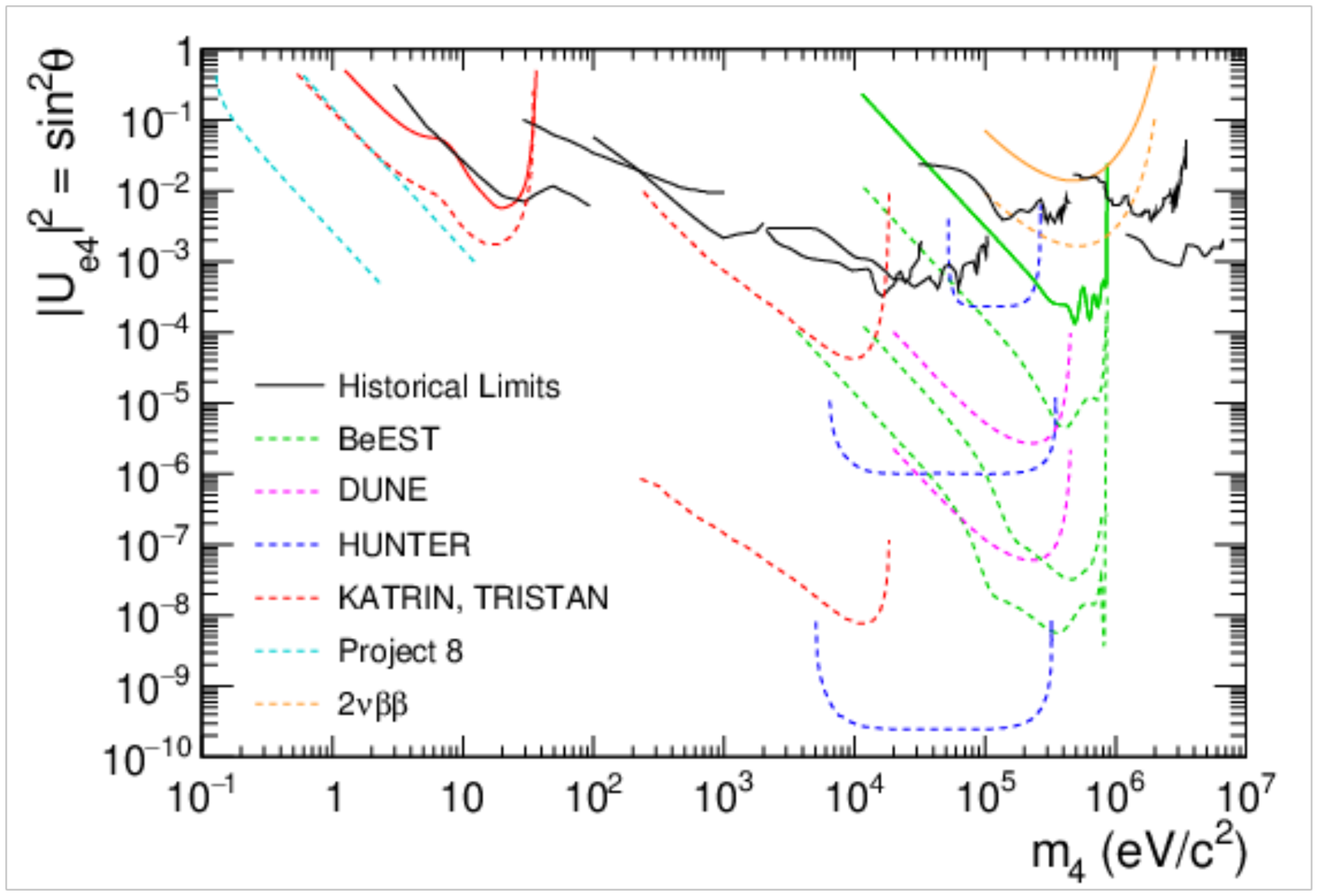}
    \caption{Achieved (solid) and projected (dashed) exclusion curves for sterile neutrinos from $\beta$-decay experiments. Figure from Ref.~\cite{Brodeur:2023eul}.}
    \label{fig:betadecay-sterile-excl}
\end{figure}

As shown in Fig.~\ref{fig:betadecay-sterile-excl}, current and next-generation measurements of $\beta$ decay provide interesting non-oscillation-based sensitivity to sterile neutrinos. For example, spectral measurements extending to sufficiently low energies are sensitive to the characteristic spectral distortion from a fourth neutrino-mass state $m_4$. This distortion would be located at $E_0 - m_4$ with an amplitude set by the mixing angle $\theta_{14}$. KATRIN has already set competitive limits on eV-scale sterile neutrinos~\cite{KATRIN:2020dpx, KATRIN:2022ith} based on its 2019 data. A planned detector upgrade, replacing the current 148-pixel silicon p-i-n diode with a 1500-pixel, high-rate TRISTAN silicon drift detector, 
will allow deep, high-rate spectral scans with sensitivity at the keV scale to begin in 2025~\cite{KATRIN:2022ayy}. The Phase III and Phase IV Project~8 experiments will be sensitive to sterile neutrinos via the same mechanism, with the $m_4$ range determined by the achievable bandwidth.

A low-$Q$ isotope like T cannot be used to search for $m_4 \gtrsim
 18$~keV. Here, the BeEST and HUNTER experiments offer an alternative path without spectral measurement: sterile-neutrino production can be probed by reconstructing momenta in a two-body electron-capture decay.  For a discussion of progress and prospects in this area, see Section~\ref{sect:beta}.

Precise probes of $\beta$ decay are sensitive to a broader range of beyond-Standard-Model signatures~\cite{Brodeur:2023eul}. For example, KATRIN's first data set has already generated improved  limits on the local relic-neutrino overdensity~\cite{KATRIN:2022kkv}, and first limits on an observable related to possible Lorentz-invariance-violating operators~\cite{KATRIN:2022qou}. Further experimental and theoretical advances will extend the reach of these precision weak-force probes.

Since the last long-range plan in 2015, $\beta$-decay experiments have entered a new era of precision. The direct, model-independent laboratory limit on the neutrino mass is now at the sub-eV level, approaching sensitivities that will test the cosmological Standard Model and the picture of neutrinoless double beta decay mediated by light-neutrino exchange. R\&D into a new tritium spectroscopic technique may allow further neutrino-mass-sensitivity improvements of up to an order of magnitude. Combining spectroscopic and momentum measurements, $\beta$-decay experiments provide competitive, non-oscillation-based probes of sterile neutrinos over a wide mass range -- from the eV scale suggested by oscillation anomalies, to keV- and 100s-of-keV candidates for warm dark matter. Continuing R\&D into new candidate isotopes and measurement technologies could result in vital cross-checks and new sensitivities.

\subsubsection{Neutrino scattering}
Understanding of neutrino interactions with nuclei is deeply intertwined with many topics in nuclear physics. 
An accurate understanding of neutrino scattering from nuclei is required to extract information on neutrino properties from measurements of neutrino oscillations, to learn about astrophysical neutrinos from supernovae and other sources, and to search for BSM physics~\cite{NuSTEC:2017hzk}. Neutrino experiments addressing these topics will provide insights on the nature of neutrino masses, the neutrino mass ordering, the presence of CP violation, and perhaps find exotic new physics in the neutrino sector (or beyond, in other sectors using neutrino detectors). 
The lack of an accurate understanding of nuclear effects constitutes an obstacle to these discoveries. Theoretical calculations with quantified uncertainties of neutrino-nucleon and neutrino-nucleus cross sections in the wide energy range probed by neutrinos are a prime requirement for progress~\cite{Ruso:2022qes}. Furthermore, neutrinos themselves are a tool to advance understanding of the properties of nuclei via scattering experiments.

\begin{figure}[t]
\includegraphics[width=0.6\linewidth]{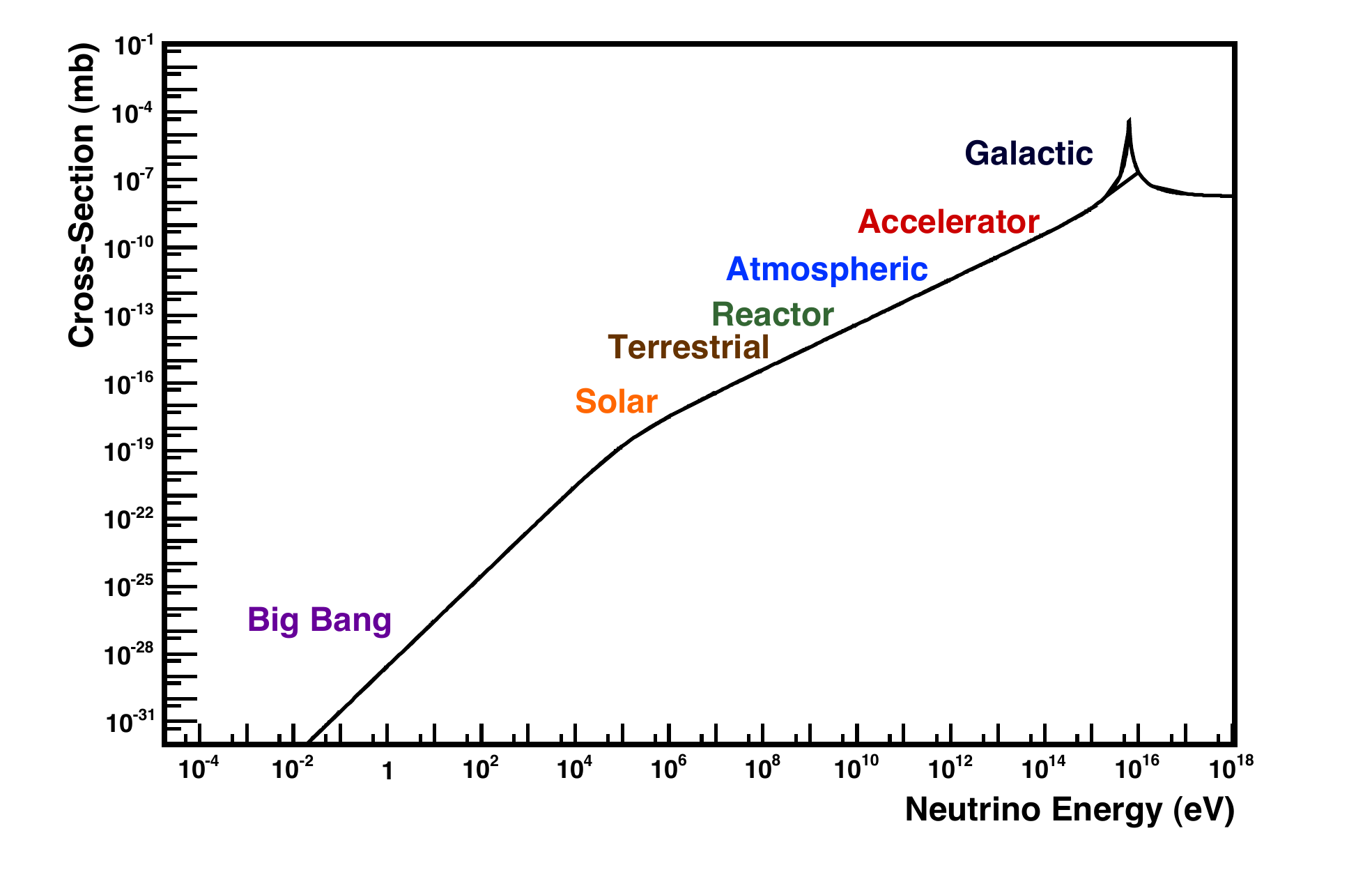}
\vspace{-0.1in}
  \caption{Representative example of the scale of neutrino interaction cross section (for $\bar{\nu}_e e-$ elastic scattering) spanning the energy regimes of various neutrino sources across decades of neutrino energy. Figure from Ref.~\protect\cite{Formaggio:2012cpf}.  
  \label{fig:nuEngy}
  }
\end{figure}

The observable phenomena and understanding of neutrino interactions with matter is highly dependent on the energy regime.   At the highest energies, neutrinos interact primarily with quarks inside nucleons and interactions can be highly destructive of the nuclear targets.  At GeV energies, interactions are typically with nucleons inside nuclei; hadro-production and final-state interactions can be disruptive.  At the MeV to few tens of MeV scale, nuclei may remain intact after neutrino interactions, but may shake off non-negligible debris.  In the lowest energy regime, where low momentum-transfer interactions dominate, the primary interaction channel is coherent elastic neutrino-nucleus scattering (CEvNS), in which a nucleus recoils as an intact entity.

Here we focus on neutrino-related physics topics of particular interest to the Nuclear Physics community, along with related experimental opportunities and needs for theoretical input.

\vspace{0.1in}
\noindent
\textit{Neutrino-nucleus interactions in the GeV regime:} Detailed understanding of neutrino interactions with nuclei in the GeV regime is especially important for interpretation of long-baseline neutrino oscillations~\cite{Ruso:2022qes}; argon,  oxygen, and carbon are of particular interest for the worldwide long-baseline program, for which current-generation experiments are T2K and NOvA and next-generation experiments are DUNE and Hyper-K. There are numerous opportunities for GeV-scale cross-section measurements using the near detectors that are part of these programs. From the theoretical point of view, this energy regime is rather challenging as several different reaction channels are in play, including quasi-elastic, multi-nucleon effects, nucleons' knock-out and resonance processes, and deep inelastic scattering. There has been a significant progress in theoretical studies of neutrino-nucleus inclusive quasi-elastic scattering within {\it ab initio} methods based on many-nucleon interactions and electroweak currents derived from effective field theories and/or phenomenological approaches. Theoretical descriptions of inelastic channels and extensions of the {\it ab initio} framework to medium-mass nuclei of experimental interest should be vigorously addressed in the forthcoming years, along with a dedicated effort to reliably assess theoretical uncertainty. Of prime importance are also LQCD calculations of electroweak elastic and transition nucleonic form factors along with single- and double-nucleon couplings which serve as input for nuclear  many-body calculations. Bridging the transition regions between medium- and high-energy theories, which use different degrees of freedom to describe neutrino-nucleus interactions, is also among the main challenges the community is facing. Additionally, simulations of neutrino-nucleus interaction through neutrino event generators play a crucial role in experimental data analysis. Given the highly interdisciplinary nature of this field, initiatives aimed at enhancing and fostering collaborations among all the involved areas of expertise, including nuclear and high energy theory, event generators, computation, and experiment, would be highly beneficial for progress~\cite{Campbell:2022qmc}. (See also Secs.~~\ref{sec:computing} and \ref{sect:theory}, on computation and theory needs.)

\vspace{0.1in}
\noindent
\textit{Electron-nucleus scattering:}
 One area of particular interest is that of electron-nucleus scattering experiments~\cite{Ankowski:2022thw,Ruso:2022qes}.  While the underlying primary interaction is different for neutrino and electron probes of nuclei, much of the nuclear and hadronic physics is shared. Limited understanding of the final-state physics is what drives much of the interaction uncertainties, and hence electron scattering measurements, which can be done with high precision and large statistics (in contrast to neutrino scattering), allow reduction of uncertainties and precision tuning of event generators. Specific experimental efforts include, for example,  E12-14-012 in JLab Hall A~\cite{Dai:2018gch,JeffersonLabHallA:2018zyx,Alcorn:2004sb}, the ongoing e4nu effort~\cite{Ankowski:2022thw}, the future LDMX experiment at SLAC~\cite{Ankowski:2019mfd}, A1 at MAMI, and eALBA~\cite{Ankowski:2022thw}.  All of these are at the GeV scale and are extremely important to precisely validate nuclear models against the bulk of accurate experimental data~\cite{Ankowski:2022thw,Ruso:2022qes}.  

\vspace{0.1in}
\noindent
\textit{Few to few tens of MeV regime:}
At lower energies, in the few to few tens of MeV range, there are different motivations for understanding of cross sections.  Interactions of neutrinos with nuclei in this range are especially important for understanding of neutrinos from core-collapse supernovae~\cite{Mirizzi:2015eza}, both for understanding of processes inside the supernova itself (including nucleosynthesis) and also for interpretation of supernova burst neutrino data in large terrestrial detectors~\cite{Scholberg:2012id}.  The materials of interest for specific detectors include argon, oxygen, carbon and lead. Interactions of relevance include charged-current interactions as well as neutral-current excitations.  For such inelastic interactions, the observable energies are of the scale of the neutrino energy. Several theoretical studies exist for these rates; however, a modern re-examination within the microscopic nuclear approach based on many-nucleon interactions and currents from effective field theories is needed. Neutrinos from stopped-pion sources are near-ideal for measuring cross sections in this energy range.  Such neutrinos come in two flavors ($e$ and $\mu$) with a precisely known spectrum.  Furthermore, pulsed beams can provide powerful background rejection.  Experimental programs with ability to measure inelastic neutrino interactions in this range are COHERENT at the Oak Ridge National Laboratory Spallation Neutron Source~\cite{Akimov:2022oyb}, and Coherent Captain Mills at Los Alamos National Laboratory~\cite{VandeWater:2022qot}.  COHERENT has made preliminary measurements of inelastic cross sections on lead and $^{127}$I~\cite{COHERENT:2022eoh,hedges}.

This energy regime is also characteristic of neutrinoless double beta decay, where the energy and momentum transfer are of the order of few MeV and hundreds of MeV/c, respectively. The same electroweak single- and many-nucleon currents entering neutrino-nucleus scattering also induce single- and double-beta decay~\cite{Cirigliano:2022oqy}. A thorough theoretical understanding of this kinematic region is essential for the computation of neutrinoless double beta decay nuclear matrix elements required to extract the neutrino absolute mass scale from the experimental rates (if observed.) See also Sec.~\ref{sect:nu-mass} on absolute neutrino mass measurements.

\vspace{0.1in}
\noindent
\textit{Coherent elastic neutrino-nucleus scattering:}
The lowest momentum transfer regime is that of coherent elastic neutrino-nucleus scattering, or CEvNS.  In CEvNS interactions, a neutrino interacts with a nucleus in such a way that the nucleus recoils with the constituent nucleons in phase.  The cross section for this process scales approximately as $N^2$, where $N$ is the neutron number, leading to significant enhancement with respect to inelastic interactions for which scattering is primarily off individual nucleons. This interaction was first observed in 2017 by COHERENT at a stopped-pion source~\cite{COHERENT:2017ipa}, and has now been observed in CsI~\cite{COHERENT:2021xmm} and Ar~\cite{COHERENT:2020iec}.  The main experimental challenge for CEvNS detection is the tiny nuclear recoil energies, for which maximum recoil scales as $2E_\nu^2/M$, where $M$ is the mass of the nucleus.\footnote{For detectors which aim also to observe inelastic neutrino-nucleus interactions, for which cross sections are typically two orders of magnitude lower but for which observable energies are a 2-3 orders of magnitude higher, a further challenge is dynamic range of energy sensitivity.}
CEvNS at reactors, where neutrino energies are an order of magnitude lower than at stopped-pion sources, has yet to be observed, due to the severe challenge of sub-keV nuclear recoil detection.  Nevertheless many experiments are underway~\cite{Abdullah:2022zue}.
Ref.~\cite{Abdullah:2022zue} provides an overview of the physics reach of CEvNS experiments.

Although the form factor uncertainties in the very low momentum transfer regime of reactor neutrinos are small, the nuclear uncertainties for predictions of CEvNS cross sections for neutrinos in the few tens of MeV regime are at the few percent level~\cite{Tomalak:2020zfh}. At experimental precision larger than the nuclear structure uncertainty, CEvNS observation serves as a SM test, and can probe non-standard interactions of neutrinos and nuclei.   CEvNS can also probe electromagnetic properties of nuclei.  Furthermore because the interaction is flavor-blind up to a few percent level~\cite{Sehgal:1985iu,Botella:1986wy,Tomalak:2020zfh}, it can be used for robust sterile oscillation searches and look for disappearance of active neutrinos.  CEvNS is sensitive to the weak charge of the nucleus, and is complementary to PVES experiments. If experimental uncertainties can be reduced to the percent level or better,  the measurements probe nuclear structure beyond current knowledge.  Precision CEvNS measurements can constrain nuclear form factors, and hence neutron radius and neutron skin.

Experiments with nuclear recoil sensitivity also have BSM physics capability, for example via search for accelerator-produced dark matter (e.g.~\cite{COHERENT:2021pvd}) and axion-like particles (e.g. ~\cite{CCM:2021lhc}) at stopped-pion neutrino sources. 

\begin{figure}[!tb]\centering
\includegraphics[width=0.6\linewidth]{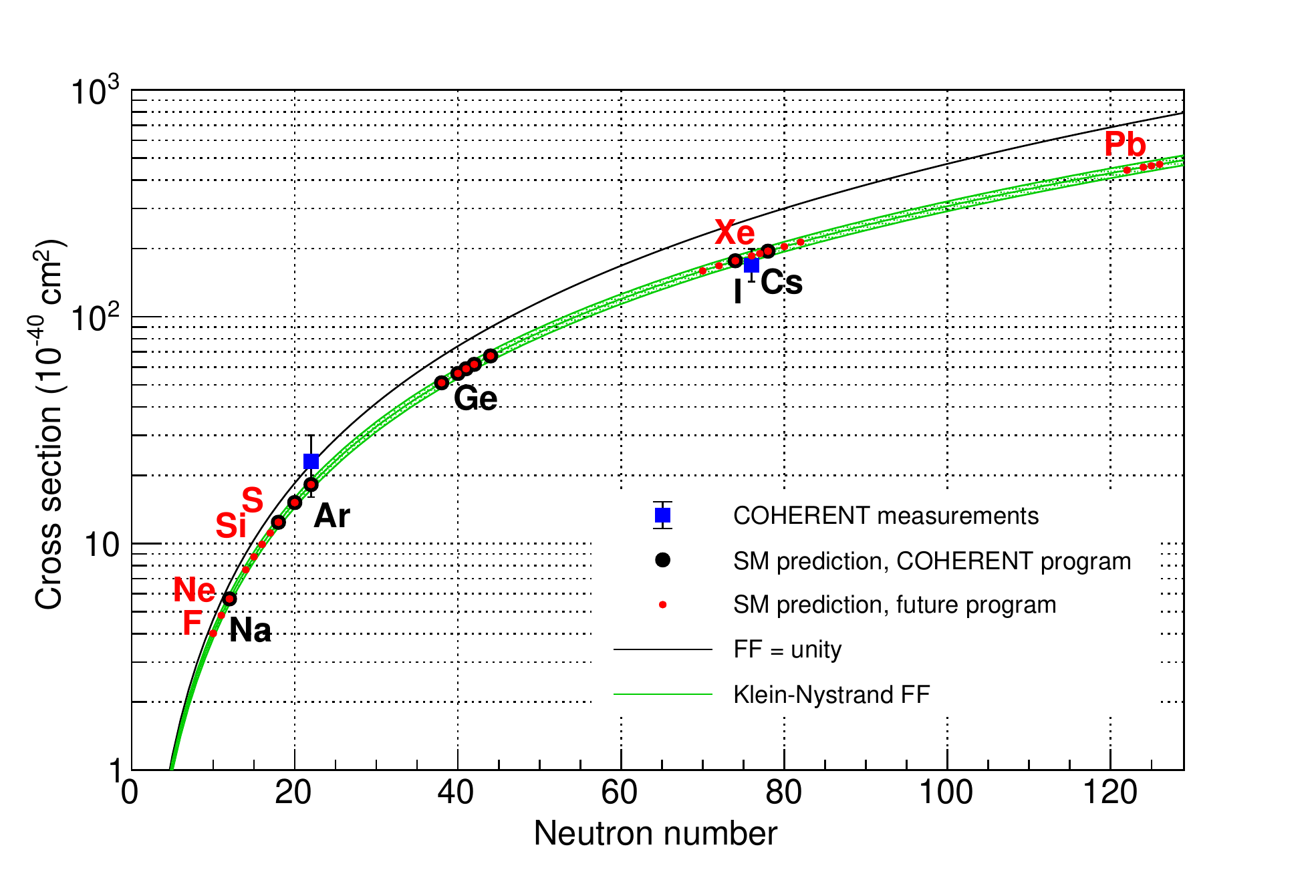}
  \caption{Flux-averaged CEvNS cross sections as a function of neutron number, from Ref.~\cite{Barbeau:2021exu}, with several potential future nuclear targets highlighted in red.  The highlighted targets are not intended to be a comprehensive set, but rather several for which particular, known-to-be-feasible detector technologies have been proposed for CEvNS measurements.  From Ref.~\cite{Asaadi:2022ojm}}
  \label{fig:targets}
\end{figure}

A new program at the ORNL Second Target Station~\cite{Asaadi:2022ojm} will provide opportunities for measurements of both inelastic and CEvNS interactions in a variety of targets, as well as a broad BSM search program.  Figure~\ref{fig:targets} shows some potential future nuclear targets.

\vspace{0.1in}
\noindent
\textit{Connections to HEP and astrophysics:}  Problems in neutrino scattering have deep and broad impacts beyond the NP program. Connections of particular relevance to the HEP community are highlighted in the 2021 Snowmass LOIs, white papers~\cite{Ruso:2022qes}, Topical Group Reports~\cite{deGouvea:2022gut,Balantekin:2022jrq} and Frontier Reports~\cite{Huber:2022lpm,Craig:2022cef}. 
In particular, understanding of neutrino scattering at GeV scale is of  great importance for interpretation of long-baseline experiments such as LBNF/DUNE.  Neutrino-nucleus interactions are furthermore relevant for BSM searches pursued within NP and HEP over a broad range of energies, via direct searches for non-standard neutrinos interactions, and as background for other new physics accessible in neutrino beams (see, e.g.,~\cite{Davidson:2003ha,Abdullah:2022zue,DUNE:2020fgq,Coloma:2022dng}).
The connections to astrophysics are furthermore numerous.   Understanding of CEvNS matters for interpretation of direct-dark-matter-search experimental data as the experimental sensitivity reaches to the ``neutrino floor" (or ``fog")~\cite{OHare:2021utq}.  Another example is understanding of interactions in the tens of MeV regime, relevant for the understanding of supernova dynamics, for interpretation of the supernova neutrino signal, and for understanding of nucleosynthesis, as discussed in more depth in the following section.

\subsubsection{Solar Neutrino Measurements}
The sun is a powerful laboratory for the study of both nuclear and particle physics. The nuclear reactions that power the sun produce more neutrinos than any other source natural or artificial. This enables precision tests of our understanding of both low-energy nuclear reactions and the interactions of neutrinos. Nuclear physics has been the steward of these studies including the Nobel Prize winning SNO experiment which proved that solar neutrinos change flavor on their journey from the sun \cite{SNO:2002tuh}. This foundational result underpins the vast neutrino program that is at the heart of major efforts in both nuclear and high energy physics.   

There are still many critical outstanding questions that remain see Ref.~\cite{Gann:2021ndb}.  Among these is an unambiguous observation of the predicted behavior from the MSW effect, which should be manifested in both an asymmetry in the solar neutrino $\nu_e$ flux between day and night, and in a transition between the matter-enhanced and vacuum-only oscillations, which occurs between about 1 and 5 MeV.  The vacuum/matter transition is particularly sensitive to new physics scenarios, particularly those involving non-standard neutrino interactions.  A more precise measurement of the CNO-cycle neutrinos than BOREXINO has made would also allow a clear discrimination between models of solar core metallicity, with implications for solar system formation. Lastly, as the founder of the field John Bahcall had always stressed, a precision measurement (at the 1\% level) of the $pp$ solar flux (or, possibly the $pep$ flux), would allow a real `unitarity' test of the solar luminosity: any deviation from the total energy measured by the Sun's photons would indicate either new energy-generating, or energy-loss mechanisms.

Experiments with potential sensitivity to the CNO neutrinos and the MSW
transition region include perhaps JUNO---depending on its background levels at its depth---and Theia, a proposed experiment that might also be a platform for a neutrinoless double beta decay experiment with sensitivity beyond the normal ordering.  The day/night effect should be measurable with precision by DUNE, which can see solar $^8$B and hep neutrinos above about 9~MeV or so.  Some of the planned LXe dark matter experiments can see the $pp$ solar neutrinos, but a 1\% measurement will be challenging, and likely requires some new ideas and dedicated R\&D.

\subsubsection{Neutrinos in astrophysics and cosmology}
\newcommand{\cm}[1]{{}}

Neutrinos from the cosmos represent a new frontier, where rapid progress is taking place. Several sources contribute to the neutrino flux from space, spanning several orders of magnitude in energy. 
Here we review the main ones, and discuss the potential of neutrinos from these natural sources to complement the laboratory program on FSNN.

\begin{figure}[htb]
\centering
\includegraphics[width=0.4\textwidth]{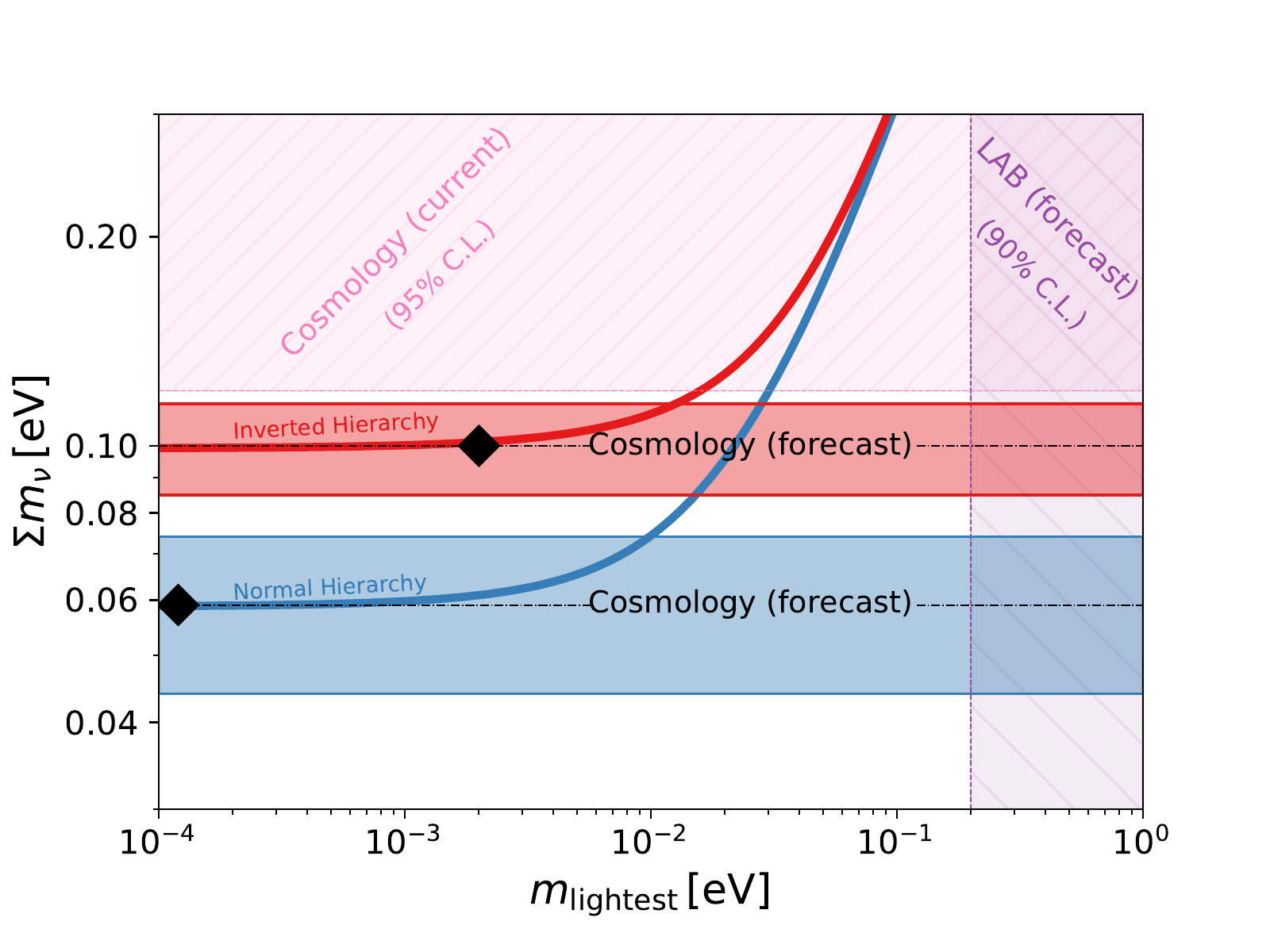}
\caption{From \cite{Dvorkin:2019jgs}: current and forecast sensitivity of cosmological surveys to $m_{tot}=\sum_{i}m_i$ (horizontal shaded regions), as well as a recent upper bound from the laboratory experiment KATRIN (vertical region). The curves show examples of the predicted value of $m_{tot}$ depending on the lightest of the three neutrino masses, for normal and inverted mass hierarchy (labels on curves).  
}
\label{fig:numasscosmo}
\end{figure}

\noindent
{\it The Cosmological neutrino Background (C$\nu$B). }
These are the relic neutrinos and antineutrinos, of all flavors, left over from the protracted epoch in the early universe when the neutrino component decoupled. This weak decoupling epoch, proceeding between temperatures $T \sim 5\,{\rm MeV}$ and $T \sim 100\,{\rm keV}$, in the standard model produces roughly relativistic Fermi-Dirac black body energy spectra for all neutrino and antineutrino flavors. Subsequent to decoupling, these free-falling neutrinos red shift, with momenta proportional to the inverse scale factor, so that at the current epoch we expect characteristic neutrino kinetic energies of the order of $\sim 10^{-4}$ eV. These neutrinos have never been detected directly. However, their participation in weak reactions, especially their isospin-changing charged-current reactions that inter-convert neutrons and protons, as well as their contribution to the radiation and matter content of the Universe are probed indirectly by several cosmological observables. These include the Big Bang Nucleosynthesis (BBN) light element abundances, the Cosmic Microwave Background (CMB) energy spectrum and anisoptropies, the spectra of Large Scale Structures (LSS), and more. Anticipated Stage-4 CMB experiments promise high precision (better than one percent)  determinations of $N_{\rm eff}$ (a measure of the radiation energy density at photon decoupling at $T \approx 0.2\,{\rm eV}$) and the primordial helium abundance. The advent of 10m-class optical telescopes likewise promise better than one percent determinations of the primordial deuterium abundance. These high precision measurements of $N_{\rm eff}$, deuterium, and helium will sharpen up constraints on BSM physics in the the neutrino sector.
Some observables, like those associated with the small-scale end of large scale structure, are sensitive to the scale of the neutrino mass, and specifically to the sum of the neutrino masses, $m_{tot}=\sum_{i}m_i$.  One of the main causes of sensitivity is the fact that neutrinos being massive alters the cosmological  evolution of matter perturbations, which are suppressed on scales smaller than the neutrino free-streaming length, which is inversely proportional to the neutrino mass, $L_{fs} \propto 1/m_{tot}$.  Recent high precision cosmological data -- mainly on the CMB and LSS spectra  --  constrain $m_{tot}$ to be below the eV scale, reaching a sensitivity that approaches 0.1 eV, where it becomes possible to test  the inverted mass hierarchy hypothesis (see fig. \ref{fig:numasscosmo}); future cosmological surveys may improve the sensitivity down to the 0.01 eV scale. On the other hand, if the neutrino mass hierarchy is determined experimentally, for example via long baseline neutrino oscillation experiments, then these cosmological/CMB observations can be turned around and in essence can probe the neutrino collision-less damping scale and, hence, the relic neutrino energy spectrum. See, e.g.,  \cite{Dvorkin:2019jgs} for a recent review.  It is expected that cosmological and laboratory tests of the neutrino mass will be complementary, perhaps facilitating distinguishing between potentially degenerate scenarios. For example, a limitation of cosmological constraints is the degeneracy with the effect of the Dark Energy density at low redshifts.
On the other hand, the mixing-independent cosmological constraints on the total mass $m_{tot}$, when combined with laboratory tests which are mixing-dependent, could help disentangle the uncertainty on the masses from those on the elements of the mixing matrix. 

\begin{figure}[htb]
\centering
\includegraphics[width=0.4\textwidth]{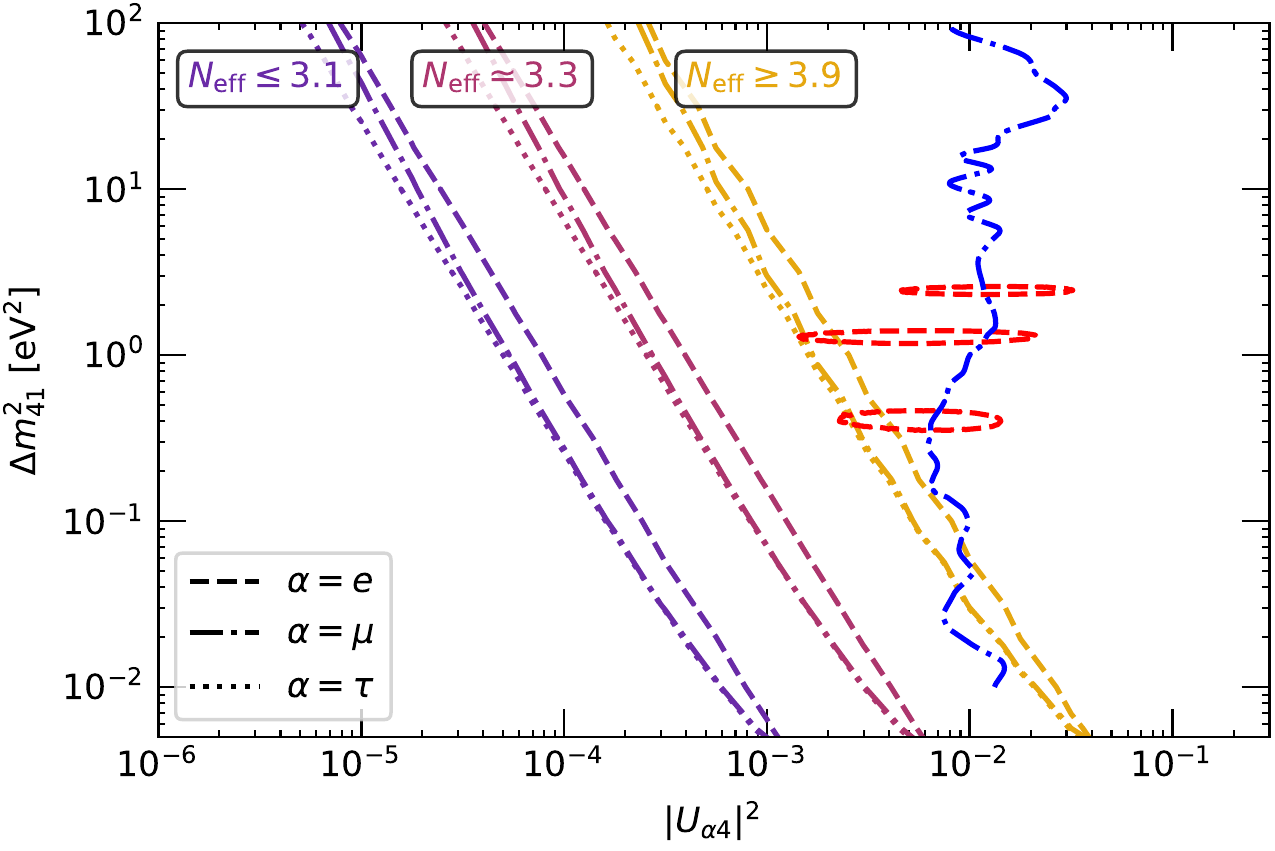}
\caption{From \cite{Gariazzo:2019gyi}: contours of constant values of $N_{eff}$  (labels of curves) for a sterile neutrino with mixing to only one of the active species ($\alpha$ in the figure), in the parameter space of the  mass squared difference between mass eigenstates, $\Delta m^2_{41}= m^2_4 - m^2_1$ and element of the flavor mixing matrix, $|U_{\alpha 4}|^2$. The normal mass hierarchy is assumed. The dashed contours show the allowed regions (99.7\% CL) from DANSS+NEOS reactor experiments; the dash-dotted lines show the exclusion constraints (99.7\% CL) from muon (anti)neutrino disappearance. }
\label{fig:nusterilecosmo}
\end{figure}

Cosmological tests of the C$\nu$B  could reveal the existence of light sterile neutrinos, like those at the eV mass scale, that are motivated by a number of laboratory anomalies. The minimal framework, where one or more such neutrinos are added to Standard Model, is disfavored by cosmological data. Indeed, for the masses and mixings favored by the anomalies, the interplay of flavor oscillations and collisions would lead to $\nu_s$ being efficiently produced in the Early Universe, leading to an $N_{eff} \simeq 4$ (see fig. \ref{fig:nusterilecosmo}), which is in tension with the measured value $N_{eff} \simeq 2.99 \pm 0.33$ from the CMB and other probes (see \cite{Dvorkin:2019jgs} and references therein). 
The tension could be resolved, however, in non-minimal scenarios where other non-Standard elements are introduced that would suppress the production of $\nu_s$. Examples are models with a large lepton asymmetry in the neutrino sector or perhaps other BSM physics. The anticipated higher precision measurements from Stage-4 CMB and extremely large telescopes discussed above may be able to constrain or probe these BSM scenarios.

Directly detecting the C$\nu$B  is possible, although difficult. The most promising method is based on the threshold-less capture of $\nu_e$  on Tritium, for which the signature consists of an excess of electrons separated by $2 m_\nu$ from the endpoint of the Tritium $\beta$ decay spectrum. This is the principle behind the proposed PTOLEMY experiment \cite{PTOLEMY:2019hkd}.  Interestingly, this process is sensitive to the nature of the neutrino as a Dirac (D) or Majorana (M) fermion: the capture rate is predicted to be larger (by factor of $\sim$2) for the Majorana case, as a consequence of the neutrinos propagating as helicity eigenstates and being (at least the heaviest one) non-relativistic today. Values of the rates  for 100 g of Tritium target are $\Gamma^{M}\simeq 8.1~{\mathrm yr^{-1}}$ and $\Gamma^{D}\simeq 4.1~{\mathrm yr^{-1}}$ respectively.

\noindent 
{\it Supernova neutrinos}. Core collapse supernovae are the most powerful known neutrino sources, with ${\mathcal O}(10^{53})$ ergs of energy emitted in neutrinos by each of them. This copious neutrino emission is a direct result of the collapse of the core of a massive star, after the star has exhausted its fission power and has become hydrodynamically unstable. A thermal population (temperatures in the ${\mathcal O}(10)$ MeV range) of neutrinos forms in the very dense region immediately surrounding the collapsed core, and then is rapidly emitted after a shockwave is launched, resulting in a $\sim 10$ s-long neutrino burst that is detectable of Earth.  Although supernovae are abundant in the Universe, modern neutrino detectors are limited in sensitivity to our galaxy and its immediate neighborhood, and therefore the detection of a supernova neutrinos burst is a rare event. It has occurred only once, in 1987, with the low-statistics detection from SN1987A. When the next galactic supernova occurs, detectors at the 10 kt mass scale will record hundreds or thousands of neutrino events, enabling high statistics tests of the physics of stellar collapse and of the properties of neutrinos.  An alternative avenue is to observe the continuous, diffuse neutrino flux from all the supernovae in the universe (the Diffuse Supernova Neutrino Background, DSNB). This flux has evaded detection so far, however low-background neutrino detectors like the current SuperK-Gd, and the upcoming DUNE and JUNO, could achieve the first observation within the next 5-10 years; see, e.g., \cite{Koshio:2022zip}. 

Supernova neutrinos  are produced and propagate in extreme conditions of temperature, density and scales of distance, which could not be replicated on Earth. In such conditions, BSM phenomena that elude laboratory searches could become evident. For example, sterile neutrinos could be efficiently produced via their mixing with the active species, and contribute to the cooling of the collapsed core at an observable level. They could also change the flavor oscillation pattern of the active neutrinos, with testable consequences on the synthesis of the heavy elements inside the star. 
Furthermore, competitive bounds could be obtained on other BSM effects that require a very long propagation baseline to appear, like neutrino decay or absorption effects due to light force mediators. \cm{add something about \n\ self-interactions, collective oscillations, NSI, etc.}

Complementary to cosmology, a supernova neutrino  burst can be used to make a time-of-flight test of the neutrino mass. The idea consists of measuring energy- and mass-dependent deviation of the neutrino velocity from the speed of light, which results in a time delay $\delta t \simeq 5 (m/eV)^2 (E/10~ MeV)^{-2} (D/10~kpc)$ ms. 
The long propagation distance ($D\sim 10$ kpc for a galactic supernova) and the presence of short time-scale features in the neutrino burst, such as the $\sim 1-10$ ms-long neutronization burst allow a sensitivity to mass values slightly below 1 eV, see  \cite{Formaggio:2021nfz} for a review.

\noindent 
{\it High energy neutrinos from cosmic accelerators. }
In the last decade, the km$^3$ antarctic facility IceCube has observed a diffuse flux of 0.1-1 PeV neutrinos from outside our galaxy, which is  consistent with a hadronic origin in cosmic ray accelerators. Recently, some of the detected neutrinos have been associated with individual 
sources, leading to the confirmation that part of the observed flux comes from the cores of active galaxies, and likely also from flares in star-shredding black holes. 
 Still, a large fraction of the flux has unidentified origin and many possibilities are open (see \cite{Halzen:2022pez,Kurahashi:2022utm} for reviews). 
 Complementary to its primary mission in multimessenger astronomy, IceCube has used this high energy astrophysical neutrinoflux to perform 
tests of the neutrino-nucleon cross sections at center-of-mass energies ccomparable to particle accelerators. Results are consistent with Standard Model predictions. 
In the near future, more advanced detectors like the planned IceCube-Gen2 could measure this cross section above the center of mass energy of the Large Hadron Collider (LHC), where signatures of BSM hypotheses like large extra dimensions, sphalerons and color-glass condensates could  be found. Similarly to supernova neutrinos, the analysis of the flavor composition and energy spectrum of the high energy astrophysical neutrinos allows several tests of BSM properties of neutrinos, like couplings to new forces or new particles, sterile neutrino species, and violations of the Lorentz invariance and of CTP symmetry; see, e.g. \cite{Arguelles:2022xxa,Adhikari:2022sve}. 

\subsubsection{Connections to dark matter}
Laboratory and astrophysical probes of dark matter overlap and leverage both the neutrino physics and fundamental symmetries issues discussed above. Revealed through motions in galaxies and clusters of galaxies, the existence of a dark component in the universe has been known for going on a century. Fundamental nuclear physics, i.e., BBN, coupled with observational determinations of the primordial deuterium abundance established that this dark component could not be baryonic. This conclusion was later confirmed by measurements of the CMB anisotopies. These revealed our most precise value for the ratio of baryons to photons, $\approx 6.1\times{10}^{-10}$. The upshot is that baryons can account for only some 4\% of the closure density, while dark energy and dark matter comprise the lion's share of the rest. In fact, the non-relativistic component of the non-baryonic closure contribution, the dark matter, makes up about a quarter of the closure density, roughly 5 times the baryon rest mass contribution. What this is remains unknown. 

The dark matter may be several entities or one, with a host of proposed dark matter candidates. These include primordial black holes (PBHs), sterile neutrinos, dark sector particles and, the long popular favorite, weakly interacting massive particles (WIMPS). The latter candidate has been attractive because a thermal freeze-out with weak-scale interaction with standard model particles could give the correct relic density for dark matter. However, these same interactions allow for collider, direct and indirect detection probes. So far these have found no compelling signatures for a dark matter particle. Variants that could have evaded detection with current experiments/observations do remain, including ideas built around super symmetry. A key frontier for nuclear theory/nuclear structure lays in calculating the WIMP-nucleus scattering cross sections for spin-independent and spin-dependent channels. These are important for interpreting the results of the large scale direct detection experiments, such as the liquid xenon experiments. The technology of these detectors is addressed above in the discussion on neutrino-less double beta decay searches in, for example, nEXO. A salient frontier in the interface of experiment and theory is understanding the response of these detectors for light dark matter candidates, where the nuclear recoil energy will be low.

The strong CP problem can be addressed with an elegant solution, Peccei-Quinn symmetry breaking and the resultant axion. That particle is an excellent cold dark matter (CDM) candidate. Moreover, the axion idea has given rise to many variants, or ALPS (axion-like particles). Again, the overlap with nuclear physics is twofold, involving a variety of detection modalities, and potential effects in compact objects resulting from capture of these particles. 

Observations and simulations of LSS evolution suggest that dark matter is more or less collision-less, at least on large scales. However, there is hot debate about whether dark matter, or a component of it, could have \lq\lq self interactions\rq\rq\ that might help explain puzzles such as the core/cusp issue in a range of halos with different masses and the velocity morphology observed in these halos. The working hypothesis for avoiding the invocation of self-interactions is that a completely collision-less dark component can be compatible with observed structure if \lq\lq baryonic feedback\rq\rq\ is taken into account. Baryons can dissipate, sink to the bottom of potential wells and dominate the local gravitational potential. If some of those baryons form stars, some of which are massive and explode and eject baryons, then some dark matter will move to higher orbits, for example softening a cusp of dark matter into a \lq\lq core.\rq\rq\  

This presents an obvious connection to nuclear astrophysics. Namely, nucleosynthesis is associated with those exploding massive stars. Nucleosynthesis could, in principle, then provide an independent constraint on the mass assembly and structure formation histories of halos. This connection between the arguments about the nature of dark matter on the one hand, and the synthesis of nuclei on the other, in effect couples in all of the discussion in the last section on neutrinos and compact object evolution \-- and nucleosynthesis, including r-process nucleosynthesis that is sensitive to frontier issues in neutrino physics as discussed above.

Though the baryonic feedback mechanism may be capable of explaining most of the issues with galaxy/structure morphology, it cannot operate without enough baryons. Ultra faint dwarf galaxies (UFDs) likely do not have enough baryons for effective feedback amelioration of the core/cusp issues. However, the UFDs are mostly dark matter, with possibly not enough stars to give a definitive observational determination of the dark matter distribution in these objects.

There is another reason to contemplate a dark sector with physics perhaps as rich as the standard model, but different: Baryon rest mass is $\sim 20\%$ of the total non-relativistic dark component. That suggests that, instead of a thermal freeze-out, perhaps the dark particles have an origin crudely similar to the origin of the net baryon number in the standard model. This asymmetric dark matter idea has give rise to a large number of possibilities for dark sector scenarios, including the production of dark composite particles, i.e., dark atoms or dark nuclei. 

The only thing we know for certain is that we share gravitation and spacetime with a putative dark sector. In some models, standard model particles could mix at small levels with corresponding dark sector particle partners. A connection with the experimental nuclear physics enterprise in this case is the same as discussed above: Detector response physics. A connection to the nuclear theory effort revolves around the many body physics of, for example, dark sector composites, and to the possible capture of dark sector particles (dark matter) in stars and compact objects. The latter possibility again couples in the discussion in the last section on compact objects and neutrinos. Dark matter-induced neutron star implosion, and dark matter capture effects in white dwarfs and other objects may provide constraints on some dark sector models, as well as suggesting new kinds of observations for multi-messenger astrophysics.

The neutrino sector, discussed at length in the last section, is replete with unknown physics, especially regarding the origin of neutrino rest mass, the nature (Majorana or Dirac) of the neutrino, and lepton number violation. All of these issues are important for compact object physics and cosmology as discussed above. They may also play a role in the dark matter puzzle.

For example, our working \lq\lq explanation\rq\rq\ for why neutrinos are so light compared to the other fermions in their respective elementary particle families is the see-saw model. That model invokes ultra-heavy sterile neutrinos. Variants of that model, for example, a \lq\lq split see-saw,\rq\rq\ could provide for lighter sterile states. Ordinary active neutrino scattering-induced decoherence, and lepton number resonantly-enhanced versions of this process, could build up a relic sea of sterile neutrino that could represent a significant dark matter component if these sterile states are mostly a mass state with rest mass $\sim {\rm keV}$ to $\sim 10\,{\rm keV}$. Alternatively, models can invoke a thermalized sea of sterile neutrinos at high temperature in the early universe that subsequently decouples and is diluted by the out-of-equilibrium decay of some other particle, maybe even another sterile state. The range of possible scenarios is breathtakingly, perhaps depressingly, large. 

However, X-ray astronomy provides our very best probe of this putative sector of particle physics. In fact, the small mixing with active neutrinos required for the simple freeze-in (scattering-induced decoherence) scenario for a significant dark matter contribution, also provides for a non-GIM suppressed radiative decay channel for these sterile neutrinos.  The X-ray observations rule out the simplest models and provide a key probe for the more baroque models. Interestingly, the next generation of proposed X-ray observatories, for example XRISM and ATHENA, will have high precision X-ray energy resolution that could tag an X-ray line from a galaxy cluster as having a dark matter origin \--- as a collision-less particle it would have a virial width and not the smaller thermal width associated with lines from heavy ions. 

The connection of this dark matter physics with nuclear theory is exotic and unique in this case: Active neutrino scattering in the high entropy and energy density (low baryon number) quark-gluon plasma of the early universe. Moreover, this process plays out mostly around the epoch of the QCD transition, making sterile neutrino relic densities and relic  energy spectra sensitive to BSM extensions in the QCD sector. In fact, to come full circle, many dark matter models, including WIMP contributions to CDM are sensitive to new physics in the QCD sector, if for no other reason than for the dilution associated with the epoch at which the quarks and gluons are incorporated into color singlets. 

All of these dark matter issues, and their connections to both laboratory and theoretical nuclear physics, were discussed in the {\it Dark Matter in Compact Objects and Low Energy Experiments} workshop at the Institute for Nuclear Theory, University of Washington, in August 2022. We refer the reader to the INT website for this meeting~\cite{INT-DM} and the talks and seminars included there.

\section{Recommendation III: Theory}
\label{sect:theory}
Assessing the implications of the experiments discussed in this whitepaper requires accurate input from nuclear theory. 
The US fundamental-symmetries/neutrinos community has played a special role in strengthening ties to particle physics
and astrophysics, helping to build appreciation for the continued relevance of the entire field.  Our
contributions to topics such as neutrino mass and lepton number conservation, the structure of neutron stars, and the dynamics of supernovae and neutron-star mergers are widely recognized.

Theoretical work in FSNN physics is thus rewarding, but also extremely challenging.  The reason is that the important problems involve physics at several different scales. 
As we've noted,
observable effects of new physics in 
hadronic and nuclear environments 
are generated at energies that range from those of nuclear levels all the way up the scale of new physics.
Understanding the effects requires expertise in areas that include 
\begin{enumerate}
\item {\it Phenomenology \& Effective Field Theory (EFT)}:
This capability is essential  
for understanding the impact of nuclear probes of BSM physics in the broader context of cosmology and high-energy physics. Work is required to compare 
the sensitivity of nuclear probes to complementary collider and cosmological probes, and to build towers of effective field theories that 
bridge theories at different energy scales, with different degrees of freedom. 
Perturbative EFTs such as the Standard Model EFT  and its low-energy version (LEFT) allow for the evolution of BSM interactions from the scale of new physics down to low-energy scales at which QCD is nonperturbative.  At nuclear scales, hadronic EFTs can be used to systematically organize nuclear interactions induced by BSM physics.

\item {\it Hadronic Physics}:
Matching the hadronic operators to quark and gluon operators requires a coordinated effort between lattice QCD and the EFTs.
Great progress has been made in recent years, with single-nucleon matrix elements determined with a full uncertainty budget of a few percent from lattice QCD.
Applications to fundamental symmetries pose a new set of challenges: the calculation of  four-point functions (for nucleon EDMs and radiative corrections to $\beta$ decays),  the complicated renormalization of higher-dimensional operators on the lattice, the calculation of nucleon-nucleon scattering amplitudes in the presence of symmetry-violating interactions (needed for studies of $0\nu\beta\beta$ decay and of time-reversal and parity violation in nuclei).

\item {\it Nuclear Structure \& Reactions}: Accurate calculations of nuclear structure and reactions with quantified theoretical uncertainties are required
 to disentangle new physics from nuclear effects.  Microscopic many-body methods based on EFT interactions and currents 
 provide us with paths for achieving this goal. 
The past few years have seen tremendous progress towards the evaluation of nuclear matrix elements relevant to the fundamental symmetries program, 
including those for single- and double-beta decay, EDMs, hadronic parity violation, and lepton-nucleus scattering.  
The challenge is to develop a comprehensive theory of nuclear dynamics in the wide range of energy and momentum probed by the experimental programs in 
FSNN. 

\end{enumerate}

In the following subsections  we provide examples of recent progress and future prospects for FSNN theory, 
followed by a discussion of serious workforce challenges and recommended actions to ensure 
a healthy FSNN theoretical workforce.

\subsection{Progress and prospects}

\subsubsection{Neutrinoless double-beta decay and LNV}

The interpretation of \BBz\ decay experiments and, in case of discovery, the identification of the underlying mechanism behind a signal require an ambitious theoretical program, with several interconnected components, ranging from lepton number violating phenomenology to the calculation of the relevant hadronic and nuclear matrix elements with quantified uncertainties. 
The breadth of this program is captured by the recent reports found in Refs.~\cite{Cirigliano:2022oqy}  and~\cite{Cirigliano:2022}, which also offer a detailed bibliography. Here we summarize some of the salient features. 

{\it LNV phenomenology:}  Here we need to explore models of LNV and neutrino mass that go beyond the high-scale see-saw paradigm, and test them against the results both of \BBz-decay experiments and of other experiments at all energy scales. The other experiments include those with low-energy neutrinos,  at high energy colliders,  in astrophysics,  and in cosmology (e.g., to connect TeV-scale LNV with leptogenesis).  
Recent highlights~\cite{Li:2020flq,Li:2021fvw,Harz:2021psp,Graesser:2022nkv} 
and future prospects are discussed in detail in Ref.~\cite{Cirigliano:2022oqy}. 

{\it Hadronic and nuclear matrix elements:} In this broad program,  the goal is to compute \BBz\-decay rates with minimal model dependence and quantified theoretical uncertainty by advancing progress in particle and nuclear EFTs, lattice QCD, and \textit{ab initio} nuclear-many-body methods. 
At the time of the 2015 LRP, nuclear matrix elements from a wide variety of many-body approaches --- the QRPA, the Shell Model, DFT and the IBM --- had been computed, but results for important nuclei varied by factors 2-3, with no guarantee that the correct matrix elements were within the spread (see \cite{Engel_2017} and a more recent update in \cite{Agostini2022MatterDiscover}).  It is difficult to assess the quality of any of these calculations and to compare them because, for example, they each use empirical interactions that are not appropriate for other methods, and they each make \textit{ad hoc} assumptions about the effects of short-range correlation on the transition. To tackle such problems, the Lattice-QCD, EFT and nuclear-structure communities launched a collaborative effort to develop a consistent, systematically improvable framework for \textit{ab initio} matrix elements: EFT to specify the form of the decay operator, a combination of lattice QCD, modeling, and fitting to determine the constants that multiply particular terms in the operator, and \textit{ab initio} nuclear-structure theory to solve the nuclear many-body problem and compute the final matrix element.  Encouraging progress has been made in the last few years on all aspects of the problem. 

On the EFT front, a comprehensive framework for \BBz\ decay was developed both
for light 
Majorana-neutrino exchange~\cite{Cirigliano:2017tvr,Cirigliano:2018hja,Cirigliano:2019vdj}
and TeV-scale mechanisms~\cite{Prezeau:2003xn,Cirigliano:2017djv,Cirigliano:2018yza}, 
with the inclusion of sterile neutrinos~\cite{Dekens:2020ttz}.
In lattice QCD good progress has been made 
for the  $\pi^- \pi^- \to e e$ process~\cite{Nicholson:2018mwc,Feng:2018pdq,Tuo:2019bue,Detmold:2022jwu} 
and towards two-nucleon amplitudes~\cite{Feng:2020nqj,Davoudi:2020gxs}. 
The error in each of these steps (EFT truncation, effective couplings, and nuclear structure) can in principle be quantified, and will eventually lead to a matrix element with a meaningful uncertainty.

An important result from the effort to develop consistent EFT interactions and transition operators is the discovery that the exchange of high-momentum virtual neutrinos between nucleons contributes non-negligibly to the decay, and in ways that cannot simply be modeled by nucleon form factors or short-range correlations between nucleons.  In the systematic EFT approach, this physics manifests itself as an additional term in the \BBz\ decay operator with zero range~\cite{Cirigliano:2018hja}.
Recently, a  calculation of the $nn \to pp$ amplitude near threshold was carried out
with dispersion-theory techniques truncated to the elastic two-nucleon channel~\cite{Cirigliano:2021hp,Cirigliano:2021gz}.
This has allowed nuclear-structure practitioners~\cite{Wirth:2021vu} 
to determine the coefficient of the contact term and implement it in calculations, where it leads to a non-negligible and robust enhancement of the NMEs.

Nuclear-structure theory itself has seen a number of important developments.  \textit{Ab inito} techniques seem to have almost fully explained the systematic over-prediction of single-$\beta$ decay rates referred to as ``$g_A$ quenching.''  A combination of correlations that have escaped phenomenological models (such as the shell model and the QRPA) and two-body weak currents (corresponding to meson exchange during the decay) are responsible.
Both mechanisms have been investigated within \BBz\ decay.  Both effects reduce those matrix elements as well, though the effects of two-body currents are still not fully quantified.  The effects we can quantify are small effects, but another zero-range term with an unknown coefficient has yet to be assessed.  

Several \textit{ab initio} methods have now been applied to the decay of $^{48}$Ca, and matrix elements for $^{76}$Ge and $^{82}$Se are starting to come in as well (see Ref.~\cite{Cirigliano:2022}).  The new matrix elements, especially in Ge, are smaller than those produced by phenomenological models, but just how much smaller is an open question because theoretical uncertainty is still significant.

What are the next steps~\cite{Cirigliano:2022} in the theory program?

\begin{itemize}
    \item The \textit{ab initio} methods must continue to improve. These approaches are defined through truncation schemes that provide systematic convergence to an exact result. However, less truncation also implies significantly greater computational costs; this is the main reason current truncation schemes are more restrictive than we would like. 

    \item The coefficients of the EFT decay operators must be fully specified, including those that appear in sub-leading order, 
such as in the two-body currents.  
This program can be carried out by studying systems of two and three nucleons 
through a combination of EFT,  dispersive methods, and ultimately lattice QCD.

\item   Hadronic and nuclear matrix elements relevant for TeV-scale LNV mechanisms 
require  more dedicated study.  The nuclear-structure community has thus far focused almost exclusively on 
light-neutrino exchange. 

\item 
The community must carry out a robust uncertainty-quantification program, as laid out in detail in Ref. \cite{Cirigliano:2022}.  
This in itself will require several steps: 

\begin{itemize}

   \item  Quantifying the EFT truncation error  by performing nuclear calculations with interactions and transition operators 
   truncated at different orders.

    \item  Developing ``emulators'' for the \textit{ab initio} methods --- surrogates that can approximate the results of the method they emulate in much less time.  That step will allow us to examine correlations between observables, vary Bayesian priors, construct posteriors, etc.  Emulators exist for some methods 
    but for others their development will require more work.  

    \item Deciding how to combine the predictions of various methods to produce a single matrix element with an uncertainty that reflects the community's confidence in each method.  Here an analysis of the ability of methods to reproduce observables correlated with the \BBz\ matrix element is essential.  Carrying it out means first quantifying the correlations, and then examining all predictions of all models, which can be ``scored'' so that one can decide how much weight to give their predictions for \BBz\ matrix elements. 
\end{itemize}

\end{itemize}

Carrying out the  multi-pronged theoretical program outlined here is an integral part of a successful US-led \BBz\-decay campaign. 
Seeing it through will require more resources than we have at present, in the form both of person power and of computational power.

\subsubsection{CP violation and EDMs}

Since the last Long Range Plan, 
a significant amount of work has been carried out to
systematically connect the EFT of BSM physics valid at the EW scale, such as the SM Effective Field Theory (SMEFT) \cite{Buchmuller:1985jz, Grzadkowski:2010es}, with the phenomenology of EDMs \cite{Brod:2013cka,Cirigliano:2016nyn,Cirigliano:2019vfc,Brod:2022bww}, in order to provide a clear picture of the possible blind directions not constrained by EDMs, and thus of the complementary probes of flavor-diagonal CPV to be investigated at the LHC.
For example, the left panel of Fig. \ref{fig:th_edm} illustrates the complementary sensitivities of present and future EDM experiments (red regions) and the High Luminosity LHC (blue region) to CP-violating couplings of the Higgs boson to photons and $Z$ bosons.
For a more detailed discussion, we refer to the Snowmass white paper  \cite{Gritsan:2022php}, and references therein. 
The framework can then be used to constrain explicit models, such as the mLRSM \cite{Bertolini:2019out,Dekens:2021bro,Ramsey-Musolf:2020ndm}, leptoquark models \cite{Dekens:2018bci,Fuyuto:2018scm}, or scenarios with new light particles, such as axions \cite{deVries:2021sxz,Dekens:2022gha}.

Integrating out BSM physics and heavy SM particles,
flavor diagonal CPV at low-energy can be described 
in terms of effective operators with photons, leptons, quarks and gluons. The minimal set involves a dimension-4 operator, the QCD $\bar\theta$ term \cite{tHooft:1976rip,tHooft:1976snw}, 
the dimension-5  lepton electric, quark electric and quark chromo-electric dipole moments
(which originate from dimension-6 operators in the SMEFT),
and several dimension-6  operators, including the Weinberg three-gluon operator, scalar, pseudoscalar and tensor semileptonic interactions 
and four-quark operators \cite{Pospelov:2005pr,deVries:2012ab,Jenkins:2017jig,Dekens:2019ept}.
Computing the neutron, atomic and molecular EDMs as a function of  quark-level couplings is a highly non trivial task, which requires nonpertubative techniques to translate quark-level interactions into CP-violating couplings of nucleons and pions, and advances in nuclear theory for the calculation of nuclear Schiff moments in terms of few-nucleon interactions.

Since the last Long Range Plan, the Lattice QCD (LQCD) community 
has invested considerable resources 
to provide calculations 
of the neutron and proton EDMs
with reliable errors \cite{Abramczyk:2017oxr,Bhattacharya:2021lol,Bhattacharya:2022whc,Dragos:2019oxn,Alexandrou:2020mds,Liang:2023jfj}.
Several calculations of the nucleon EDM induced by the QCD $\bar\theta$ term have appeared \cite{Dragos:2019oxn,Alexandrou:2020mds,Bhattacharya:2021lol,Liang:2023jfj}. These calculations turned out to be extremely challenging, because of the small signal, which gets even smaller as the quark masses are decreased towards their physical values, and because of sizable
lattice artifacts, as for example the
contamination from nucleon-pion excited states \cite{Bhattacharya:2021lol}. 
The two calculations at the physical point yield a neutron EDM compatible with zero, but with uncertainties that are approaching the  values expected if the
``chiral logarithm'' identified in Ref. \cite{Crewther:1979pi} dominates the neutron EDM.    
Ref. \cite{Dragos:2019oxn,Liang:2023jfj} used larger pion masses, but extrapolating to the physical point, they find a non-zero neutron EDMs (at the $2\sigma$ and $4\sigma$ level, respectively), of a size compatible with Ref. \cite{Crewther:1979pi}.
A compilation of the most recent rLQCD results is showed in Fig. \ref{fig:th_edm}.

In the case of dimension-5 and dimension-6 operators, a further complication is the involved mixing structure of higher-dimensional operators on the lattice. Since the last LRP, the matching between the $\overline{\rm MS}$
scheme and schemes that can be implemented on the lattice has been worked out for the quark chromo-electric dipole moment and the Weinberg three-gluon operator  
\cite{Constantinou:2015ela,Bhattacharya:2015rsa,Cirigliano:2020msr,Rizik:2020naq,Kim:2021qae,Mereghetti:2021nkt}. Work remains to be done for the Weinberg operator in the gradient flow, and for four-fermion operators.
Concerning the calculations of lattice matrix elements,
the contribution of the $u$ and $d$ quark EDMs to the neutron EDM have been determined with 8\,\% and 4\,\% uncertainties
\cite{Bhattacharya:2015esa,Gupta:2018qil,Gupta:2018lvp}. Preliminary calculations for the chromo-electric dipole moment and the Weinberg operator also exist  \cite{Abramczyk:2017oxr,Bhattacharya:2022whc},
but they still do not have full control over all systematics.
For more details, we refer to
Ref. \cite{Alarcon:2022ero}.

Building on these extremely promising preliminary results, the primary goal of the EDM LQCD effort in the next LRP will be to produce controlled   
calculations for the neutron and proton EDMs induced by the QCD $\bar\theta$ term, and by the quark and gluon chromo-EDM operators, and to start the study of four-quark operators. Moving beyond single nucleon EDMs, 
the contribution of semileptonic CPV operators to atomic and molecular EDMs is mediated by the nucleon scalar, pseudoscalar and tensor form factors \cite{Chupp:2017rkp}, which are precisely computed on the lattice \cite{FlavourLatticeAveragingGroupFLAG:2021npn}.
LQCD can play an important role in the determination of CPV pion-nucleon couplings \cite{deVries:2016jox},
and, once two-nucleon techniques are mature, CPV couplings in the nucleon-nucleon sector, necessary to make contact with EDMs of light ions  \cite{Song:2012yh,deVries:2020loy,Yang:2020ges}
and nuclear Schiff moments \cite{Chupp:2017rkp,Engel:2013lsa}.

A further step towards a solution of the ``inverse problem" lies in understanding the 
complementarity between measurements or bounds on the nEDM and
atomic and molecular EDMs. 
Good progress has been achieved in 
the calculation of EDMs of light nuclei,
which can be carried out using \textit{ab initio} methods
\cite{Stetcu:2008vt,deVries:2011an,Song:2012yh,Bsaisou:2014zwa,Yamanaka:2015qfa,Gnech:2019dod,Yang:2020ges,Froese:2021civ,Yamanaka:2019vec}.
Atomic EDMs, such as $^{199}$Hg,  $^{129}$Xe and $^{225}$Ra,  are, on the other  hand, affected by large theoretical 
uncertainties, due to the complicated nuclear structure entering nuclear Schiff moments.
In the last few years 
there have been new calculations for $^{199}$Hg and $^{225}$Ra \cite{Engel:2013lsa,Dobaczewski:2018nim,Yanase:2020agg}. The great progress in the application of \textit{ab initio} techniques to medium mass and heavy nuclei promises the first 
\textit{ab initio} calculation of Schiff moments in the near future
\cite{Alarcon:2022ero}. 

More theoretical work is required for a deeper understanding of the
connection between EDMs and weak scale baryogenesis. 
Open questions exist in two main areas:
(i) The study of the electroweak phase transition: here it is necessary to identify scenarios that admit a first order phase transition
or sufficiently sharp crossover
(needed to provide sufficient departure from equilibrium) and study their falsifiable signatures at the Large Hadron Collider and possible future colliders~\cite{Ramsey-Musolf:2019lsf,Wang:2022dkz}.
(ii) The generation of CP asymmetries at the phase boundary through CP-violating particle transport: 
this requires identifying and solving an appropriate set of quantum kinetic equations~\cite{Cirigliano:2009yt,Cirigliano:2011di} (QKEs), 
to track both the coherent evolution necessary for CP violating phases to manifest themselves, as well as the incoherent interactions of particles with the thermal bath. The main challenge here concerns a systematic field-theoretic formulation of QKEs for massive fermions that mix through the Higgs vacuum expectation value(s) and an efficient computational scheme to obtain numerical solutions and scan the parameter space.

\subsubsection{Precision measurements}

Since the last LRP, many exciting theoretical developments have taken place in the area of precision measurements related 
to the FSNN program 
~\cite{Chang2018,Walker-Loud2020,Seng:2018yzq,Du:2019evk,Gorchtein:2018fxl,Seng:2022wcw,Cirigliano:2022hob,Pastore:2017uwc,Gysbers:2019uyb},
as summarized in the appropriate topical sections of this white paper.
Here we discuss  future prospects, focusing mostly on beta decays.  

Further efforts in high-precision SM theory calculations and understanding the impact of $\beta$ decays on new physics are necessary.
This requires collaborations between theorists (phenomenology, lattice, nuclear structure) as well as 
engagement with experimentalists in the design of new experiments. 

In the $|V_{ud}|$ extraction, there are on-going efforts to compute the single-nucleon axial $\gamma W$-box diagram using lattice QCD, which may fully pin down the inner radiative corrections in the nucleon sector.  A proof-of-principle study based on lattice computations of four-point correlation functions was successful on the simpler pion system~\cite{Feng:2020zdc,Yoo:2022lmt}, but to extend the method to the neutron requires more computational resources and independent studies from multiple lattice groups for cross-checking. Alternative approaches
are also possible~\cite{Seng:2019plg,Endres:2015gda}.

For nuclear decays, significant progress in the determination of nuclear structure corrections using as theoretical methods with a rigorous analysis of the uncertainties, such as nuclear \textit{ab initio} methods will be essential to maximize the potential of the superallowed global data set. In particular, a benchmarking effort centered around low mass nuclei with high precision experimental data ($^6$He, $^{10, 11}$C, $^{14}$O, $^{19}$Ne) that are accessible to nuclear many-body methods with a minimum number of approximations \cite{Heiko:20a} (No Core Shell Model, Quantum Monte Carlo, Lattice Effective Field Theory, $\ldots$) and methods with a wider mass reach (Coupled Cluster, In Medium Similarity Renormalization Group, and hybrid models) will allow one to more reliably compute corrections for the full data set. Supplemented by focused experimental measurements of $0^+ \to 0^+$ and mirror decays, the community foresees a synergistic approach with maximal impact.  Given that the uncertainties for the value of $V_{ud}$ determined from $0^+ \to 0^+$ decays are dominated by the uncertainty in theoretical corrections due to nuclear structure effects in the electroweak radiative corrections, we can anticipate significant progress in the precision with which these are calculated and a reduction in the uncertainty budget for the superallowed data set.

A $0.1\,\%-0.2\,\%$ precision for lattice calculation of the isospin-symmetric $g_A$ may be expected in a $\sim$ 5 year time scale, which approaches the current experimental precision of $\lambda$~\cite{WalkerLoud}. However, in order to compare with experimental measurements, one needs to understand the radiative corrections to $g_A$. 
While the $\gamma W$-box  contribution to $g_A$ is well under control~\cite{Hayen:2020cxh,Gorchtein:2021fce},  it was recently argued that a potentially much larger contribution comes from the vertex correction to the neutron charged weak form factor,  associated to the pion mass splitting~\cite{Cirigliano:2022hob}. An appropriate combination of effective field theory~\cite{Tomalak:2023xgm} and lattice QCD techniques will allow one to achieve the precision needed to perform stringent hadronic right-handed current searches.

Coordinating the development of BSM analysis, SM electroweak radiative corrections, and nuclear structure will be possible for part of the LRP period through the just-established Nuclear Theory for New Physics topical collaboration~\cite{NTNP}, but a need for broader coordination with experimental groups suggests a collaboration or center with the broad mission of coordinating these multiple threads of investigation to ensure effort is focused most effectively on key systems and observables.

\subsection{Challenges}

The growing complexity of the problems the subfield is tackling and the absence of an institutional center to support the field 
present 
challenges to theory workforce development for Fundamental Symmetries, Neutrons, and Neutrinos (FSNN).  There are two major issues:
\begin{enumerate}[leftmargin=*]
\item {\it Multi-component workforce:}
The success of the FSNN field relies on the synergy of three theoretical components, as outlined above: 
EFT/phenomenology; hadronic physics and lattice QCD; nuclear structure. 
All have  strong overlap with other important areas of nuclear science, namely cold QCD~\cite{ColdQCD} and nuclear structure and astrophysics~\cite{NSRA}.
Synergies with theory efforts in high-energy physics and astrophysics
(in the form of collaborative projects and career opportunities for students and postdocs)
have been and will  continue to be critical for the health of the field.
Constructing synergistic programs can be challenging because of barriers that range from non-homogeneous training backgrounds 
to separate funding streams.

\item {\it Lack of an institutional center:}
Other subfields in nuclear physics are generally built around the community's major user facilities,
JLab, RHIC, FRIB, ATLAS, 
and the planned EIC, 
and thus have institutional centers that recognize their importance and
support theory-workforce development.
In addition to serving as hubs for experimental activities, 
these national facilities bring experimentalists and theorists together on a recurring schedule.
The FSNN community plays a role in the major user facilities, but it has not been a central focus of any of them.
As a consequence, the community has had to depend on general-purpose theory centers to support its collaborative activities; these have included  the Institute for Nuclear Theory (INT)
and focused organizations such as
the Amherst Center for Fundamental Interactions (ACFI) and the Network for Neutrinos, Nuclear Astrophysics, and Symmetries (N3AS). 
But in contrast
to FRIB, JLab, and RHIC, these university-based theory centers provide fewer opportunities for interaction with experimentalists and do not
have workforce-development programs that can help young researchers find faculty positions.

\end{enumerate}
These challenges mean that the FSNN theory community, while successful, is more  fragmented than other communities. 
Fragmentation limits the community's ability to fully realize the broad FSNN experimental program, creating 
an obstacle to progress for the whole FSNN field, not just theory.
Opportunities to remedy these problems are outlined in the next section.

\subsection{Recommended actions}
A robust theoretical research program is essential for taking full advantage of the
FSNN experimental program. 
 We identify below a set of initiatives aimed at strenghtening and growing the nuclear theory
program in FSNN  to keep pace with the growing experimental effort.
Here are important elements of this program:
\begin{enumerate}[leftmargin=*]
    \item The funding agencies have recognized the importance of supporting collaborative work on
    high-impact 
    multifaceted problems, such as those described here, that require the integration of
    phenomenology \& EFT, lattice QCD, and nuclear structure \& reactions.   
     New opportunities have been created through   Hubs,
    Topical Collaborations \cite{DBD,NTNP}, Physics Frontier Centers \cite{N3AS}, and SciDAC programs.  
    These programs,  of particular importance to FSNN,  would greatly benefit from   increased support, including more opportunities to sustain successful
    collaborations beyond the five-year periods common in these programs;
     \item
    
The collaborative opportunities described above have significantly increased
 the participation of graduate students and postdocs in FSNN.  Because FSNN draws talent from boundaries shared
 with particle and astrophysics, the pool of young scientists is unusually broad and diverse.  For the same reason, these young people can compete in broad faculty searches. Yet the lack of faculty bridging programs suited to FSNN is inhibiting success.
 We thus call for the timely identification and implementation of
mechanisms to enlarge and support the FSNN
theoretical workforce at universities and national laboratories, with procedures and best practices that
develop and sustain a diverse, equitable, welcoming, and inclusive
workforce and culture.
Effective mechanisms include the creation of a faculty bridging program for FSNN theory. 

\end{enumerate}

Strong endorsements for increased efforts in FSNN theory along these lines  have appeared in 
several 2022 FSNN white papers~\cite{LRPWPs}.
An NSAC subcommittee could be charged with considering these endorsements and the two recommendations above and
finding appropriate ways to address the subfield's needs.
NSAC subcommittee recommendations could also spur the formation of a national consortium with elements analogous to those of the 
NP FRIB Theory Alliance~\cite{FRIBTA} and
HEP Neutrino Theory Network (NTN)~\cite{NTN}.  This consortium could work with the funding
agencies to administer the FSNN faculty bridge program.  It could also help existing and future Hubs, Topical
Centers, and Frontier Centers coordinate their activities in workforce development at all levels.  In partnership with
the INT and other visitor centers, these organizations could extend their coordination to the field's
workshops, including those promoting experimental participation.

\section{Recommendation IV: DEI, outreach, workforce development}
The nuclear-physics research program serves an important role in developing a diverse STEM workforce capable of fulfilling the critical needs of the nation. The small to medium-scale projects that are the hallmark of FSNN area of nuclear physics are a particularly strong training ground, allowing students to take part in all aspects of their experiment. The recruiting and maintenance of a diverse workforce requires treating all community members with respect and dignity, so that everyone has the opportunity to succeed in physics. The scientific community benefits, as diversity ultimately leads to stronger teams. Furthermore, a more diverse community allows us to communicate the importance and excitement of nuclear research to the broader public. 

Nuclear physics has benefited from larger initiatives including the long-running NSF ADVANCE program\cite{advance} and the recent DOE RENEW program\cite{renew}, and has led the way within the greater physics community by founding the DNP Allies Program\cite{allies}. However, nuclear physics has a ways to go to be fully representative of the nation. Fortunately, there are unifying solutions to our current DEI, outreach, and workforce development challenges that will benefit from over-arching commitments shared by all participants in the LRP process. The FSNN community puts forward the following global recommendation:

\textbf{We recommend enhanced investment in the growth and development of a diverse workforce to maximize our opportunities for scientific discovery and increase its impact in society.}

To achieve this important goal requires building new bridges to reach a broader and more diverse workforce, providing support through the professional stages of their careers, and enforcing standards of conduct in the broader community. 
Therefore the FSNN community also makes the following more detailed recommendations:

\begin{itemize}
\item  Resources and training programs, curated by social psychology professionals, are needed to minimize the impact of bias and create a more inclusive comumnity. Codes of conduct should be established and enforced.
\end{itemize}

The first step in building an inclusive and diverse workplace is ensuring the community is trained to provide equitable access to opportunities in physics. Training programs to understand the impact of bias and how to counter it are still needed for much of the community. To make a stronger impact, these programs should also include topics of inclusion in the workforce and in the scientific community, with a goal to normalize these discussions as a component of research. We stress that resources are needed to engage with experts in the fields of social science and organizational psychology, as recommended by the Snowmass Diversity and Inclusion Topical Group of the Community Engagement Frontier (CEF3)~\cite{Bonifazi:2022wdc}, and to avoid overburdening underrepresented groups with these tasks. Additionally, workforce, material, and financial resources are needed to create community networks, like the Multimessenger Diversity Network~\cite{MDN}, which bring together community representatives to develop resources and spread best practices. 

Codes of conduct must be reviewed to be consistent with new recommendations from funding agencies to ensure they meet expectations for enforcement and disciplinary actions. The NP community should be expected to take individual and collective action towards such enforceable conduct standards, and to prepare to enforce these community standards.  We must ensure that each of our institutions and individual community members take full ownership of that responsibility.

\begin{itemize}
\item  Programs that encourage participation in research by students and faculty from under-represented communities and cultures should be developed and expanded. Researchers at minority-serving institutions, primarily undergraduate serving institutions, research universities, and national labs should be supported.
\end{itemize}

Several initiatives are already taking place which will have an impact on our field, including APS IDEA, Change-Now, TEAM-UP, and others, and should be supported. Expansion of funding to make NP events accessible is also needed, so that a more diverse set of community members may fully participate and engage in the work being done across the community. This includes captioning services, childcare, and mobility-friendly buildings.

\begin{itemize}     
\item  To improve recruitment and retainment of diverse junior faculty and staff at universities and national laboratories, bridge positions, fellowships, traineeships, and other incentives should be developed and expanded. 
\end{itemize}

Programs to support junior faculty should also include support for career development, such as travel funds. Considerations for accessibility of events is also an important factor for retention. 

\begin{itemize}     
\item Federal grants should include resources to support living wages for graduate research assistants and postdoctoral researchers.
\end{itemize}

\begin{itemize}  
\item To understand gaps, monitor progress, and improve recruitment, resources for improved data collected is needed, including membership statistics and career trajectories.
\end{itemize}

Workforce and financial/material resources are needed for a dedicated, central organization to professionally collect and store this information, maintain records, and analyze the data for a comparison study in the next NSAC LRP.

The recommendations are in general agreement with those provided by the other areas of nuclear physics. These are in line with the goals of DEI, outreach, and workforce development, and these efforts should be woven throughout the NSAC LRP because they are critical to the success of any individual project and of nuclear physics more generally.

\section{Cross-cutting initiatives and needs for FSNN}
\label{sect:crosscutting}

\subsection{Initiatives and applications}

\subsubsection{Computing}
\label{sec:computing}
\paragraph{\bf Background:} Computing plays an essential role in both the theoretical and experimental efforts in the FSNN program. Particularly for theoretical efforts, high performance computing (HPC) resources have become essential for carrying out the calculations necessary for the prediction and/or interpretation of experimental results. With the advent of exascale computing and advancements in algorithms and methods, new regimes of nuclear physics are beginning to be accessed and properly quantified.

The FSNN experimental program probes nuclei in a wide range of kinematics. This includes the low energy regime of beta decay used for absolute neutrino mass measurements, the intermediate momentum regime of neutrinoless double beta decay, and the quasieleastic regime relevant to precision short- and long-baseline experiments. Maximizing the discovery potential of these experiments will require a theoretical understanding of neutrino-nucleon and neutrino-nucleus interactions and cross sections over a wide range of energy and momentum transfer where different reaction mechanisms are at play. This is a challenging problem whose solution requires a combination of expertises, including lattice QCD, nuclear many-body methods, nuclear effective theories, phenomenology, and neutrino event generators to make reliable theory predictions relevant to the experimental programs~\cite{Ruso:2022qes}. 

Lattice QCD, a numerical method for solving QCD directly, provides information about these interactions, as well as interactions with non-standard BSM currents, at the level of quarks and gluons. These calculations form the basis for theory input for heavier nuclei, give direct Standard Model predictions to be confronted with experimental data in searches for new physics, and provide potential BSM interactions which have not yet been experimentally measured. Lattice QCD is at its core an HPC endeavor: even the simplest of calculations require significant allocations of time on the largest machines in the world. Currently, precision single-nucleon calculations are routinely being performed, while two-nucleon calculations at the physical point with all systematics in check have not yet been performed (though enormous progress toward this goal has been made). In particular, the need for computing resources scales exponentially with both lighter quark mass and number of nucleons, necessitating access to ever-larger machines and dedicated research into new techniques and algorithms to tackle this massive endeavor.

Similarly, HPC, applied mathematics, computer science, and statistics are required to advance precision calculations of nuclear structure and reactions with quantified uncertainties. {\it Ab inito} or microscopic approaches describe nuclei as a collection of nucleons interacting via two-, three-, and many-nucleon forces \cite{Hergert:2020am}. External probes, such as electrons, neutrinos, and photons interact with single-nucleon and clusters of correlated nucleons via standard and BSM many-nucleon  currents. Computational methods used to solve the many-body problem of strongly correlated nucleons scale at best polynomially with the atomic number, making access to sustained and substantial computational resources along with storage capabilities critical for progress. 

As outlined above, access to and availability of exascale computing and beyond is crucial for the advancement of nuclear physics across all of the energy regimes of relevance for FSNN.  As HPC hardware continues its evolution, scientists in the U.S. must be prepared to take full advantage of it. The on-boarding of new machines promises disruptive progress for this field, but only if the necessary software has been created and optimized for use on these machines. As an example, Early Science time on Livermore's Sierra machine was used to produce a five-fold increase in statistics on a calculation of the nucleon axial charge in 2.5 weekends~\cite{Walker-Loud:2019cif}, a calculation which originally took 1 year on Titan combined with more than 2 years on Livermore clusters. This window of opportunity was short, as the machine is now utilized for classified research.

\paragraph{\bf Recommendations:} A recent workshop on ``Computational Nuclear Physics and AI/ML" was held on September 6-7, 2022 at SURA headquarters in Washington, DC. There, they set forth a resolution for the future of computational nuclear physics, which was endorsed at the 2022 Hot \& Cold QCD and Nuclear Structure, Reactions and Astrophysics Town Hall meetings. The key elements of their recommended program are to:
\begin{enumerate}
    \item Strengthen and expand programs and partnerships to support immediate needs in HPC and AI/ML, and also to target development of emerging technologies, such as quantum computing, and other opportunities.
    \item Take full advantage of exciting possibilities offered by new hardware and software and AI/ML within the nuclear physics community through educational and training activities.
    \item Establish programs to support cutting-edge developments of a multi-disciplinary workforce and cross-disciplinary collaborations in high-performance computing and AI/ML.
    \item Expand access to computational hardware through dedicated and high-performance computing resources.
\end{enumerate}
We further endorse these resolutions, supplemented with recommendations particular to the goals of the FSNN community. Our recommendations are outlined below.

Preparing for new hardware and making full use of the hardware currently available requires a properly educated, dedicated workforce, allocation of resources with community goals in mind, and shared, readily available software for the variety of purposes discussed in this report. Software must be optimized for a variety of available and planned architectures. There are many strategies for performing this task, and utilizing synergies between the various fields that make up the FSNN community, as well as partnering with industry, will aid in the endeavor. For example, many former lattice QCD theorists leave the academic track to work with companies such as NVIDIA and Intel, and can provide assistance in building libraries specific to our needs for optimal use on a given platform. Efficiency in developing the tools necessary for achieving the broader goals of the community must also be pursued. Topical collaborations, DOE SciDAC awards, training programs for the use of software, and workshops designed for the sharing of ideas for software building will be necessary to meet these goals. Community averages for the various quantities of interest for FSNN are also crucial to these efforts, and should be carried out by a designated group of leaders from the various collaborations performing the calculations, in the spirit of the Particle Data Group and Flavour Lattice Averaging Group. Additionally, both DOE and NSF provide critical computing hardware and data storage and retrieval capabilities to support experimental and theoretical programs. For example, the Advanced Scientific Computing Research (ASCR) provides capability computing resources at leadership-class computing centers, through
programs such as the ASCR Leadership Computing Challenge (ALCC).

These investments in the production, support, and modification of code, as well as the maintenance of community averages, must be sustained over time, typically on a time scale of at least a decade.  Thus, there is an urgent need for long-term, software-focused positions through, e.g., permanent staff at national labs, SciDAC positions, and joint lab-university tenure-track appointments.

Computing allocations at leadership-class facilities are limited and highly competitive, but adequate access to these HPC resources is critical for the success of the FSNN program. Equally important is a concerted community effort for acquiring and maintaining this access. 
Successful examples are the USQCD collaboration of US-based lattice QCD theorists and the NUCLEI SciDAC collaboration of \emph{ab initio} nuclear many-body practitioners, who collectively apply for large computing allocations.
The resulting time is then distributed amongst the various collaborations with the broader scientific goals in mind.

\subsubsection{AI / ML}   
\label{sect:ai-ml}
\paragraph{\bf Background:}
In the past decade, there has been an ever-accelerating growth in the field of Artificial Intelligence and Machine Learning that offers tremendous opportunities to the FSNN program and its related fields. Applications of AI/ML encompass instrumentation, experiment and theory \cite{Boehnlein:2022wd,Shanahan:2022ifi}: ML techniques offer powerful tools for modeling and controlling complex devices like accelerators and detectors; for classifying data and enhancing signal-to-noise ratios, which is particularly relevant for the rare processes the FSNN community is interested in; exploring correlations and patterns in data; and last but not least, for propagating and quantifying the uncertainties of theoretical simulations.

Within the FSNN experimental community, initial AI/ML efforts focused primarily on event classification. An early use of Artificial Neural Networks was to demonstrate discernment between neutral-current and charged-current solar neutrino interactions in the SNO (Sudbury Neutrino Observatory) experiment \cite{Brice:1996pm}. Many \BBz\ experiments have demonstrated event classifiers based on Boosted Decision Trees, Artificial Neural Networks, Convolutional Neural Networks, and Recurrent Neural Networks \cite{KamNet, GERDA_ANN, anton2019search, MJD_BDT, NEXT_ML, CUORE_PileUp}. In several of these cases, the use of neural network architectures led to improved sensitivity relative to the traditional analysis methods. 
Examples of event classification implementations can also be found in other settings, such as CRES (cyclotron radiation emission spectroscopy)-based neutrino mass measurements \cite{Project8_ML}. 

More recently, FSNN experiments have begun using machine learning techniques for a broader range of applications. Methods for detector design optimization that can replace some aspects of computationally-expensive simulations with ML/AI emulators have been used for the g-2 and KATRIN experiments \cite{g2_emulator, KATRIN_emulator}. The MODE (Machine-learning Optimized Design of Experiments) Collaboration is seeking to develop a fully-differentiable model for every aspect of information extraction from experiments, with the goal of allowing automatic exploration of design choices \cite{MODE}. Another major area of interest has been the use of AI/ML for real-time control and optimization of accelerators and detectors \cite{Boehnlein:2022wd}. These techniques are a focus of Jefferson Laboratory computing activities, and will likely play an important role in the MOLLER and SoLID experimental programs \cite{HOW19_JLab}. 

As detailed in previous section,
Lattice QCD continues to advance problems of relevance to the FSNN program, like the determination of specific reaction amplitudes as well as structure and response functions at the most fundamental theoretical level possible. Results may be directly relevant for experimental
efforts, or they can provide important guidance for the construction of EFT descriptions (see, e.g., \cite{Cirigliano:2018hja,Cirigliano:2021qko,Davoudi:2020gxs}. The lattice-gauge-theory community has been continuously exploring the use of ML techniques to overcome the sign problems and signal-to-noise ratio degradation that pose significant challenges in LQCD computations. New developments are occurring at a significant pace, and ML nowadays is deployed in all aspects of the lattice-gauge-theory program, from accelerating the generation of gauge-field configurations in Monte-Carlo-sampling-based methods, to assisting the computation of observables from hadronic correlation functions, to enhancing data-analysis techniques and observable estimations, including in inverse problems (see Refs.~\cite{Boyda:2022nmh,Shanahan:2022ifi,Davoudi:2022bnl,Constantinou:2022yye} for details and references.)

Significant progress is still needed to achieve broad and reliable applications of ML in lattice gauge theory, with methods that need to be generalized to lattice-QCD problems of relevance to the NP and FSNN programs, and results that need to be statistically rigorous and predictive~\cite{psaros2023uncertainty}. Many cross-disciplinary connections exist between the lattice-gauge-theory and ML communities, as well as with Applied Math and Computer Science communities, and collaborative developments must be emphasized and supported over the next decade. Physics-inspired algorithmic advances, such the incorporation of symmetries, developed and applied by lattice gauge theorists, have found their way in the broader ML community~\cite{Bogatskiy:2022hub}, and therefore, ML and lattice gauge theory can mutually benefit one another.

In nuclear many-body theory, ML techniques have seen significant use in the construction of computationally efficient surrogate models, also known as emulators. At a small fraction of the computational cost, they accurately predict the outcomes of computationally expensive applications like large-scale Density Functional Theory (DFT) calculations of nuclear masses or other bulk properties, or sophisticated \textit{ab initio} many-body calculations. This makes it possible to explore the sensitivity of nuclear observables to the parameters of the underlying interactions, to propagate uncertainties of these parameters, and to ultimately perform a comprehensive statistical uncertainty quantification.

\paragraph{\bf Recommendations:}
In Section \ref{sec:computing}, we presented the recommendations from the workshop on “Computational Nuclear Physics and AI/ML” that was held on September 6-7, 2022 at SURA headquarters in Washington, DC, which were endorsed by the FSNN Town Hall after previous endorsements from the Hot \& Cold QCD and Nuclear Structure, Reactions and Astrophysics Town Halls. 
AI and ML, in particular, were the focus of recommendation 3.

The AI/ML field is evolving at a rapid pace, and now is an excellent time to establish connections that ensure that the FSNN community can attract AI/ML talent to join collaborative efforts for reaching our goals, as well as benefit from cutting-edge developments. The observables and processes of highest interest to the FSNN community are more
challenging than those targeted by prior applications, hence they offer exciting opportunities for exploring new ideas and approaches, and to achieve high-impact results. One of the key requirements for FSNN program is high precision in experiment and theory, which crucially depends on the statistically rigorous uncertainty quantification and propagation that can only be provided with the aid of emulation (see, e.g., \cite{Bonilla:2022wd,Melendez:2022br,psaros2023uncertainty} and references therein). Many cross-disciplinary connections already exist between the lattice gauge-theory, nuclear physics and ML communities, as well as with Applied Math and Computer Science communities, and collaborative developments must be emphasized and supported over the next decade, e.g., as key components in FSNN centers or a SciDAC collaboration. Moreover, physics-inspired algorithmic advances, such the incorporation of symmetries, developed and applied by lattice gauge theorists, have found their way into the broader ML community~\cite{Bogatskiy:2022hub}, and therefore there is a high potential for generating mutual benefit and interest.

While model reduction is a key thrust of the AI/ML efforts for FSNN and our affiliated communities, access to HPC resources, including leadership-class assets, remains crucially important: The quality of surrogate models for LQCD or nuclear many-body calculations relies on the availability of sufficient training data that must be generated with the full simulation codes. Likewise, the training of AI models for experimental applications like signal refinement or online control of devices can be a numerical challenge in itself.

\subsubsection{Quantum computing, quantum sensing and R\&D for the future}

\label{sect:QIS}
\label{sect:quantum}
Nuclear physics has much to gain from, and contribute to, quantum information science (QIS) and quantum sensing. The precision measurements that are at the heart of the FSNN experimental effort are natural test beds for cutting edge quantum sensing technologies.  While quantum simulation has great potential to revolutionize our simulation capabilities to tackle difficult theoretical problems relevant to FSNN. 

To follow-up on the NSAC Subcommitttee on QIS report from 2019, a whitepaper on QIS for the greater nuclear physics has been assembled~\cite{Beck:2023xhh}. It reiterates  the need to increase investment in the areas of quantum sensing and simulation to capitalize on the rapid worldwide developments in these areas to meet the needs of nuclear physics and the nation at large. The recommendations tailored for the FSNN community are as follows:
\\

{\it We recommend increased investment to capitalize on the rapid worldwide development of quantum sensor technology and its timely implementation in NP.}
\\

Advances made over the past two decades in material science and cryogenic infrastructure have accelerated the development of quantum sensors and quantum integrated systems, and in some cases provide revolutionary approaches to historically inaccessible problems. Quantum sensors are already in use in: neutrinoless double-beta decay, neutrino mass measurements, sterile neutrino searches, precision tests of fundamental symmetries, permanent electric dipole moment searches, and as probes to rare and exotic processes.  Their targeted use in NP continues to grow and expanding R\&D in this area, including through investments in national and university facilities, is essential.   
\\

{\it We recommend investment in exploratory research directions that aim to develop, integrate, and apply quantum-based simulation and computation techniques in NP.}
\\

Current classical-computation techniques are well advanced, providing—and will continue to provide—input to the broad range of NP problems. Nonetheless, they face challenges in simulating real-time dynamics of matter created in heavy-ion collisions or after the Big Bang, coherent neutrino oscillations in astrophysical environments, and of relevance to the FSNN program, a wealth of dynamical response functions, e.g., for neutrino-nucleus scattering, accurate rate of nuclear-reaction processes, and more. Quantum simulation has great potential to revolutionize our simulation capabilities in these problems. Furthermore, quantum information tools can guide the design of more efficient classical NP simulations, and quantum entanglement can serve as a new principle to enhance our understanding of NP phenomena and the underlying theory. The community needs to engage in the co-design of quantum-simulating devices dedicated to the NP program and be provided sustained access to quantum hardware. Programs and partnerships that enable collaborations across NP in QIS would be valuable. 
\\

{\it We recommend strengthening the QIS expertise in NP training to create a diverse, quantum-ready, nuclear-capable workforce.}
\\

A diverse, quantum-ready workforce is a necessity for both communities. QIS can attract young talent from a variety of backgrounds to our programs, and empower them with skills in emerging quantum technology and computing trends. Increased investment in recruiting and training these young researchers will accelerate the development and integration of QIS technology in NP, and will improve sustainability of the field. 
\\

{\it We recommend further development of research that leverages the knowledge base of NP that can help inform and solve outstanding problems in QIS, and to encourage cross-cutting research between the two research communities.}
\\

The nuclear community’s knowledge of the interactions between particles and matter serves as a valuable asset in the development of a future quantum computer and a robust quantum workforce.  For example, the expertise developed by nuclear physicists in shielding against cosmic rays, and in the development of radio-pure materials for rare event searches, can play an important role in increasing the coherence times of next-generation qubits for a range of computing and sensing applications. We encourage support of these cross-disciplinary efforts that can potentially strengthen collaboration between the two communities.

\subsubsection{Isotope Science and Accelerator Science}
The nuclear component of the FSNN field has had a long and successful symbiotic relationship with facilities built for nuclear structure/nuclear astro (and neutron) physics. Nuclear beta decay is used to search for BSM currents and to determine precisely the up-down quark mixing element of the Cabibbo-Kobayashi-Maskawa matrix and test its unitarity. This work requires access to light short-lived isotopes around the $N=Z$ line, where the symmetries and simple nuclear structure allow for selective decays that can isolate the effects of interest and be calculated accurately. The lightest of these isotopes can be produced at a number of facilities, including some of the ARUNA facilities, while the heavier $N=Z$ systems, up to roughly mass 100, require more powerful facilities such as ATLAS and FRIB. The combination of these diverse facilities and their constant upgrades address the present needs of most of the nuclear part of the FSNN program. It is important that they remain available to ensure continued progress and leadership in this field. 

As we've discussed earlier, the community has begun to to take advantage of the sensitivity of heavy nuclear systems to EDMs and other manifestations of CPV or the violation of other symmetries. The heaviest nuclei benefit from large enhancements in sensitivity because of the high Z, large octupole deformation present in the region, and the possibility of extremely high effective electric fields if they are embedded in polar molecules. Long-lived isotopes of radium (or their thorium progenitors), of francium (or their actinium progenitors), and iof other nuclides in the region can be obtained from the Isotope Program to pursue the development of these promising approaches. The ability to produce these long-lived isotopes is not widespread and we often face competition from other areas to for this limited resource. The full potential of our new approaches will only be reached if these isotopes remain available to the FSNN program. Future isotope-harvesting capabilities at FRIB should alleviate some of these needs, but the highest sensitivity measurements will still require access to the higher intensities available through the Isotope Program.

\subsubsection{Nuclear Data}
\label{sect:ND}
High-accuracy nuclear data for particle-induced reactions, nuclear decay, and
nuclear structure are essential for benchmarking high-fidelity models and
theoretical calculations and for simulating particle interactions and transport
in detectors, experimental setups and applications.   In simulations of nuclear
physics experiments, code packages such as Geant4 \cite{GEANT4:2002zbu} and
FLUKA \cite{Ahdida:2022gjl} are commonly used.  These packages rely on nuclear
data to simulate detector response.  For example, the data used in Geant4 for
photon evaporation, radioactive decay, and nuclide properties are taken directly
from the Evaluated Nuclear Structure Data File (ENSDF) \cite{Tepel:1984},
maintained at the National Nuclear Data Center at Brookhaven National Laboratory
\cite{NNDC}.  Neutron cross sections and final states are based on nuclear data
libraries such as JEFF-3.3 \cite{Plompen:2020due} and ENDF/B-VII.1
\cite{Chadwick:2011xwu} while the TENDL library \cite{Koning:2019qbo} is used
for interactions of incident protons with matter.  The SAID database is used for
proton, neutron and pion elastic, inelastic and charge exchange reaction cross
sections for interactions with nucleons below 3 GeV \cite{Arndt:2007qn}.
Nuclear shell effects are based on the liquid drop model of the nucleus,
including ground state deformations.  Nuclear data are also required for the
nuclear density profiles, photoelectric interactions, impact ionization, and
optical reflectance, see Ref.~\cite{GEANT4:2002zbu} for more references and
details.

In the case of fundamental symmetries, neutrons and neutrinos, experiments such
as neutron and atomic electric dipole moment measurements, double beta decay
experiments from modest prototypes to ton scale and beyond, and reactor
antineutrino experiments all require nuclear data to simulate data rates and
detector responses and efficiencies.  Specific examples are given in the next
paragraphs. 

One proposed atomic electric dipole moment measurement involves studying nearly
degenerate parity doublet nuclei with large octupole deformation.  These
pear-shaped nuclei have large intrinsic Schiff moments \cite{Auerbach:1996}.
Nuclear data help inform which nuclei are good candidates for these
measurements, such as $^{223}$Ra \cite{Dobaczewski:2005}.  These deformed nuclei
are significantly more sensitive to CP-violation in the nuclear medium.  In
addition, when these nuclei are part of molecules \cite{GarciaRuiz,PRL126},
their sensitivity to new physics is enhanced by several orders of magnitude
beyond the state of the art $^{199}$Hg EDM experiments \cite{Ben:2010}.

Double beta decay experiments rely on calculations of the matrix elements for
$A(Z,N) \rightarrow A(Z+2,N-2) + e^- e^-$.  These calculations in turn rely on
accurate nuclear data, e.g., nuclear masses. Uncertainties in the nuclear data
input to models used to compute the nuclear matrix elements are manifested as
uncertainties in the calculations.
When designing experiments for double beta decay, nuclear data are used to guide
decisions on which isotopes are most promising  
in terms of the decay rate, the required size of the detector to obtain a
sufficiently high rate for a discovery measurement, and which measurement
techniques 
result in the smallest systematic errors \cite{Adams:2022jwx}.  For example,
$^{136}$Xe is a promising isotope to search for neutrinoless double beta decay
but building a ton-scale and beyond experiment based on Xe requires production
of sufficient Xe for the vessel.  Current Xe production from the steel industry
is insufficient.  Other ways of acquiring more Xe include reprocessing from
nuclear fuel, since Xe is one of the most abundant fission products and is
directly captured in air \cite{Xe:2021}.  Xenon production rates via fission can
be computed from fission product yield and neutron-induced fission cross section
data.

Reactor antineutrinos are also of great interest since they can be employed to
probe neutrino oscillations, determine the existence of a sterile neutrino,
study coherent $\nu N$ scattering and also have implications for
nonproliferation.  The KamLAND reactor experiment provided the first terrestrial
observation of neutrino oscillations \cite{Eguchi:2003,Eguchi:2004,Araki:2005}.
Such experiments are also a good example of how nuclear data can be employed to
help resolve more fundamental physics questions.  The antineutrino flux from the
Daya Bay \cite{DayaBay}, Double Chooz \cite{DoubleChooz}, and RENO \cite{RENO}
reactor experiments, designed to study the $\theta_{13}$ mixing angle, was 5\%
lower than predicted by the Huber-Mueller method \cite{Huber:2011,Mueller:2011}
for calculating the expected antineutrino flux, referred to as the ``reactor
antineutrino anomaly". In addition, a significant excess in the measured
antineutrino spectra at 5~MeV was observed compared to the Huber-Mueller
predictions.  Oscillations with a sterile neutrino were proposed as a potential
solution to the anomaly: it could be caused by exotic neutrino oscillations with
a fourth “sterile” neutrino.  However, updated fission fragment yields from
$^{235}$U and $^{239}$Pu point to the anomaly being due to an overestimate of
certain fission fragment yields in the Huber-Mueller model \cite{Sonzogni:2018}.
Data from experiments such as PROSPECT \cite{Andriamirado:2022} and STEREO
\cite{Almazin:2021}, located near reactors using highly-enriched uranium fuel,
could prove definitive.  \\

{\bf Endorsement -- } Recognizing the relevant of nuclear data for nuclear science, 
the FSNN working group on nuclear data prepared the following statement, which the larger community endorses:

Nuclear data play an essential role in all facets of nuclear science. Access to reliable, complete and up-to-date recommended nuclear data is crucial for the fundamental nuclear physics research enterprise, as well as for the successes of applied missions in the areas of defense and security, nuclear energy, space exploration, isotope production, and nuclear medicine diagnostics and treatments. It is imperative to maintain an effective US role in the stewardship of nuclear data.

\begin{itemize}
\item We endorse support for identifying and prioritizing opportunities to advance and enhance the stewardship of nuclear data and efforts to build a diverse, equitable and inclusive workforce that maintains the currency and reliability of the nuclear databases.
\item We recommend prioritizing opportunities that enhance the currency and quality of recommended nuclear data and its utility for propelling scientific progress in fundamental symmetry, neutrino and neutron projects and the broader nuclear science program.
\item We endorse identifying interagency-supported crosscutting opportunities for nuclear data with other programs that enrich the utility of nuclear data in both science and society.
\end{itemize}

\subsection{Facilities and Infrastructure}
\subsubsection{Overview} 

The field of Fundamental Symmetries, Neutrons, and Neutrinos (FSNN) is unique, as compared to other fields within the nuclear physics portfolio, in that there is no mission-centered nuclear physics-owned facility for FSNN \cite{Ito_Talk}.  Instead, as described below, FSNN research employs a broad range of different facilities with unique capabilities and challenges.  In addition, small-scale FSNN experiments not connected to a dedicated facility play a valuable role in the community and present their own unique challenges. 

\subsubsection{Underground Facilities in the U.S.} 

The Sanford Underground Research Facility (SURF) \cite{SURF_White_Paper} is the U.S.’s deepest underground laboratory and has been operating since 2007 as a dedicated scientific laboratory supporting underground research in rare-process physics, as well as offering research opportunities in other disciplines.  The 4850-foot-level (i.e., 1500 meters, or 4300 m.w.e.) hosts a range of experiments, including those studying dark matter, neutrino properties, and nuclear astrophysics topics.  In addition, the SURF facility offers several support capabilities, including low-background assays for materials as managed through the Black Hills State University (BHSU) underground campus, production of electroformed copper, management of up to 1.5 million liters of xenon, and through the holding of a Nuclear Regulatory Commissioning broad scope license for radioactive materials, with various gamma-ray and neutron survey instruments and a liquid scintillator counting system.  As part of SURF’s strategic plan, underground expansion possibilities are being explored, including expansion designs at the 4850L and a concept for laboratory space on the 7400L (i.e., 2300 meters, or 6500 m.w.e.), for which the cosmic ray muon flux is projected to be a factor of 30 times lower than at the 4850L.

Continued access to SURF is critically important for U.S. leadership in the worldwide neutrinoless double beta decay community, especially for material characterizations (i.e., radioassays), R\&D on fabrication and handling methods, and material production, fabrication, and storage.  Although the proposed ton-scale neutrinoless double decay experiments will likely be sited outside the U.S. (i.e., SNOLAB, LNGS), it is clear that it will be important to maintain successful partnerships with these international facilities as the ton-scale experiments move forward.

Nuclear physics, through its access to underground facilities with reduced environmental backgrounds and its expertise in low background techniques, can also have an impact on quantum information science \cite{Formaggio_Talk}.  Techniques for detecting and removing unwanted particle interactions, as has already been done in rare event searches (such as in solar neutrino and neutrinoless double beta decay experiments, etc.) can prove to be of great use to large quantum systems. 
\subsubsection{Parity-Violating Electron Scattering at Jefferson Laboratory} 

The energy, luminosity, and stability of the electron beam at Jefferson Laboratory are uniquely suited to carry out parity-violating electron scattering (PVES) measurements \cite{PVES_White_Paper}, including the MOLLER experiment and the SoLID PVDIS experiment.  With the parity-violating asymmetries on the order of $\sim 10$'s to $\sim 1000$'s of ppb, sub-nrad, sub-nm, and sub-ppb levels of beam control are required.  The high-intensity, high-polarization electron source is optimized for minimizing beam asymmetries.  Required upgrades for the future PVES program include an upgrade to the injector for better control of beam asymmetries, an upgrade to the End Station Refrigerator (ESR-2) for increased cryopower, and an upgrade to the polarimetry for robust, high-precision measurements of the beam polarization. 

\subsubsection{Fundamental Symmetries at FRIB} 

An opportunity exists at FRIB \cite{Ito_Talk, Singh_Talk} to carry out isotope harvesting of several symmetry-violating ``enhancer'' pear-shaped isotopes, such as $^{225}$Ra and $^{229}$Pa, that will be produced in the water beam dump while beams are delivered to other experiments (i.e., ``commensal operation'').  However, support is required to transform the harvested isotopes into a chemical form needed for fundamental symmetries experiments.  A dedicated beamline fed with harvested isotopes that then delivered beams insto a variety of experimental stations would more readily allow for long-integration-time precision measurements of fundamental symmetries. 

\subsubsection{Neutron Sources for Fundamental Neutron Physics} 

Within the U.S., research in fundamental neutron physics is currently carried out at three locations with complementary capabilities \cite{Ito_Talk}: (a) the Fundamental Neutron Physics Beamline (FNPB) at the Spallation Neutron Source (SNS) at Oak Ridge National Laboratory (ORNL); (b) the NG-C Cold Neutron Beamline at the National Institute of Standards and Technology (NIST); and (c) the Ultracold Neutron (UCN) Source at Los Alamos National Laboratory (LANL).  Brief descriptions of these facilities are as follows.

\textit{FNPB at SNS at ORNL} -- The FNPB is one of 22 beam lines at the SNS, the world's most intense source of pulsed cold neutrons.  The program of hadronic parity violation experiments NPDGamma and n$^3$He have been completed on the FNPB, and the Nab neutron $\beta$ decay experiment is currently commissioning.  Construction of the nEDM@SNS experiment is projected to continue through the late-2020's.

\textit{NG-C at NIST} -- The NG-C Cold Neutron Beamline offers the highest-flux cold neutron beam in the U.S. for fundamental neutron physics.  As a result of the recent (2023) Cold Source upgrade, the beamline's performance is comparable to the best in the world.  This national resource is operated in service to the FSNN community, with a broad range of experiments in hadronic parity violation, neutron $\beta$ decay angular correlations, and the neutron lifetime.

\textit{UCN Source at LANL} -- The UCN source at LANL is currently the only operating UCN source hosting fundamental neutron physics experiments in North America.  The UCN source has hosted the UCNA and UCN$\tau$ experiments, and is also a R\&D test-bed for the nEDM@SNS experiment.  Experiments under development include the LANL nEDM experiment, and the $\beta$ decay experiments UCN$\tau +$, UCNA$+$, and UCNProBe.  The LANSCE accelerator, delivering the proton beam for production of spallation neutrons which are moderated to the UCN regime in solid deuterium, is expected to continue operating for the next few decades.

All three of these locations leverage resources funded by other agencies.  Investments are needed to support research and beamline operations at these three locations, including additional personnel, in order to realize the full potential of their physics programs, improve their capabilities, and provide continuity to the workforce of researchers \cite{NeutronWhitePaper}.

Given the growing importance of U.S.-based UCN research to the worldwide fundamental neutron physics research community, there are several ongoing efforts to improve existing UCN sources and develop new UCN sources.  To maintain the U.S.'s leadership in this field, it is critical that an investment be made in the development of next-generation UCN sources, to ensure the future of a vibrant community with capabilities for high-impact science.  The U.S.-based UCN community has already developed ideas for next-generation UCN sources \cite{NeutronWhitePaper}, which range from the conceptual R\&D stage to those ready to be implemented soon.  These ideas include: (a) a uranium neutron multiplier for the LANL UCN source; (b) a liquid-helium converter coupled to a spallation target at LANL or the SNS; (b) a liquid-helium converter coupled to HFIR (taking advantage of a planned upgrade to replace the HFIR pressure vessel) at ORNL; and (d) a future UCN source at NIST.  The UCN community is currently discussing and comparing these options.  Modest support in the next decade for evaluating these possibilities will be crucial to avoid missing a key opportunity for a world-leading next-generation UCN source in the U.S. \cite{NeutronWhitePaper}.

\subsubsection{Small Scale Experiments} 

Small scale experiments \cite{Mumm_Talk}, although not easily defined, can be considered to be those experiments which are not connected to a dedicated facility, are not a central component of a large project, are ``table top'' in size, and/or are innovative with higher risk.  One such example of small-scale experiments is the robust program of nuclear $\beta$ decay studies probing fundamental symmetries which occur at universities, the ARUNA laboratories \cite{ARUNA_Link}, and also at national laboratories. Another example are the precision measurements performed using neutron interferometry at NIST~\cite{NeutronWhitePaper} and proposed with UCN at LANSCE~\cite{UCNInt-wp2022}.

Such small scale experiments play a central role in the FSNN community through their coverage of a wide range of science and through their range of technologies, where new ideas are explored, cross-pollinated, and slowly gain traction in the community.  These experiments are often cost effective, leveraging existing investments, equipment, and user programs, and also collectively offer a quicker turnaround time.  Further, by their nature, these small scale experiments offer opportunities for workforce training in all aspects of an experiment: design, engineering, data analysis, publication, and leadership.  However, much of this small scale work occurs in spaces not directly funded by nuclear physics, and as such, faces unique challenges \cite{Mumm_Talk}, such as a lack of facility-level advocates, additional costs for infrastructure, engineering requirements, challenges for the small scale community to speak with one voice, etc.  Support for these small scale experiments must continue, and solutions to these challenges must be identified.

\subsection{Connections with High Energy Physics \& Atomic and Molecular Physics} 
\label{sect:hep}
 The DOE Office of Nuclear Physics and the NSF Nuclear Physics (NP) program both strongly support portfolios in fundamental symmetries and neutrinos.  These are generally discovery-oriented campaigns, which similarly drive the traditional field of high-energy particle (HEP) physics; that is, they address questions related to the Standard Model (SM) and beyond (BSM).  Both communities include experimental and theoretical efforts that establish the nature of the SM itself.  This includes determination of the masses of the fundamental particles, the couplings of the forces, the structure of the interactions, the generational structure of the quarks and their hadronic manifestations, and the properties of the neutrinos. 

The primary tools used in the HEP approach to probe BSM physics involve lepton or hadron colliders.  The interaction regions are surrounded by generic, multi-purpose detectors that can reconstruct a wide variety of event types to search for new physics. The recent discovery of the Higgs, and the earlier discoveries of the $W$ and $Z$ bosons are wonderful examples. The tools used in NP are rather different. Instead of the direct approach to producing new high-mass particles,``Precision Frontier" experiments employ the phenomenon of quantum  
fluctuations 
to tease out the effects of very high mass candidate BSM particles as they might impact a delicate measurement. Very often, the mass scales probe greatly exceed anything that can be produced directly using a collider.  Examples of this approach include the measurement of the Muon anomalous magnetic moment, the running of the weak proton charge, the violation of lepton flavor universality, the unitarity of the CKM mixing matrix, probes of CP violation, and tests of scalar or tensor components in the weak interaction. In each case, a unique, purpose-built experiment is needed; rarely do they provide a broad range of investigations. The HEP and NP approaches are complementary and both are required to investigate potential BSM models, providing needed clues and setting stringent limits.  In this Long Range Plan (LRP) period, several neutron decay experiments will provide data to test CKM unitarity, MOLLER will provide a precision determination of $\sin^{2}\Theta_{W}$, the neutron electron dipole moment (see below) will be probed, atoms and molecules will be used to study CP-violation in the nucleus, and new initiatives in beta and pion decay will be explored.

The NP and HEP communities also overlap in the field of neutrino physics where the dividing line is somewhat differently drawn.  Neutrinos from accelerators and reactors are typically the venue of HEP where they study oscillation parameters, the mass hierarchy, sterile neutrinos, possible CP and electroweak physics, and coherent {\it nuclear} scattering; the latter is also an emerging NP topic.  Neutrinos from the sun and from natural radioactive sources tend to drive the NP program. While the SNO experiment and others led to the oscillation model and proof of neutrino mass, a current NP focus is on directly measuring that mass.  This campaign is ongoing with KATRIN and Project\,8.  One of the most exciting initiatives in neutrino physics is the test of the Majorana nature of the neutrino.  A campaign of neutrinoless double beta decay experiments, with every increasing sensitivity, will directly probe lepton number violation, an important possible route needed to explain the baryon asymmetry of the universe.  In this Long Range Plan, the major ton-scale experiments LEGEND, CUPID, and nEXO each hope to proceed along their design plans to build experiments that probe the complete inverted hierarchy.  

It would be remiss if one did not recognize the significant technological overlap of these fields.  These include advances in accelerator physics, Monte Carlo modeling using GEANT, high-speed micro-electronics and ASIC readouts, wire chambers, calorimetry, detector physics, superconducting magnets, ROOT data analyses, just to mention a few.  One can hardly find an area of experimental technique that is solely owned or used by only one community.

This support has also started to extend into the realm of atomic, molecular, and optical (AMO) physics, including for example searches for the electron EDM, hadronic CP-violating moments, parity violation, axions and axion-like particles, ultralight dark matter, and new forces.  These efforts are of the precision measurement variety, and have both benefited from and contributed to the advances in the ability of AMO systems to leverage quantum control for maximum sensitivity.

Since AMO-based searches for the electron EDM became the most sensitive probe in the early 1960s, their sensitivity has improved by ten orders of magnitude, including two orders of magnitude since molecular experiments surpassed atomic ones in the last decade, driven by advances in AMO science.  Many of these same advances are currently being translated to a much broader range of NP science, including hadronic CP- and P-violation searches.  At the same time, much of the motivation for advancing AMO science, particularly in the areas of controlling complex species (including molecules), extending coherence times in applied fields, coherent control, new trapping technologies, controlling environmental interactions, and more, came from the need of these techniques for precision measurement of fundamental symmetries.   There are also areas of research where the synergy between AMO and NP is not merely beneficial, but necessary.  For example, AMO experiments using short-lived radioactive nuclei, which includes a broad range of hadronic CP- and P-violation searches, precision $\beta-$decay studies, and tailoring nuclear structures to make improved qubits, rely on the NP community.  The production, handling, and study of these species requires NP experimental expertise, and the experimental design and interpretation of results requires NP theoretical expertise.  

This strong synergy between precision measurement motivated by NP science goals, and the broader AMO community, has been extremely beneficial and is poised to grow significantly as interest in AMO-based precision measurements increases. In terms of fraction of AMO or NP communities, their overlap is still small, but has been rapidly growing both in numbers and impact.  Much of what is discussed in this LRP period is focused on actively fostering the interactions between those communities to foster and accelerate the science.

\newpage


\providecommand{\noopsort}[1]{}\providecommand{\singleletter}[1]{#1}%

\end{document}